%% file: main.tex
\documentclass{sig-alternate}

\usepackage{latexsym, mathrsfs}
\usepackage{graphicx}
\usepackage[T1]{fontenc}
\usepackage[usenames,dvipsnames,svgnames,table]{xcolor}
\usepackage[linesnumbered,ruled,procnumbered]{algorithm2e}
\usepackage{times}
\usepackage{paralist}
\usepackage{multirow}
\usepackage{xspace}
\usepackage{url}
\usepackage{subcaption} 
\usepackage{hhline}
\usepackage{algorithm2e}
\usepackage[noend]{algpseudocode}
\usepackage{balance}
\usepackage{float}
\usepackage[normalem]{ulem}
\usepackage{caption}
\usepackage{wrapfig}

\usepackage{amsmath}

\newcommand{\stitle}[1]{\vspace{0.5em}\noindent\textbf{#1}}

\numberwithin{equation}{section}
\newtheorem{theorem}{\textbf{Theorem}}
\newtheorem{lemma}{Lemma}

\newtheorem{example}{Example}

\newtheorem{observation}{\textbf{Observation}}

\newtheorem{problem}{\textbf{Problem}}

\newcommand{\orpheus}{{\sc OrpheusDB}\xspace}
\newcommand{\postgres}{{PostgreSQL}\xspace}

\newcommand{\sql}{{\sc SQL}\xspace}
\newcommand{\appr}{{\sc LyreSplit}\xspace}
\newcommand{\aggl}{{\sc Agglo}\xspace}
\newcommand{\kmeans}{{\sc Kmeans}\xspace}

\newcommand{\vv}{\mathcal{V}\xspace}

\newcommand{\ee}{\mathcal{E}\xspace}
\newcommand{\cc}{\mathcal{C}\xspace}
\newcommand{\rr}{\mathcal{R}\xspace}
\newcommand{\calS}{\mathcal{S}\xspace}
\newcommand{\cavg}{\cc_{avg}\xspace}

\newcommand{\pp}{\mathcal{P}\xspace}
\newcommand{\ebb}{\mathbb{E}\xspace}
\newcommand{\gbb}{\mathbb{G}\xspace}
\newcommand{\vbb}{\mathbb{V}\xspace}
\newcommand{\tbb}{\mathbb{T}\xspace}
\newcommand{\cvd}{{\sc cvd}\xspace}
\newcommand{\cvds}{{\sc cvd}s\xspace}

\newcommand{\eat}[1]{}
\newcommand{\papertext}[1]{} 
\newcommand{\techreport}[1]{#1}
\newcommand{\later}[1]{}

\newcommand{\code}[1]{{\small {\sf #1}}}
\newcommand{\para}[1]{\smallskip \noindent {\bf #1.}}
\newcommand{\vlist}{{\em vlist}\xspace}
\newcommand{\rlist}{{\em rlist}\xspace}

\newcommand{\rid}{{\em rid}\xspace}
\newcommand{\vid}{{\em vid}\xspace}
\newcommand{\rids}{{\em rid}s\xspace}

\makeatletter
\algrenewcommand\ALG@beginalgorithmic{\scriptsize}
\makeatother

\newcommand{\techreporttext}[1]{}

\newenvironment{denselist}{
    \begin{list}{\small{$\bullet$}}%
    {\setlength{\itemsep}{0ex} \setlength{\topsep}{0ex}
    \setlength{\parsep}{0pt} \setlength{\itemindent}{0pt}
    \setlength{\leftmargin}{1.5em}
    \setlength{\partopsep}{0pt}}}%
    {\end{list}}

\makeatletter
\def\@copyrightspace{\relax}
\makeatother

\begin{document}

\title{{\LARGE \orpheus}: Bolt-on Versioning for Relational Databases}

\numberofauthors{5}


\author{Silu Huang$^1$, Liqi Xu$^1$, Jialin Liu$^2$, Aaron J.~Elmore$^3$, Aditya Parameswaran$^1$ \\
\affaddr{\hspace{2em}$^1$University of Illinois (UIUC) \hspace{5.5em} $^2$Peking University \hspace{4em} $^3$University of Chicago} \\
\affaddr{\{shuang86,liqixu2,adityagp\}}@illinois.edu\; \hspace{1em} 1300012936@pku.edu.cn\; \hspace{1em}
\affaddr{aelmore}@cs.uchicago.edu
}

\maketitle


\input{abstract}
\input{intro}

\input{syst_arch}
\input{storage}

\input{optimization}
\input{evaluation}
\input{related}

\input{conclusion}




\balance
{\scriptsize
\bibliographystyle{abbrv}
\bibliography{all}
}

\techreport{
\appendix}

\techreport{\input{extension}

\input{extension_2}

\section{Additional Experiments} \label{sec:extra_exp}
\input{evaluation_cost_model}
\input{estimated-vs-actual}
}






\end{document}

%% file: abstract.tex

\begin{abstract}

Data science teams often collaboratively analyze
datasets, generating dataset versions at each stage
of iterative exploration and analysis.
There is a pressing need for a system
that can support dataset versioning,
enabling such teams to efficiently store, track,
and query across dataset versions. \techreport{While {\tt git} and {\tt svn} are 
highly effective at managing code,
they are not capable of managing
large unordered structured datasets
efficiently, nor do they support analytic (\sql)
queries on such datasets.}
We introduce \orpheus,
a dataset version control system
that ``{\em bolts on}''
versioning capabilities to a 
traditional relational database system,
thereby gaining the analytics capabilities 
of the database ``for free''\techreport{, while the database itself 
is unaware of the presence of dataset versions}.
We develop and evaluate multiple
data models for representing
versioned data, as well as a light-weight partitioning
scheme, \appr, to further optimize the models for reduced query latencies.
With \appr,
\orpheus is on average $10^3\times$ faster in finding
effective (and better) partitionings than competing approaches,
while also reducing the latency of version retrieval
by up to $20\times$ relative to schemes without
partitioning.
\appr can be applied in an online fashion
as new versions are added, alongside an intelligent migration scheme
that reduces migration time by $10\times$ on average.
\end{abstract}

%% file: intro.tex

\section{Introduction}\label{sec:intro}

When performing data science,
teams of data scientists repeatedly transform their datasets
in many ways, by normalizing, cleaning, editing,
deleting, and updating one or more data items at a time;
the New York Times defines data science as a
{\em step-by-step process of experimentation on data}~\cite{nyt-janitor}.
The dataset versions generated, often into the hundreds or thousands,
are stored in an ad-hoc manner,
typically via copying and naming conventions
in shared (networked) file systems.
This makes it impossible to effectively manage, make sense of,
or query across these versions.
One alternative is to use a source
code version control system like {\tt git} or {\tt svn}
to manage dataset versions.
However, source code version control systems
are both inefficient at
storing unordered structured datasets,
and do not support advanced querying capabilities,
e.g., querying for versions that satisfy some predicate,
performing joins across 
versions,
or computing some aggregate statistics across versions~\cite{bhardwaj2014datahub}.
Therefore, when requiring
advanced (SQL-like) querying capabilities,
data scientists typically store each of the
dataset versions as independent tables in a
traditional relational database. 
This approach results
in massive redundancy and inefficiencies in storage,
as well as manual supervision and maintenance to track versions.
As a worse alternative, they only store the
most recent versions---thereby
losing the ability to retrieve the original
datasets or trace the provenance of the new versions.

A concrete example of this phenomena occurs with biologists
who operate on shared datasets, such as a gene annotation dataset~\cite{gene2015gene} or
a protein-protein interaction dataset~\cite{szklarczyk2011string}, both of which are
rapidly evolving, by
periodically checking out versions, performing local analysis,
editing, and cleaning operations,
and committing these versions into a branched network of
versions.
This network of versions is also often repeatedly
explored and queried for global statistics and differences (e.g., the aggregate
count of protein-protein tuples with confidence in interaction greater than $0.9$, for each version)
and for versions with specific properties (e.g., versions with
a specific gene annotation record, or versions with ``a bulk delete'',
ones with more than 100 tuples deleted from their parents).


While recent work has outlined a vision for collaborative
data analytics and versioning~\cite{bhardwaj2014datahub},
and has developed solutions for dataset
versioning from the ground up~\cite{maddox2016decibel,bhattacherjee2015principles},
these papers offer partial solutions, require redesigning the entire database stack,
and as such cannot benefit from the querying capabilities that exist in current database systems.
Similarly, while temporal databases~\cite{tansel1993temporal,jensen1999temporal,ozsoyoglu1995temporal,kulkarni2012temporal} offer functionality to revisit instances at various
time intervals on a linear chain of versions, they do not support the full-fledged branching and merging
essential in a collaborative data analytics context,
and the temporal functionalities offered and concerns are very different.
We revisit related work in Section~\ref{sec:related}.



The question we ask in this paper is: {\em can
we have the best of both worlds---advanced
querying capabilities, plus effective and efficient
versioning in a mature relational database}? More specifically,
{\bf \em can traditional relational databases be made
to support versioning for collaborative data analytics?}

To answer this question we develop a system, \orpheus\footnote{\small Orpheus is a
musician and poet from ancient Greek mythology with the ability to
raise the dead with his music, much like \orpheus has
the ability to retrieve
old (``dead'') dataset versions on demand.},
to {\em ``bolt-on''} versioning capabilities to a
traditional relational
database system that is unaware of the existence
of versions.
By doing so, we seamlessly leverage
the analysis and querying capabilities that
come ``for free'' with a database system,
along with efficient versioning capabilities.
Developing \orpheus comes with a host
of challenges, centered around the choice of the representation
scheme or the data model used to capture versions
within a database, as well as effectively balancing
the storage costs with the costs for querying and operating
on versions.
We describe the challenges associated with the data model first.

\begin{figure*}[t!]
\vspace{-10pt}
\includegraphics[width=\linewidth]{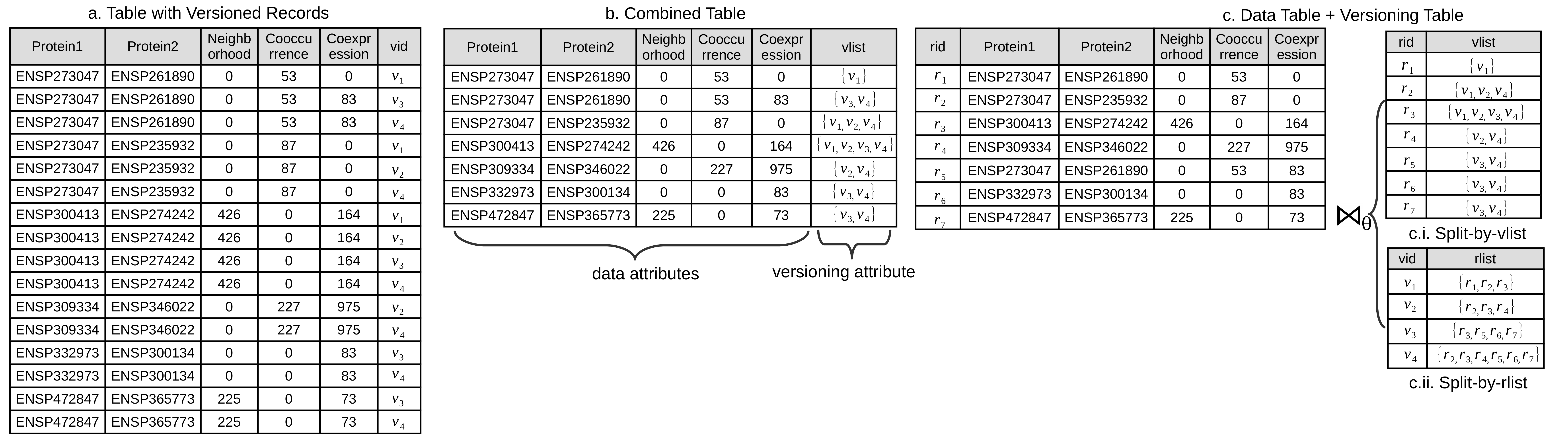}
\vspace{-30pt}
 \caption{Data models for protein interaction data~\cite{szklarczyk2011string}}
\label{fig:datamodels}
\vspace{-15pt}
\end{figure*}

\stitle{Challenges in Representation.}
One simple approach of capturing dataset versions
would be to
represent the dataset as a relation in a database, and
add an extra attribute corresponding to the version number,
called \vid, as shown in Figure~\ref{fig:datamodels}(a) for 
simplified protein-protein interaction data~\cite{szklarczyk2011string};
the other attributes will be introduced later.
The version number attribute allows us to apply selection
operations to retrieve specific versions.
However, this approach is extremely wasteful as each record
is repeated as many times as the number of versions it
belongs to.
It is worth noting that a timestamp is not sufficient here,
as a version can have multiple parents (a merge)
and multiple children (branches).
\techreport{Therefore, a scalar timestamp value cannot capture
which versions a tuple belongs to.}
To remedy this issue, one can use the {\tt array}
data type capabilities offered in current database systems,
by replacing the version number attribute
with an array attribute {\em vlist} containing all of the versions that
each record belongs to, as depicted in Figure~\ref{fig:datamodels}(b).
This reduces storage overhead from replicating tuples.
However, when adding a new version (e.g.,
a clone of an existing version) this approach leads to
extensive modifications across the entire relation,
since the array will need to be updated
for every single record that belongs to the new version.
Another strategy is to separate the data
from the versioning information into two tables as in Figure~\ref{fig:datamodels}(c),
where the first table---the data table---stores
the records appearing in any of the versions,
while the second table---the versioning table---captures
the versioning information, or which version contains
which records.
This strategy requires us to perform a join
of these two tables to retrieve any versions.
Further, there are two ways of recording the versioning information:
the first involves using an array of versions,
the second involves using an array of records; we illustrate this in
Figure~\ref{fig:datamodels}(c.i) and Figure~\ref{fig:datamodels}(c.ii) respectively.
The latter approach allows easy insertion of new
versions, without having to modify existing version information,
but may have slight overheads relative to the former approach when
it comes to joining the versioning table and the data table.
Overall, as we demonstrate in this paper, the latter approach
outperforms other approaches
(including those based on recording deltas)
for most common operations.

\stitle{Challenges in Balancing Storage and Querying Latencies.}
Unfortunately, the previous approach still requires a full theta join
and examination of all of the data to reconstruct any given version.
Our next question is
if we can improve the efficiency
of the aforementioned approach, at the cost of possibly additional storage.
One approach is to partition the
versioning and data tables such that we limit
data access to recreate
versions, while keeping storage costs bounded.
However, as we demonstrate in this paper,
the problem of identifying
the optimal trade-off between the storage
and version retrieval time is {\sc NP-Hard},
via a reduction from the {\sc 3-Partition}
problem.
To address this issue, we develop
an efficient and light-weight
approximation algorithm, \appr,
that enables us to trade-off
storage and version retrieval
time, providing
a guaranteed
$((1+\delta)^{\ell}, \frac{1}{\delta})$-factor
approximation
under certain reasonable assumptions---where the storage
is a $(1+\delta)^{\ell}$-factor of optimal,
and the average version retrieval time is $\frac{1}{\delta}$-factor
of optimal,
for any value of parameter $\delta\leq 1$
that expresses the desired trade-off.
The parameter $\ell$ depends on
the complexity of the branching structure of the
version graph.
In practice, this algorithm
always leads to lower retrieval times
for a given storage budget,
than other schemes
for partitioning, while being
 about $1000\times$
faster than these schemes.
Further, we adapt \appr
to an online setting that incrementally maintains partitions 
as new versions arrive, 
and develop an intelligent migration approach to minimize the time taken for migration (by up to $10\times$). 



\stitle{Contributions.}
The contributions of this paper are as follows:
\begin{denselist}
        \item We develop a dataset version control system, titled \orpheus, with the ability to support both git-style version control commands and \sql-like queries. (Section~\ref{sec:overview})
        \item We compare different data models for representing versioned datasets and evaluate their performance in terms of storage consumption and time taken for querying. (Section~\ref{sec:models})
        \item To further improve query efficiency, we formally develop the optimization problem of trading-off between the storage and version retrieval time via partitioning
        and demonstrate that this is {\sc NP-Hard}.
        We then propose a light-weight approximation algorithm for this
        optimization problem, titled \appr, providing a $((1+\delta)^{\ell}, \frac{1}{\delta})$-factor guarantee. (Section~\ref{one_to_many_alg} and \ref{ssec:prob})
        \item We further adapt \appr to be applicable to an online setting with new versions coming in, and develop an intelligent migration approach. (Section~\ref{ssec:inc})
        \item We conduct extensive experiments using a
        versioning benchmark~\cite{maddox2016decibel} and
        demonstrate that \appr is on average 1000$\times$ faster
        than competing algorithms and performs better in balancing
        the storage and version retrieval time. We also demonstrate that our intelligent migration scheme reduces the migration time by $10\times$ on average. (Section~\ref{sec:evaluation})
\end{denselist}

%% file: syst_arch.tex
\section{\large{\orpheus} \textbf{Overview}} \label{sec:overview}

\orpheus is a dataset version management system
that is built on top of
standard relational databases. 
It inherits much of the same benefits of
relational databases, while also
compactly storing, tracking, and recreating versions
on demand. \orpheus has been developed as open-source software (\url{orpheus-db.github.io}). 
We now describe fundamental version-control
concepts, followed by the \orpheus APIs, and finally, the design of \orpheus.

\subsection{Dataset Version Control}
The fundamental unit of storage within \orpheus
is a {\em collaborative versioned dataset}
({\em\cvd}) to which
one or more users can contribute.
Each \cvd corresponds to a relation
and implicitly contains many {\em versions} of that
relation.
A {\em version} is an instance
of the relation, specified by the user
and containing a set of records.
Versions within a \cvd are
related to each other via a {\em version graph}---a
directed acyclic graph---representing
how the versions
were derived from each other:
a version in this graph with two or
more parents is defined to be a {\em merged version}.
Records in a \cvd are {\em immutable}, i.e.,
any modifications to any record attributes
result in a new record, and are
stored and treated separately within the \cvd.
Overall, there is a many-to-many
relationship between records and versions:
each record can belong to many versions,
and each version can contain many records.
Each version has a unique version id, {\it vid},
and each record has its unique record id, {\it rid}.
The record ids are used to identify immutable records
within the \cvd and are not visible to end-users of \orpheus.
In addition, the relation corresponding
to the \cvd may have primary key attribute(s);
this implies that for any version
no two records can have the same values for the
primary key attribute(s). \techreport{However, across versions, this need not be the case.}
\orpheus can support multiple \cvds at a time.
However, in order to better convey the core ideas of \orpheus,
in the rest of the paper, we focus our discussion
on a single \cvd.


\subsection{{\large{\orpheus}} APIs}\label{ssec:interface}
Users interact with \orpheus via the command line,
using both SQL queries, as well as git-style
version control commands.
In our companion demo paper, we also describe
an interactive user interface depicting the version graph,
for users to easily explore and operate on dataset versions~\cite{orpheusdemo}.
To make modifications to versions, users
can either use SQL operations issued to the
relational database that \orpheus is built on top of,
or can alternatively operate on them
using programming or scripting languages.
We begin by describing the version control commands.

\para{Version control commands}
Users can operate on \cvds much like they would with
source code version control. The first 
operation is
{\em checkout}: this command materializes
a specific version of a \cvd as a newly created
regular table within a relational
database that \orpheus is connected to.
The table name is specified within the checkout command, as follows:
\vspace{-5pt}
\begin{equation*}
{\sf checkout ~~[\cvd] ~~ \text{-}v ~ [vid] ~~\text{-}t ~[table~ name]}
\vspace{-5pt}
\end{equation*}
Here, the version with id {\vid} is materialized
as a new table {\sf [table name]}
within the database, to which standard SQL
statements can be issued, and
which can later be added to the \cvd as a new version.
The version from which this table was derived (i.e., {\vid}) is
referred to as the {\em parent version} for the table.

Instead of materializing one version at a time,
users can materialize multiple versions, by listing
multiple {\em vid}s in the command above,
essentially {\em merging} multiple versions
to give a single table.
When merging, the records in the versions
are added to the table
in the precedence order listed after {\sf -v}:
for any record being added, if another record
with the same primary key has already been added, it is
omitted from the table.
This ensures that the eventual materialized table
also respects the primary key property.
There are other conflict-resolution strategies,
such as letting users resolve conflicted records manually;
for simplicity, we use a precedence based approach.
Internally, the {checkout} command
records the versions that this table was derived from\techreport{~(i.e., those listed after {\sf -v})},
along with the
table name.
Note that only the user who performed the checkout operation is
permitted access to the materialized table, so
they can perform any analysis and modification on this
table without interference from other users, only making
these modifications visible when they use the {\em commit} operation,
described next.

The {\em commit} operation adds a new version to the
\cvd, by making the local changes made by the user on their
materialized table visible to others.
The commit command has the following format:
\vspace{-5pt}
\begin{equation*}
{\sf commit ~~ \text{-}t ~ [table~ name] ~~ \text{-}m ~[commit ~ message]}
\vspace{-5pt}
\end{equation*}
The command does not need to specify
the intended \cvd since \orpheus internally
maintains a mapping between the table name
and the original \cvd.
In addition, since the versions that
the table was derived from originally during checkout
are internally known to \orpheus, the table is added to the
\cvd as a new version with those versions as parent versions.
During the commit operation, \orpheus \techreport{checks the primary key constraint if PK is specified, and }compares the
(possibly) modified materialized table to
the parent versions.
If any records were added or modified
these records
are treated as new records and added to the \cvd.
\techreport{(Recall that records are immutable within a \cvd.)}
An alternative is to compare the new records
with all of the existing records in the \cvd
to check if any of the new records have existed
in any version in the past,
which would take longer to execute.
At the same time, the latter approach would identify
records that were deleted then re-added later.
Since we believe that this is not a common case,
we opt for the former approach, which would only lead
to modest additional storage at the cost of much less
computation during commit.
We call this the {\em no cross-version diff} implementation rule.
Lastly, if the schema of the table that is being committed
is different from the \cvd it derived from,
we alter the \cvd to incorporate the new schema; we discuss this in Section~\ref{ssec:version_graph}\papertext{.}\techreport{, but for most of the paper we consider the static schema case.}



In order to support data science workflows, we additionally
support the use of {\em checkout} and {\em commit} into
and from {\sf csv}
(comma separated value)
files via slightly different flags:
{\sf -f} for {\sf csv} instead of {\sf -t}.
The {\sf csv} file can be
processed in external tools and programming
languages such as Python or R,
not requiring that users perform the modifications and analysis
using {\sf SQL}.
However, during commit, the user is expected to also provide
a schema file via a {\sf -s} flag
so that \orpheus can make sure that the columns are mapped
in the correct manner.
\techreport{An alternative would be to use schema inference tools,
e.g., \cite{miller2001clio,fisher2008dirt}, which could be seamlessly
incorporated if need be.}
Internally, \orpheus also tracks the name of the {\sf csv}
file as being derived from one or more versions of the \cvd,
just like it does with the materialized tables.

In addition to checkout and commit,
\orpheus also supports other commands, described
very briefly here:
\begin{inparaenum}[\itshape (a)\upshape]
\item {\em diff:} a standard differencing operation
that compares two versions
and outputs the records in one but not the other.
\item{\em init:} initialize either an external csv file
or a database table as a new \cvd in \orpheus.
\item{\em create\_user, config, whoami:} allows users to register, login,
and view the current user name.
\item{\em ls, drop}: list all the \cvds
or drop a particular \cvd.
\item {\em optimize:} 
as we will see later, 
\orpheus can benefit from intelligent incremental
partitioning schemes (enabling operations
to process much less data).
Users can setup the corresponding parameters 
(e.g., storage threshold, tolerance factor, described later)
via the command line; the \orpheus backend will periodically invoke the
partitioning optimizer
to improve the versioning performance. 
\end{inparaenum}

\para{SQL commands}
\orpheus supports the use of SQL commands
on \cvds via
the command line using the {\em run} command,
which either takes a SQL script as input or the SQL statement as a string.
Instead of materializing a version (or versions) as a table
via the checkout command
and explicitly
applying SQL operations on that table,
\orpheus also allows users to directly execute SQL
queries on a specific version, using special
keywords \code{VERSION}, \code{OF}, and \code{CVD}
via syntax \vspace{-5pt}
\begin{quote}
\code{SELECT ... FROM VERSION [vid] OF CVD [cvd], ...} \vspace{-5pt}
\end{quote}
without having to materialize it.
Further, by using renaming,
users can operate directly on multiple versions
(each as a relation) within
a single SQL statement,
enabling operations such as joins across
multiple versions.

However, listing each version individually
as described above
may be cumbersome for some types of queries
that users wish to run, e.g.,
applying an aggregate
across a collection of versions,
or identifying versions that
satisfy some property. For this, \orpheus also supports constructs that enable users to issue aggregate queries across \cvds grouped by version ids, or select version ids that satisfy certain constraints.  Internally, these constructs are translated into regular SQL queries that can be executed by the underlying database system.
In addition, \orpheus provides shortcuts for several types of queries
that operate on the version graph,
e.g., listing the descendant or ancestors
of a specific version,
or querying the metadata,
e.g., identify the last modification (in time)
to the \cvd.
The details of the query syntax, translation, as well as
 examples can be found in our companion demo paper~\cite{orpheusdemo}.

\subsection{System Architecture}\label{ssec:sysop}
\begin{figure}[h!]
\vspace{-10pt}
\centering
\includegraphics[width=0.8\linewidth]{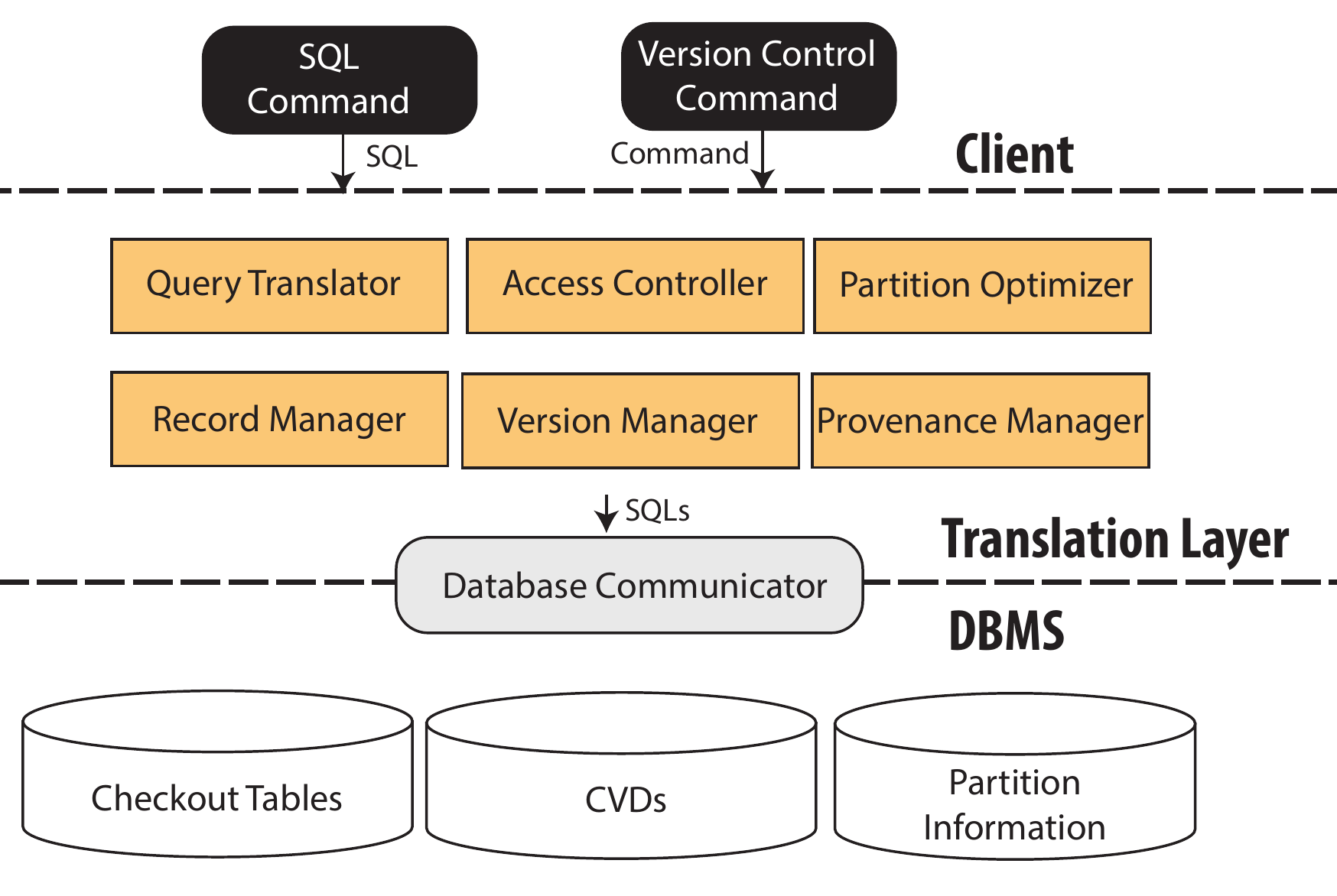}
\vspace{-12pt}
\caption{\orpheus Architecture}
\label{fig:system_arch}
\vspace{-15pt}
\end{figure}

We implement \orpheus as a middleware layer
or wrapper between end-users (or application programs)
and a traditional relational database system---in our case, \postgres.
\postgres is completely unaware of the existence
of versioning, as versioning is handled entirely
within the middleware.
Figure~\ref{fig:system_arch} depicts the overall architecture
of \orpheus.
\orpheus consists of six core modules:
the \textit{query translator} is responsible
for parsing input and translating it into SQL statements
understandable by
the underlying database system;
the \textit{access controller}  monitors user permissions
to various tables and files within \orpheus;
the \textit{partition optimizer} is responsible
for periodically
reorganizing and optimizing the partitions via
a partitioning algorithm \appr along with a {\em migration engine} to migrate data from one partitioning scheme to another, 
and is the focus of Section~\ref{sec:partition};
the \textit{record manager} is in charge of recording and retrieving information about records in \cvds;
the \textit{version manager} is in charge of recording and retrieving versioning information,
including the \rids each version contains as well as the metadata for each version;
and the \textit{provenance manager} is responsible for
the metadata of uncommitted tables or files, such as their parent version(s) and the creation time.
At the backend, a traditional DBMS, we
maintain \cvds that consist of
versions, along with the records they contain,
as well as metadata about versions.
In addition, the underlying DBMS
contains a temporary staging
area consisting of all of the materialized tables
that users can directly manipulate via SQL without
going through \orpheus.
Understanding how to best represent and operate
on these \cvds within the underlying DBMS 
is an important challenge---this is the 
focus of the next section.



\techreport{
In brief, we now describe how these components
work with each other for the basic checkout and commit commands,
once the command is parsed.
For checkout,
the query translator generates SQL queries to retrieve
records from the relevant versions, which are
then handled and materialized in the temporary staging area
by the record manager;
the provenance manager logs
the related derivation information
and other metadata;
and finally the access
controller to grant permissions to the relevant user.
On commit, the record manager
appends new records to the \cvd,
also performs cleanup by removing the table from the staging area; the version manager updates the metadata
of the newly added version.
}





%% file: storage.tex

\begin{table*}[!t]
\centering
\scriptsize
\vspace{-6pt}
\begin{tabular}{|l|l|l|l|}
  \hline
   Command & \sql Translation with combined-table & \sql Translation with Split-by-vlist &\sql Translation with Split-by-rlist \\
  \hline
  \hline
  CHECKOUT & \begin{tabular}{@{}l@{}} \texttt{SELECT * into T' FROM T } \\ \texttt{WHERE ARRAY[$v_i$] <@ vlist} \end{tabular} &  \begin{tabular}{@{}l@{}l@{}l@{}} \texttt{SELECT * into T' FROM dataTable, } \\ \texttt{(SELECT rid AS rid\_tmp } \\ \texttt{FROM versioningTable} \\ \texttt{WHERE ARRAY[$v_i$] <@ vlist) AS tmp} \\ \texttt{WHERE rid = rid\_tmp} \end{tabular} &
  \begin{tabular}{@{}l@{}l@{}l@{}} \texttt{SELECT * into T' FROM dataTable, } \\ \texttt{(SELECT unnest(rlist) AS rid\_tmp } \\ \texttt{FROM versioningTable} \\ \texttt{WHERE vid = $v_i$) AS tmp} \\ \texttt{WHERE rid = rid\_tmp} \end{tabular}\\
  \hline
  COMMIT & \begin{tabular}{@{}l@{}} \texttt{UPDATE T SET vlist=vlist+$v_j$} \\ \texttt{WHERE rid in } \\ \texttt{(SELECT rid FROM T')} \end{tabular} & \begin{tabular}{@{}l@{}} \texttt{UPDATE versioningTable } \\ \texttt{SET vlist=vlist+$v_j$} \\ \texttt{WHERE rid in } \\ \texttt{(SELECT rid FROM T')} \end{tabular} & \begin{tabular}{@{}l@{}} \texttt{INSERT INTO versioningTable} \\ \texttt{VALUES ($v_j$, } \\ \texttt{ARRAY[SELECT rid FROM T'])}\end{tabular} \\
  \hline
\end{tabular}
\vspace{-10pt}
\caption{\sql Queries for {Checkout} and {Commit} Commands with Different Data Models}
\label{table:SQL}
\end{table*}

\section{Data Models for {\Large \cvds}}\label{sec:models}

In this section,
we consider and compare methods to represent
and operate on \cvds
 within
a backend relational database,
starting with the data within versions,
and then the metadata about versions.

\subsection{Versions and Data: The Models}\label{ssec:RM}

To explore alternative storage models,
we consider the array-based data models,
shown in
Figure~\ref{fig:datamodels}, and compare them to a delta-based data model, which we describe later.
The table(s) in Figure~\ref{fig:datamodels} displays 
simplified protein-protein interaction data~\cite{szklarczyk2011string}, 
and has a composite primary key {\em <protein1, protein2>},
along with numerical attributes indicating sources and strength of interactions: 
{\em neighborhood} represents how frequently the two proteins occur close 
to each other in runs of genes,
{\em cooccurrence} reflects how often the two proteins 
co-occur in the species, 
and {\em coexpression} refers to the
level to which genes are co-expressed in the species.

One approach, as described in the introduction, is to augment
the \cvd's relational schema
with an additional versioning attribute.
For example, in Figure~\ref{fig:datamodels}(a) the
tuple of <ENSP273047, ENSP261890, 0, 53, 83> exists in two versions: $v_3$ and $v_4$.
(Note that even though <protein1, protein2> is the primary key,
it is only the primary key for any single version and
not across all versions.)
\techreport{There are two records with <ENSP273047, ENSP261890> that have different
values for the other attributes: one with (0, 53, 83) that is present in $v_3$ and $v_4$,
and another with (0, 53, 0) that is present in $v_1$.}
However, this approach implies that
we would need to duplicate each record as many times
as the number of versions it is in, leading to
severe storage overhead due to redundancy, as well as
inefficiency for several operations, including
checkout and commit.
We focus on alternative approaches that are more space efficient
and discuss how they can support
the two most fundamental operations---commit and checkout---on
a single version at a time.
Considerations for multiple version checkout is similar to
that for a single version; our findings generalize to that case as well.

\para{Approach 1: The Combined Table Approach}
Our first approach of representing the data and versioning
information for a \cvd is the \textit{combined table approach}.
As before, we augment the schema with an additional
versioning attribute, but now, the versioning attribute
is of type \code{array}
and is named \vlist (short for version list) as shown
in Figure~\ref{fig:datamodels}(b). For each record the \vlist
is the ordered list of version ids
that the record is present in, which
 serves as an inverted index for each record.
Returning to our example, there are two
versions of records corresponding to
<ENSP273047, ENSP261890>, with coexpression 0 and 83 respectively---these two versions
are depicted
as the first two records,  with an array corresponding
to $v_1$ for the first record, and $v_3$ and $v_4$
for the second.

Even though \code{array} is a non-atomic
data type, it is commonly supported in many database
systems~\cite{postgres9.5intarray,db2Array,mysqlAddArray};
thus
\orpheus can be built with any of these systems
as the back-end database.
As our implementation uses \postgres,
we focus on this system for the rest of the discussion,
even though similar considerations apply to the
rest of the databases listed.
\techreport{\postgres provides a number of useful functions
and operators
for manipulating arrays, including append
operations, set operations, value containment operations,
and sorting and counting functions.}


For the combined table approach,
committing a new version to the \cvd is
time-consuming due to the expensive append operation
for every record present in the new version.
Consider the scenario where the user checks out version $v_i$
into a materialized table $T'$ and then immediately
commits it back as a new version $v_j$.
The query translator parses the user commands and generates the
corresponding \sql queries for {checkout} and {commit}
as shown in Table~\ref{table:SQL}.
\techreport{In the checkout statement, the containment operator \code{`int[] <@ int[]'} returns true
if the array on the left is contained within the array on the right.}
When checking out $v_i$ into a materialized table $T'$,
the array containment operator \code{`ARRAY[$v_i$] <@ vlist'}
first examines whether $v_i$ is contained in \vlist for each record in \cvd,
then all records that satisfy that condition are added to the materialized table $T'$.
Next, when $T'$ is committed back to the \cvd as a new version $v_j$,
for each record in the \cvd, if it is also present in $T'$ (i.e., the \code{WHERE} clause),
we append $v_j$ to the attribute \vlist (i.e., \code{vlist=vlist+$v_j$}).
In this case, since there are no new records that are added to the \cvd,
no new records are added to the combined table.
However, even this process of appending $v_j$ to \vlist
can be expensive especially when the number of records
in $v_j$ is large, as we will demonstrate.

\para{Approach 2: The Split-by-vlist Approach}
Our second approach addresses the limitations
of the expensive commit operation for the combined table approach.
We store two tables,
keeping the versioning information separate from the data
information,
as depicted in Figure~\ref{fig:datamodels}(c)---the \emph{data table}
and the {\em versioning table}.
The data table contains all of the original data attributes
along with an extra primary key \rid,
while the versioning table maintains the mapping between versions
and \rids.
The \rid attribute was not needed in the previous
approach since it was not necessary to associate
identifiers with the immutable records.
\techreport{Specifically, the relation primary key--- <protein1, protein2> ---is not
sufficient to distinguish between multiple copies of the same record.
For example, $r1$ and $r5$ are two versions of the same record
(i.e., the record with a given <protein1, protein2>).}
There are two ways we can store the versioning data.
The first approach is to store the \rid
along with the \vlist, as depicted in Figure~\ref{fig:datamodels}(c.i).
We call this approach \emph{split-by-vlist}.
Split-by-vlist has a similar \sql translation as combined-table for {commit},
while it incurs the overhead of joining the data table with the versioning table
for {checkout}. Specifically, we select the \rids
that are in the version to be checked out
and store it in the table \code{tmp},
followed by a join with the data table.
\techreport{For example, when checking out version $v_1$,
\code{tmp} will comprise the relevant \rids $r_1, r_2, r_3$, which
are identified by looking at the \vlist for each record in the versioning table
and checking if $v_1$ is present, which is then joined with the data table
to extract the appropriate results into the materialized table $T'$.}

\begin{figure*}[t!]
\vspace{-13pt}
\centering
\includegraphics[width=0.8\linewidth]{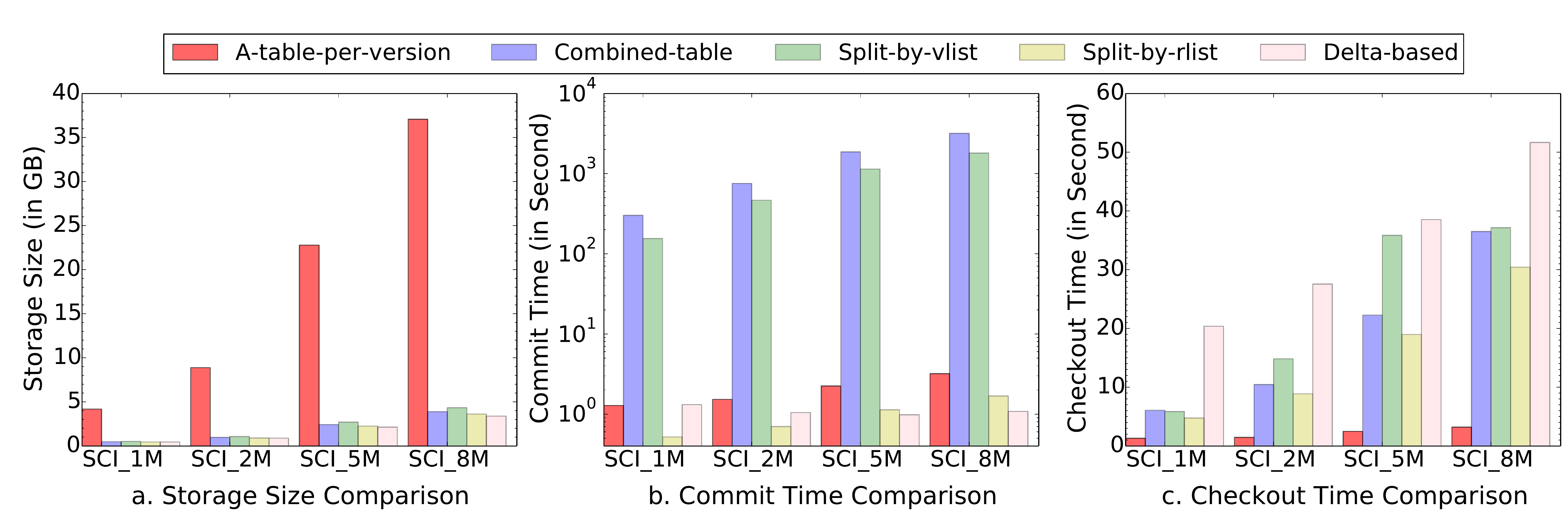}
\vspace{-8pt}
\caption{Comparison Between Different Data Models}
\label{fig:model_exp}
\vspace{-16pt}
\end{figure*}

\para{Approach 3: The Split-by-rlist  Approach}
Alternatively, we can organize the versioning table with
a primary key as \vid (version id),
and another attribute \rlist, containing the array of the records
present in that particular version, as in
Figure~\ref{fig:datamodels}(c.ii).
We call this approach the \emph{split-by-rlist} approach.
When committing a new version $v_j$ from the materialized table $T'$,
we only need to add a single tuple in the versioning table
with \vid equal to $v_j$,
and \rlist equal to the list of record ids in $T'$.
This eliminates the expensive array appending operations
that are part of the previous two approaches,
making the {commit} command much more efficient.
For the {checkout} command for version $v_i$,
we first extract the record ids associated
with $v_i$ from the versioning table,
by applying the unnesting operation: \code{unnest(\emph{rlist})},
following which we join the \rids
with the data table to identify all of the relevant records.
\techreport{For example, for checking out $v_1$, instead of examining
the entire versioning table, we simply
need to examine the tuple corresponding to $v_1$,
unnest those \rids---$r_1, r_2, r_3$, followed by a join.
}

\smallskip
So far, all our models support convenient rewriting
of arbitrary and complex versioning queries into SQL queries understood
by the backend database; see details in our demo paper~\cite{orpheusdemo}.
However, our delta-based model, discussed next,
does not support convenient rewritings for
some of the more advanced queries, e.g.,
``find versions where the total count of tuples with {\em protein1} as ENSP273047 is greater than 50'':
in such cases, delta-based model essentially needs to recreate
all of the versions, and/or perform extensive and expensive
computation outside of the database.
Thus, even though this model does not support
advanced analytics capabilities ``for free'',
we include it in our comparison to contrast its
performance
to the array-based models.

\para{Approach 4: Delta-based Approach}
Here,
each version records the modifications (or deltas) from
its precedent version(s).
Specifically,
each version is stored as a separate table,
with an added tombstone boolean attribute
indicating the deletion of a record.
In addition, we maintain a precedent metadata table
with a primary key \vid
and an attribute {\em base} indicating from
which version \vid stores the delta.
When committing a new version $v_j$,
a new table stores the delta from its previous version $v_i$.
If $v_j$ has multiple parents,
we will store $v_j$ as the modification from the parent
that shares the largest common number of records with $v_j$.
(Storing deltas from multiple parents would make
reconstruction of a version complicated, since we
would need to trace back multiple paths
in the version graph\techreport{, or alternatively materialize
each version in the version graph in a top-down manner,
merging versions based on conflict resolution mechanisms}.
Here, we opt for the simpler solution.)
A new record is then inserted into the metadata table,
with \vid as $v_j$ and {\em base} as $v_i$.
For the {\em checkout} command for version $v_i$,
we trace the version lineage (via the {\em base} attribute)
all the way back to the root.
If an incoming record has occurred before,
it is discarded; otherwise, if it is marked as ``insert'',
we insert it into the checkout table $T'$.

\para{Approach 5: The A-Table-Per-Version Approach}
Our final array-based data model is impractical
due to excessive storage, but is useful from a comparison standpoint.
In this approach, we store each version as a separate table.
We include a-table-per-version in our comparison;
we do not include the approach in Figure~\ref{fig:datamodels}a,
containing a table with duplicated records,
since it would do similarly
in terms of storage and
commit times to a-table-per-version, but worse in terms of checkout times.

\vspace{-1.5mm}
\subsection{Versions and Data: The Comparison}\label{ssec:datamodelExp}
We perform an experimental evaluation
between the approaches described in the previous
section
on storage size, and {commit} and {checkout} time.
We focus on the commit and checkout times
since they are the primitive versioning operations on which the
other more complex operations and queries are built on.
It is important that these operations are efficient,
because data scientists checkout a version to start working on it
immediately, and often commit a version to have their changes
visible to other data scientists who may be waiting for them.


In our evaluation, we use
four versioning benchmark datasets \code{SCI\_1M}, \code{SCI\_2M},
\code{SCI\_5M} and \code{SCI\_8M},
each with $1M$, $2M$, $5M$ and $8M$ records respectively,
that will be described in detail in Section~\ref{exp:setup}.
For split-by-vlist, a physical primary key index
is built on \rid in both the data table and the versioning table;
for split-by-rlist, a physical primary key index
is built on \rid
in the data table and on \vid in the versioning table.
When calculating the total storage size,
we count the index size as well.
Our experiment involves first
checking out the latest version $v_i$ into a
materialized table $T'$ and then committing $T'$ back
into the \cvd as a new version $v_j$.
We depict the experimental results in Figure~\ref{fig:model_exp}.


\stitle{Storage.} From Figure~\ref{fig:model_exp}(a),
we can see that a-table-per-version
takes $10\times$ more storage than the other data models.
This is because each record exists on average in 10 versions.
Compared to a-table-per-version and combined-table, split-by-vlist and
split-by-rlist deduplicate the common records across versions
and therefore have roughly similar storage.
In particular, split-by-vlist and split-by-rlist
share the same data table, and thus the difference
can be attributed to the difference in the size of the versioning table.
For the delta-based approach, the storage size is similar to or even slightly
smaller than split-by-vlist and split-by-rlist.
This is because our versioning benchmark contains only a few deleted tuples
(opting instead for updates or inserts);
in other cases, where deleted tuples are more prevalent, the
storage in the delta-based approach is worse than split-by-vlist/rlist,
since the deleted records will be repeated.
\techreport{We also remark that the storage size for array-based appoaches can be further reduced by applying compression techniques like range-encoding~\cite{buneman2004archiving}.}

\stitle{Commit.} From Figure~\ref{fig:model_exp}(b),
we can see that the combined-table and split-by-vlist
take multiple orders of magnitude more time than
split-by-rlist for {commit}.
We also notice that the {commit} time when
using combined-table is almost $10^4 s$ as the dataset size increases:
when using combined-table,
we need to add $v_j$ to the attribute \vlist
for each record in the \cvd that is also present in $T'$.
Similarly, for split-by-vlist, we need to perform an
append operation for several tuples in the versioning table.
On the contrary, when using split-by-rlist,
we only need to add one tuple to the versioning table,
thus getting rid of the expensive array appending operations.
A-table-per-version also has higher
latency for commit than split-by-rlist
since it needs to insert all the records in $T'$ into the \cvd.
For the delta-based approach, the commit time is small since
the new version $v_j$ is exactly the same as its precedent version
$v_i$. It only needs to update the precedent metadata table,
and create a new empty table.
The commit time of the delta-based approach is not small in general
when there are
extensive modifications to $T'$, as illustrated by other experiments (not displayed);
For instance, for a committed version with 250K records of which
30\% of the records are modified,
delta-based takes 8.16s, while split-by-rlist takes 4.12s.

\stitle{Checkout.} From Figure~\ref{fig:model_exp} (c),
we can see that split-by-rlist is a bit faster
than combined-table and split-by-vlist for {checkout}.
Not surprisingly, a-table-per-version
is the best for this operation since
it simply requires retrieving all the records in
a specific table (corresponding to the desired version).
\techreport{We dive into the query plan for the other data models.}
Combined-table requires one full scan over
the combined table to check whether each
record is in version $v_i$.
On the other hand, split-by-vlist
needs to first scan the versioning table
to retrieve the \rids in version $v_i$,
and then join the \rids with the data table.
Lastly, split-by-rlist retrieves
the \rids in version $v_i$ using
the primary key index on \vid in the versioning table,
and then joins the \rids with the data table.
For both split-by-vlist and split-by-rlist,
we used a hash-join, which was the most efficient\footnote{\scriptsize We also tried
alternative join methods---the findings were unchanged;
we will discuss this further in Section~\ref{ssec:prob}.
We also tried using an additional secondary
index for \vlist for split-by-vlist
which reduced the time for checkout but increased
the time for commit even further.}, where
a hash table on \rids is first built,
followed by a sequential scan on the data table
by probing each record in the hash table.
Overall, combined-table, split-by-vlist, and split-by-rlist
all require a full scan on the combined table
or the data table, and even though split-by-rlist
introduces the overhead of building a hash table,
it reduces the expensive array operation for containment
checking as in combined-table and split-by-vlist.
For the delta-based approach, the checkout time is large
since it needs to probe into a number of tables,
tracing all the way back to the root,
remembering which records were seen.


\stitle{Takeaways.}
Overall, considering the space consumption, 
the {commit} and {checkout} time,
plus the fact that delta-based models are inefficient in
supporting advanced queries as discussed in Section~\ref{ssec:RM},
we claim that split-by-rlist is preferable to the
other data models in supporting versioning within
a relational database.
Thus, we pick split-by-rlist as our
data model for representing \cvds.
That said, from Figure~\ref{fig:model_exp}(c),
we notice that the {checkout} time for
split-by-rlist grows with dataset size.
For instance, for dataset \code{SCI\_8M} with $8M$
records in the data table, the {checkout} time is as high as 30 seconds.
On the other hand, a-table-per-version has very low {checkout}
times on all datasets; it only needs to access the relevant records
instead of all records as in split-by-rlist.
This motivates the need for the partition optimizer module
in \orpheus, which tries to attain the best of both worlds
by adopting a hybrid representation of split-by-rlist and a-table-per-version,
described in Section~\ref{sec:partition}.

\input{metadata}

%% file: metadata.tex
\subsection{Version Derivation Metadata}\label{ssec:version_graph}
\para{Version Provenance} 
As discussed in Section~\ref{ssec:sysop},
the version manager in \orpheus keeps track
of the derivation relationships among versions
and maintains metadata for each version.
We store version-level provenance information
in a separate table called the \emph{metadata table};
Figure~\ref{fig:metaTable_versionG}(a) depicts the
metadata table for the example in Figure~\ref{fig:datamodels}.
It contains attributes including version id, 
parent/child versions, creation time, commit time, 
a commit message,
and an array of attributes present in the version.
Using the data contained in this table,
users can easily query for the provenance of versions 
and for other metadata.
In addition, using the attribute {\em parents}
we can obtain each version's derivation information
and visualize it as directed acyclic graph
that we call a \emph{version graph}.
Each node in the version graph is a version and each directed edge
points from a version to one of its children version(s).
An example is depicted in 
Figure~\ref{fig:metaTable_versionG}(b),
where \techreport{version $v_2$ and $v_3$ are both
derived from version $v_1$,
and }version $v_2$ and $v_3$ are merged into version $v_4$.
\techreport{We will return to this concept in Section~\ref{one_to_many_alg}.}

\papertext{
\begin{figure}[h!]
\centering
\vspace{-3mm}
\includegraphics[width=0.85\linewidth]{fig/meta_attribute_table.pdf} 
\vspace{-10pt}
\caption{Metadata Table and its Corresponding Version Graph}
\label{fig:metaTable_versionG}
\vspace{-3mm}
\end{figure}
}


\techreport{
\begin{figure}[h!]
\centering
\vspace{-3mm}
\includegraphics[width=0.85\linewidth]{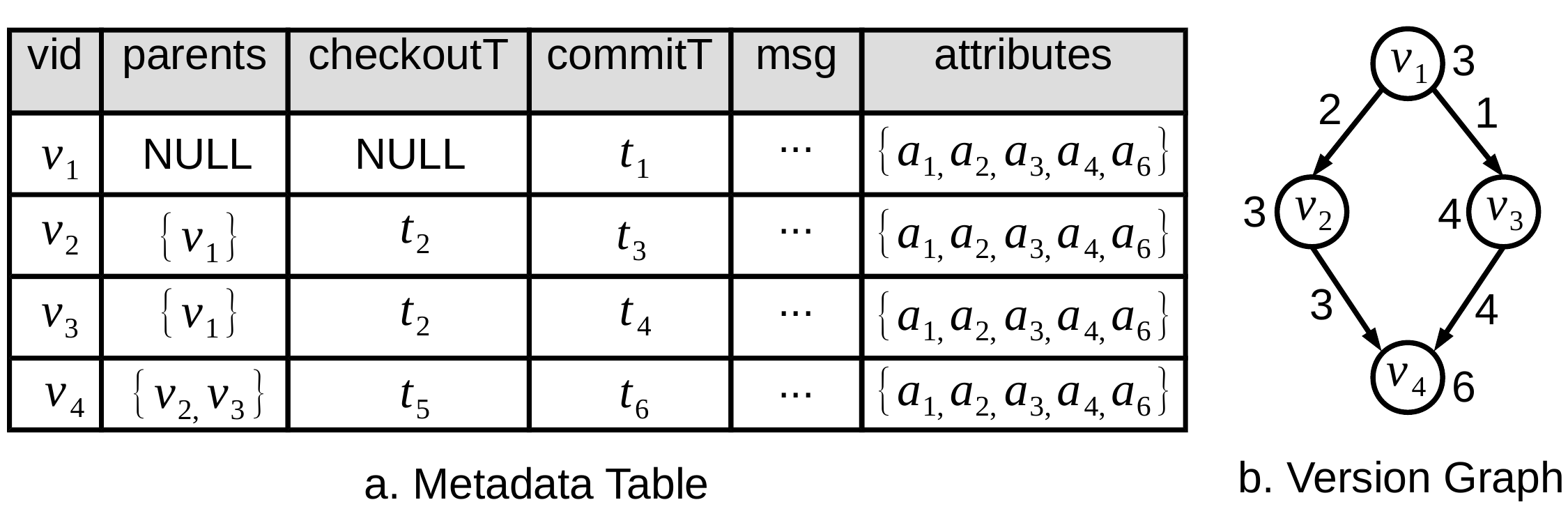} 
\vspace{-10pt}
\caption{Metadata Table and Version Graph (Fixed Schema)}
\label{fig:metaTable_versionG}
\vspace{-7mm}
\end{figure}
\begin{figure}[h!]
\centering
\includegraphics[width=0.8\linewidth]{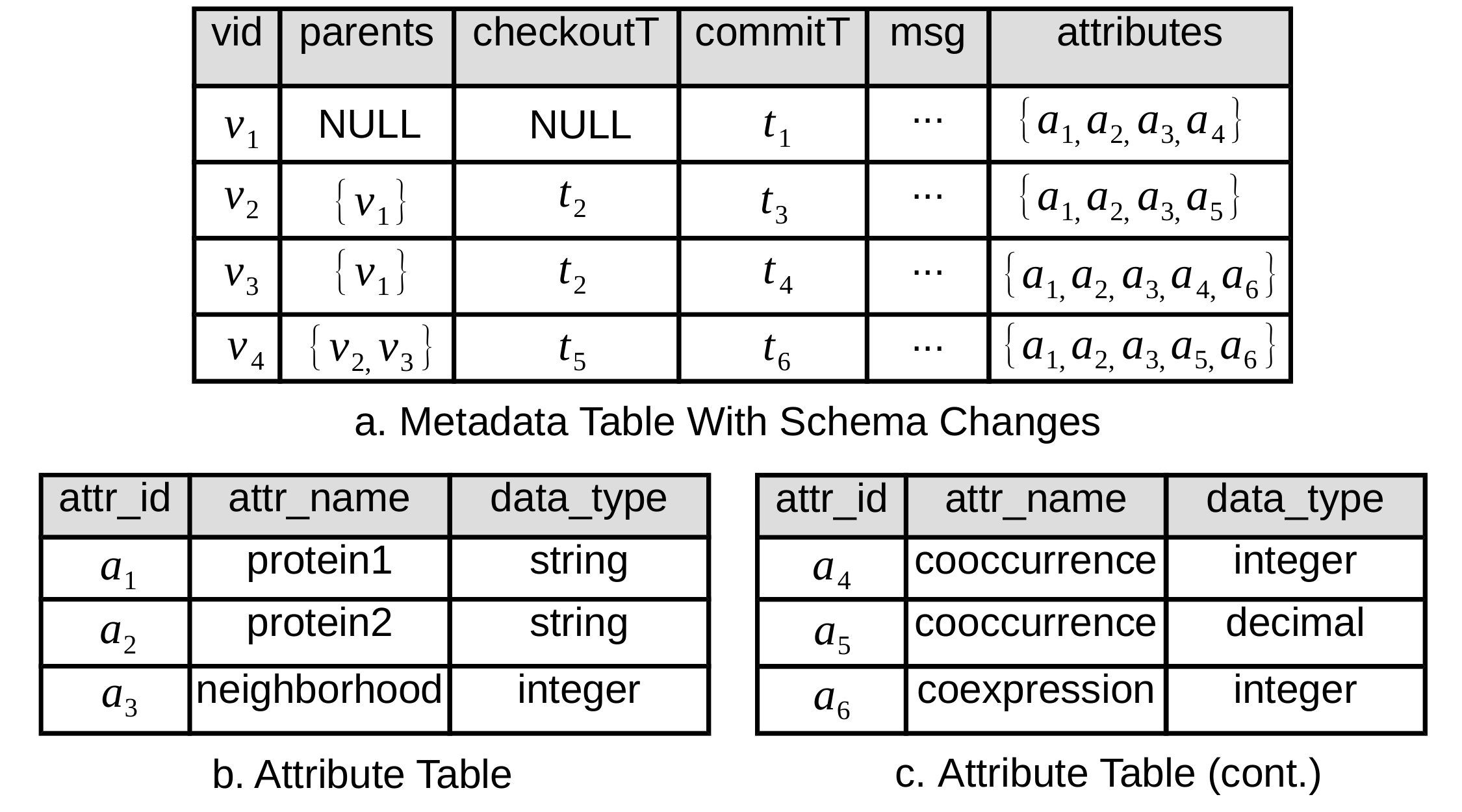} 
\vspace{-9pt}
\caption{Metadata Table and Attribute Table (Schema Changes)}
\label{fig:attributeTable}
\vspace{-4mm}
\end{figure}
}

\para{Schema Changes} 
During a commit, if the schema of the table being committed
is different from the schema of the \cvd 
it was derived from, we update the schema
of \cvd to
incorporate the changes. 
\papertext{We describe details in our technical report~\cite{orpheustr}.}\techreport{More precisely, in \orpheus, we
maintain an attribute table (as in Figure~\ref{fig:attributeTable})
where each
tuple represents an attribute with a unique identifier,
along with the corresponding attribute name and data type; any
change of a property of an attribute results in 
a new attribute entry in the table. 
If the data type of any attribute changes,
we transform the attribute type to a more general data
type (e.g., from integer to string as in Jain et al.~\cite{jain2016sqlshare}),
and insert a new tuple into the attribute table with the updated
data type.}
All of our array-based models can adapt to changes
in the set of attributes: a simple solution 
for new attributes is 
so use the {\sf ALTER} command to add any new
attributes to the model,
assigning NULLs to the records from the previous versions
that do not possess these new attributes. 
Attribute deletions only require an update in the version
metadata table.
\techreport{To illustrate, we modify the previous example in Figure~\ref{fig:metaTable_versionG} (which showed a static schema) to a dynamic one.
For example, as shown in Figure~\ref{fig:attributeTable}, initially version $v_1$ has four attributes: protein1, protein2, neighborhood and cooccurrence.
When a user commits version $v_2$,
with the data type of the {cooccurrence} attribute ($a_4$)
changed from integer to decimal, 
within \orpheus, we create another attribute ($a_5$)
in the attribute table with data type decimal, 
log $a_5$ in the metadata table for $v_2$ and 
alter the cooccurrence attribute to decimal within the \cvd. 
Moreover, when a new {coexpression} attribute is added in $v_3$, 
we generate a corresponding attribute ($a_6$) in the attribute table, 
add $a_6$ in the metadata table for $v_3$, and add the {coexpression} attribute to the \cvd. 
During the merge, the resulting version includes all attributes from its parents and 
contains the more general data type for conflicting attributes (e.g., attributes in $v_4$).}
This simple mechanism is similar to the 
single pool method proposed in a temporal schema versioning
context by De Castro et al.~\cite{de1995schema}. 
Compared to the multi pool method where any
schema change results in the new version being stored 
separately, the single pool
method has fewer records with duplicated attributes and
therefore has less storage consumption overall. 
Even though ALTER TABLE is indeed a costly operation, due to the
partitioning schemes we describe later, we only need to ALTER a
smaller partition of the \cvd rather than a giant \cvd,
and consequently the cost of an ALTER operation is substantially mitigated.
\papertext{In our technical report}\techreport{In Appendix~\ref{ssec:schema_change}}, we describe 
how our partitioning schemes (described next\techreport{~in Section~\ref{sec:partition}}) can adapt to the single pool mechanism
with comparable guarantees;
for ease of exposition, for the rest of this paper, we focus
on the static schema case, which is still important and challenging.
There has been some work on developing schema versioning 
schemes~\cite{de1997schema,moon2008managing, moon2010scalable}
and we plan to explore these and other schema evolution mechanisms 
(including hybrid single/multi-pool methods) as future work.

%% file: optimization.tex
\section{Partition Optimizer}\label{sec:partition}
\techreport{Recall that Figure~\ref{fig:model_exp}(c)
indicated that as the number of records
within a \cvd increases,
the checkout latency of our data
model (split-by-rlist) increases---this
is because the number of ``irrelevant''
records, i.e., the records
that are not present in the
version being checked out,
but nevertheless require processing
increases. Even with index on \rid, the checkout latency is still high since records are scattered across the whole data table, and hundreds of thousands of random accesses are eventually reduced to a full table scan as demonstrated in Appendix~\ref{ssec:exp_cost_model}.}
In this section, we introduce
the concept of partitioning
a \cvd by breaking up the data
and versioning tables,
in order to reduce
the number of irrelevant records
during checkout.
\papertext{All of our proofs can be found in our technical report~\cite{orpheustr}.}\techreport{We formally
define our partitioning problem,
demonstrate that this problem
is {\sc NP-Hard}, and identify a
light-weight approximation algorithm.}
\techreport{We provide a convenient table of notation
in the Appendix (Table~\ref{tbl:notation}).}

\subsection{Problem Overview}\label{ssec:prob}

\begin{figure}[h!]
\centering
\vspace{-3mm}
\includegraphics[width=0.6\linewidth]{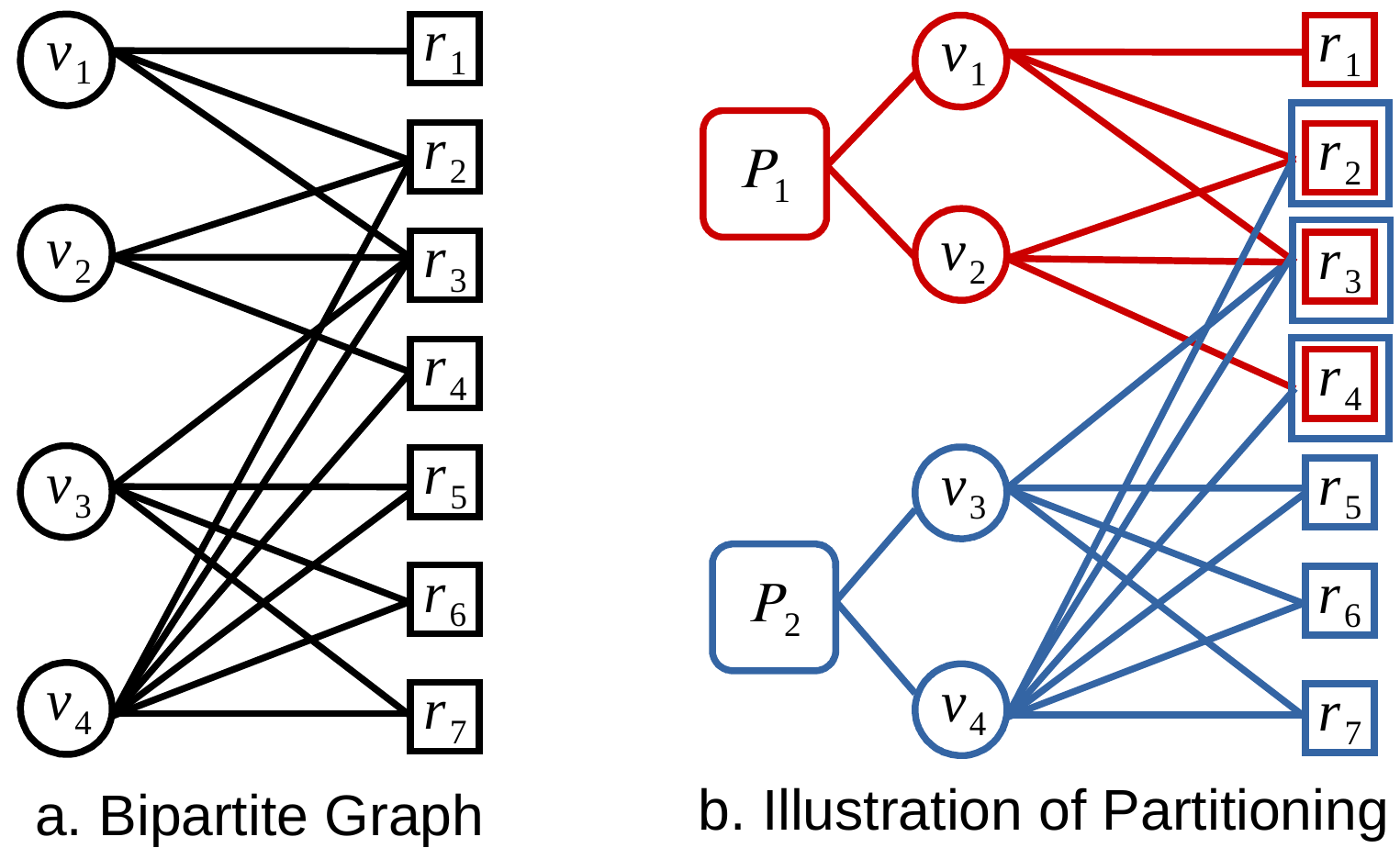}
\vspace{-2mm}
\caption{Version-Record Bipartite Graph \& Partitioning}
\label{fig:biPartite}
\vspace{-10pt}
\end{figure}

\vspace{-5pt}
\stitle{The Partitioning Notion.}
Let $V=\{v_1,v_2,...,v_n\}$ be the $n$ versions
and $R=\{r_1,r_2,...,$ $r_m\}$ be the $m$ records in a \cvd.
We can represent the presence of records in versions
using a version-record bipartite graph $G=(V,R,E)$,
where $E$ is the set of edges---an edge between $v_i$ and $r_j$ exists
if the version $v_i$ contains the record $r_j$.
The bipartite graph in Figure~\ref{fig:biPartite}(a)
captures the relationships between records and versions
in Figure~\ref{fig:datamodels}.

The goal of our partitioning problem
is to partition $G$ into smaller
subgraphs, denoted as $\pp_k$.
We let $\pp_k=(\vv_k,\rr_k,\ee_k)$,
where $\vv_k$, $\rr_k$ and $\ee_k$ represent the
set of versions, records and bipartite graph edges
in partition $\pp_k$ respectively.
Note that $\cup_k \ee_k = E$,
where $E$ is the set of edges in the
original version-record bipartite graph $G$.
We further constrain each version in the \cvd
to exist in only one partition,
while each record
can be duplicated
across multiple partitions.
In this manner, we
only need to access one partition when checking
out a version, consequently
simplifying the {checkout} process
by reducing the overhead from accessing multiple
partitions.
\techreport{(While we do not consider it in this paper,
in a distributed setting, it is even more important
to ensure that as few partitions are consulted
during a checkout operation.)}
Thus, our partition problem
is equivalent to partitioning $V$,
such that each partition ($\pp_k$)
stores all of the records corresponding
to all of the versions assigned to that partition.
Figure~\ref{fig:biPartite}(b) illustrates a possible
partitioning strategy for Figure~\ref{fig:biPartite}(a).
Partition $\pp_1$ contains version $v_1$ and $v_2$,
while partition $\pp_2$ contains version $v_3$ and $v_4$.
Note that records $r_2,r_3$ and $r_4$ are duplicated in $\pp_1$ and $\pp_2$.


\stitle{Metrics.}
We consider two criteria while partitioning:
the storage cost and the time for checkout.
Recall that the time for commit is fixed
and small---see Figure~\ref{fig:model_exp}(b), so
we only focus on checkout.

The overall storage costs involves
the cost of storing all of the partitions
of the data and the versioning table.
However, we observe that the
versioning table simply encodes
the bipartite graph, and
as a result, its cost is fixed.
Furthermore, since all of the records
in the data table have the same (fixed) number of attributes,
so instead of optimizing the actual storage
we will optimize for the number of records
in the data table across all the partitions.
Thus, we define the {\em storage cost}, $\calS$,
to be the following:
\begin{equation}\label{eqn:storage_cost}\small
\calS=\sum_{k=1}^{K} |\rr_k|
\end{equation}
Next, we note that the time taken for checking out
version $v_i$ is proportional to
the size of the data table in the
partition $\pp_k$
that contains version $v_i$,
which in turn is proportional to
the number of records present in
that data table partition.
We theoretically and empirically justify
this
\techreport{in Appendix~\ref{ssec:exp_cost_model}.}\papertext{in our technical report \cite{orpheustr}.}
So we define the {\em checkout cost of a version} $v_i$,
$\cc_i$, to be $\cc_i =  |\rr_{k}|$, where $v_i \in \vv_k$.
The {\em checkout cost}, denoted as $\cc_{avg}$,
is defined to be the average of $\cc_i$,
i.e., $\cc_{avg} = \frac{\sum_i \cc_i}{n}$.
While we focus on the average case,
which assumes that each version is checked out
with equal frequency\techreport{---a reasonable
assumption when we have no other information about the
workload}, our algorithms
generalize to the {\em weighted} case\papertext{~\cite{orpheustr}.}\techreport{ as described in~Appendix~\ref{ssec:weighted}.} 
\techreport{(The weighted case can help represent the workload in real world settings, where recent versions may be checked out more frequently.)}
On rewriting the expression for $\cc_{avg}$ above,
we get:
\techreport{\vspace{4mm}}
\begin{equation}\label{eqn:avg_checkout_cost_transform}\small
\cc_{avg} = \frac{\sum_{k=1}^K |\vv_{k}||\rr_{k}| }{n}
\techreport{\vspace{4mm}}
\end{equation}

The numerator is simply sum of the number of records in each partition,
multiplied by the number of versions in that partition,
across all partitions\techreport{---this is
the cost of checking out all of the versions}.

\stitle{Formal Problem.}
Our two metrics $\calS$
and $\cavg$ interfere with each other:
if we want a small $\cavg$,
then we need more storage,
and if we want the storage to be
small, then  $\cavg$ will be large.
Typically, storage is under our control;
thus, our problem can be stated as:
\vspace{-5pt}
\begin{problem}[Minimize Checkout Cost]\label{prob:min_rec}
Given a storage threshold $\gamma$ and a
version-record bipartite graph $G=(V,R,E)$,
find a partitioning of $G$ that minimizes $\cavg$ such that
$ \calS \leq \gamma$.
\vspace{-5pt}
\end{problem}
We can show that Problem~\ref{prob:min_rec}
is {\sc NP-Hard} using a reduction from the  {\sc 3-Partition} problem,
whose goal is to decide whether a given set of $n$ integers
can be partitioned into $\frac{n}{3}$ sets with equal sum.
{\sc 3-Partition} is known to be strongly {\sc NP-Hard}\techreport{,
i.e., it is {\sc NP-Hard} even when its numerical
parameters are bounded by a polynomial in the length of
the input}.
\vspace{-5pt}
\begin{theorem}\label{theorem:one_np}
Problem~\ref{prob:min_rec} is {\sc NP-hard}. 
\vspace{-5pt}
\end{theorem}
\techreport{The proof for this theorem can be found
in Appendix~\ref{sec:proof}.}

We now clarify one complication between our formalization so far
and our implementation.
\orpheus uses the {\em no cross-version diff rule}:
that is, while performing a commit operation,
to minimize computation,
\orpheus does not compare the committed version
against all of the
ancestor versions, instead only
comparing it to its parents.
Therefore, if some records are deleted and then re-added later,
these records
would be assigned different \rids, and are treated
as different.
As it turns out, Problem~\ref{prob:min_rec} is still
{\sc NP-Hard} when the space of instances
are those that can be generated when this rule is applied.
For the rest of this section, we will
use the formalization with the no cross-version diff
rule in place, since that relates more closely
to practice.

\subsection{Partitioning Algorithm}\label{one_to_many_alg}
\begin{wrapfigure}{rt}{0.23\textwidth}
\papertext{\vspace{-10pt}}
\hspace{-15pt}
\includegraphics[width=0.25\textwidth]{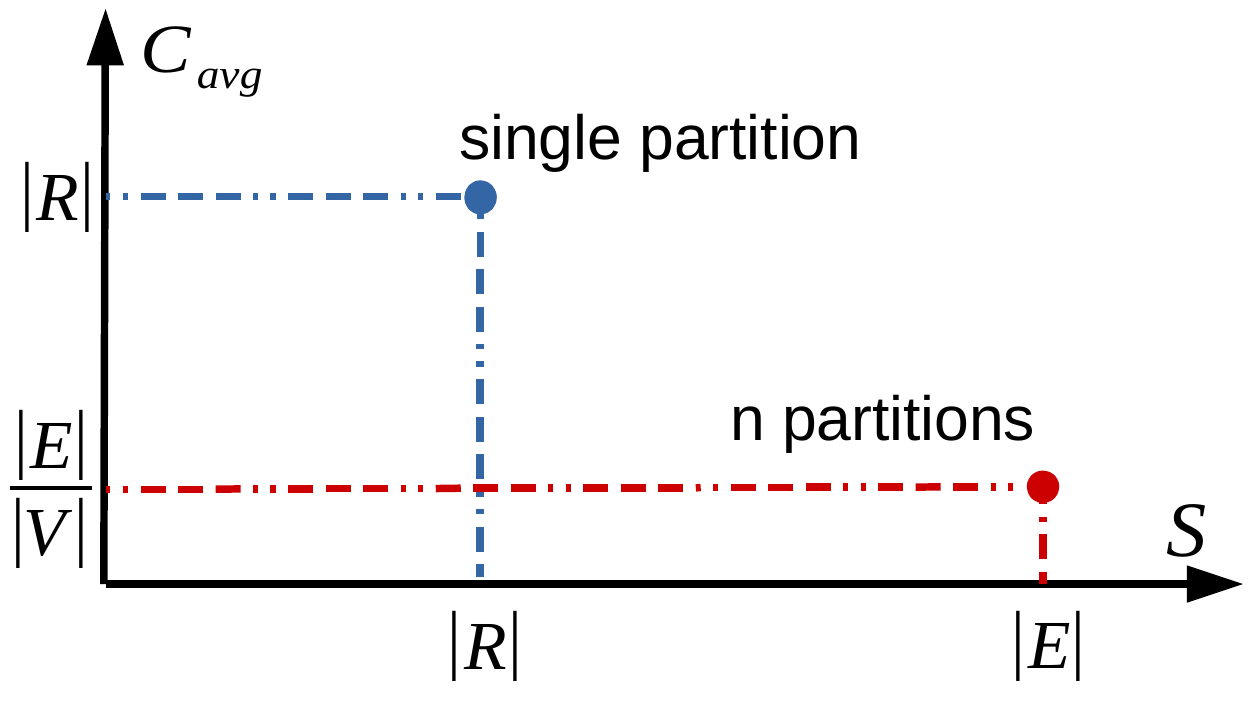}
\vspace{-10pt}
\caption{Extreme Schemes}
\label{fig:extreme_cases}
\vspace{-10pt}
\end{wrapfigure}
Given a version-record bipartite graph $G=(V,R,E)$,
there are two extreme cases for partitioning.
At one extreme, we can minimize the checkout cost
by storing each version in the \cvd as one partition;
there are in total  $K=|V|=n$ partitions, and
the storage cost is $\mathcal{S}=\sum_{k=1}^n |\rr_k|=|E|$
and the checkout cost is $\cc_{avg} = \frac{1}{n}\sum_{k=1}^n$ $(|\vv_k| |\rr_k|) =\frac{|E|}{|V|}$.
At another extreme, we can minimize the storage by
storing all versions in one single partition;
the storage cost is $\mathcal{S}=|R|$ and
$\cc_{avg}=|R|$.
We illustrate these schemes in Figure~\ref{fig:extreme_cases}.
\techreport{We also list them as formal observations below:
\begin{observation}\label{obs:min_checkout}
\vspace{-5pt}
Given a bipartite graph $G=(V,R,E)$, the
checkout cost $\cc_{avg}$ is minimized by
storing each version as one separate partition: $\cc_{avg}=\frac{|E|}{|V|}$.
\end{observation}
\vspace{-8pt}
\begin{observation}\label{obs:min_storage}
Given a bipartite graph $G=(V,R,E)$, the
storage cost $\mathcal{S}$ is minimized by storing
all versions in a single partition: $\mathcal{S}=|R|$.
\end{observation}
\vspace{-5pt}
}

\stitle{Version Graph Concept.}
Our goal in designing our partitioning algorithm, \appr\footnote{\scriptsize A lyre was the musical instrument of choice for Orpheus.},
is to trade-off between these two extremes.
Instead of operating on the version-record bipartite graph,
which may be very large,
\appr operates on the much smaller version graph
instead, which makes it a lot more lightweight.
\techreport{We recall the concept of a \emph{version graph}
from Section~\ref{ssec:version_graph},
and depicted in Figure~\ref{fig:metaTable_versionG}.}
We denote a version graph as $\gbb=(\vbb,\ebb)$,
where each vertex $v\in \vbb$ is a version
and each edge $e\in \ebb$ is a derivation relationship.
Note that $\vbb$ is essentially the same as $V$
in the version-record bipartite graph.
An edge from vertex $v_i$ to a vertex $v_j$
indicates that $v_i$ is a parent of $v_j$;
this edge has a weight $w(v_i, v_j)$
equals the number of records in common
between $v_i$ and $v_j$.
We use $p(v_i)$ to denote the parent versions
of $v_i$.
For the special case when there are no merge operations,
$|p(v_i)| \leq 1, \forall i$, and
the version graph is a tree, denoted as $\tbb = (\vbb, \ebb)$.
Lastly, we use $R(v_i)$ to be
the set of all records in version $v_i$,
and $l(v_i)$ to be the depth of
$v_i$ in the version graph $\gbb$ in a
topological sort of the graph---the root
has depth 1.
For example,
in Figure~\ref{fig:metaTable_versionG},
version $v_2$ has $|R(v_2)| = 3$ since
it has three records, and is at
level $l(v_2) = 2$.
Further, $v_2$ has a single parent
$p(v_2) = v_1$,
and shares two records with its parent,
i.e., $w(v_1, v_2) = 2$.
Next, we describe the algorithm for \appr
when the version graph is a tree
(i.e., no merge operations).
\techreport{We then naturally extend our algorithm
to other settings, as we will describe next.}

\stitle{The Version Tree Case.}
Our algorithm is based on the following lemma,
which intuitively states
that if every version $v_i$
shares a large number of records with its parent
version, then the checkout cost
is small, and bounded by some factor of
$\frac{|E|}{|V|}$, where $\frac{|E|}{|V|}$
is the lower bound on the optimal checkout cost\techreport{~(from Observation~\ref{obs:min_checkout})}.

\begin{lemma}\label{le:checkout_cost}
Given a bipartite graph $G=(V,R,E)$,
a version tree $\mathbb{T}=(\mathbb{V},\mathbb{E})$,
and a parameter $\delta\leq 1$,
if the weight of every edge in $\mathbb{E}$
is larger than $\delta|R|$,
then the checkout cost $\mathcal{C}_{avg}$
when all of the versions are in one single partition
is less than $\frac{1}{\delta}\cdot\frac{|E|}{|V|}$.
\end{lemma}

\techreport{
\begin{proof}
Consider the nodes of the
version tree $\mathbb{T}$ level-by-level,
starting from the root.
That is, all of a version's ancestors are considered
before it is evaluated.
Now, given a version $v_i$,
the number of new records added by $v_i$ is
$R(v_i) - w(v_i, p(v_i))$.
Thus, we have:
\begin{equation*} \small
\begin{split}
|R| &=|\cup_{i=1}^{|V|} R(v_i)| \\
 & = R(v_1)+ \sum_{l(v_i)=2}(R(v_i)-w(v_i,p(v_i))) \\
 & + \sum_{l(v_i)=3}(R(v_i)-w(v_i,p(v_i))) + \cdots \\
\implies |R|& = \sum_{i=1}^{|V|} R(v_i) - \sum_{i=2}^{|V|}(w(v_i,p(v_i))) \\
\end{split}
\end{equation*}
Since each edge weight is larger than $\delta|R|$, i.e., $w(v_i,p(v_i)) > \delta|R|, \forall 2\leq i\leq |V|$, we have:
$$ |R| < |E| - \delta (|V|-1) |R| \leq |E| - \delta |V||R| + |R|$$
where the last inequality is because $\delta \leq 1$.
Thus, we have $|R| < \frac{1}{\delta}\cdot\frac{|E|}{|V|}$.
Since $\cavg=|R|$ when we have only one partition, the result follows.
\end{proof}
}

Lemma~\ref{le:checkout_cost} indicates
that when $\cc_{avg}\geq  \frac{1}{\delta}\cdot\frac{|E|}{|V|}$,
there must exist some version $v_j$
that only shares a small number of common records
with its parent version $v_i$, i.e., $w(v_i,v_j) \leq \delta |R|$;
otherwise $\cc_{avg}< \frac{1}{\delta}\cdot\frac{|E|}{|V|}$.
Intuitively, such an edge $(v_i,v_j)$ with $w(v_i,v_j) \leq \delta |R|$
is a potential edge for splitting
since the overlap between $v_i$ and $v_j$ is small.

\techreport{
\begin{algorithm}[!t]
\DontPrintSemicolon
\SetKwInOut{input}{Input}\SetKwInOut{output}{Output}
{\small
\input{Version tree $\gbb=(\mathbb{V},\mathbb{E})$ and parameter $\delta$}
\output{Partitions $\{\pp_1,\pp_2,\cdots,\pp_K\}$}
 \If {$|R|\times |V| < \frac{|E|}{\delta}$} {
    return $V$
    }
 \Else{
    $\Omega \leftarrow \{e|e.w \leq \delta \times |R|, e\in \mathbb{E}\}$ \;
    $e^* \leftarrow \text{PickOneEdgeCut}(\Omega)$ \;
    Remove $e^*$ and split $\gbb$ into two parts $\{\gbb_1, \gbb_2\}$ \;
    Update the number of records, versions and bipartite edges in $\gbb_1$, denoted as $|R_1|$, $|V_1|$ and $|E_1|$ \;
    Update the number of records, versions and bipartite edges in $\gbb_2$, denoted as $|R_2|$, $|V_2|$ and $|E_2|$ \;
    $\pp_1$=\appr($\gbb_1, |R_1|, |V_1|, |E_1|, \delta$) \;
    $\pp_2$=\appr($\gbb_2, |R_2|, |V_2|, |E_2|, \delta$) \;
    return $\{\pp_1,\pp_2\}$
    }
}
\caption{\appr($\gbb,|R|,|V|,|E|,\delta$)}
\label{alg:divisive_partition}
\end{algorithm}
}

\begin{figure}[h!]
\centering
\vspace{-2mm}
\includegraphics[width=0.8\linewidth]{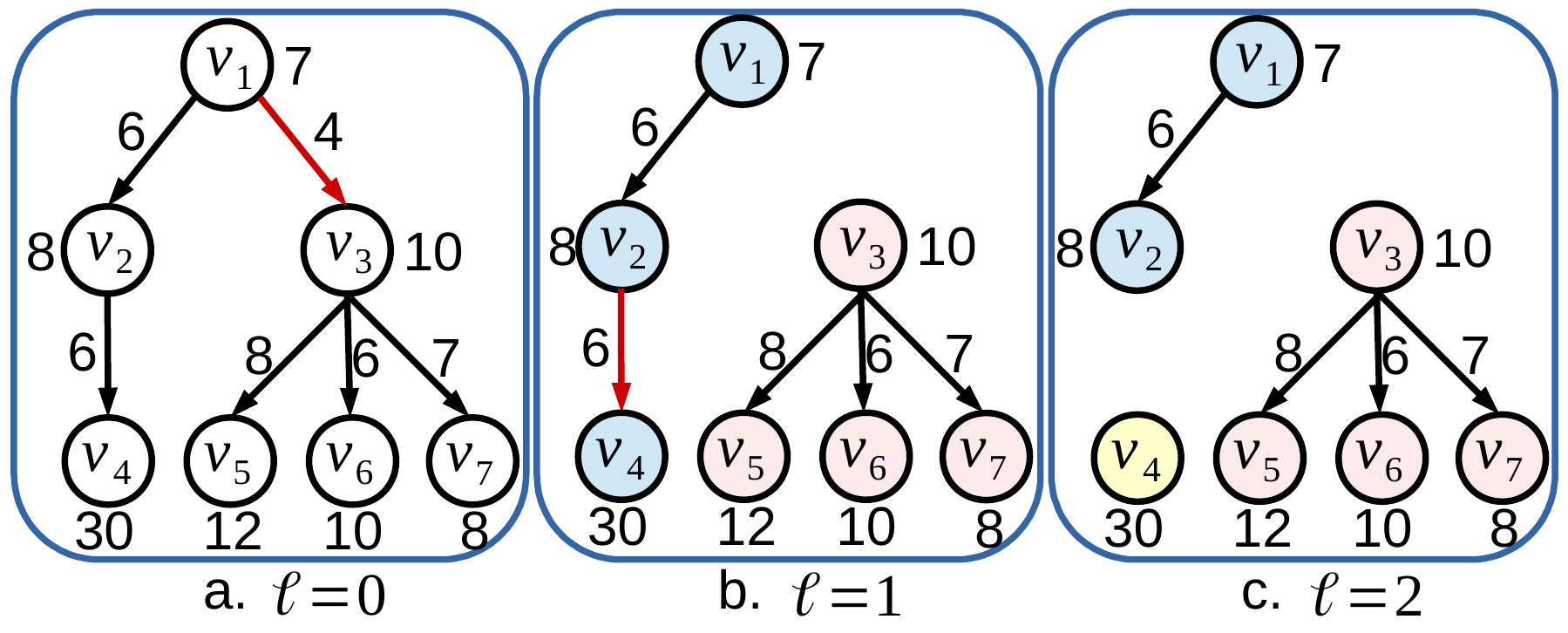}
\vspace{-3mm}
\caption{Illustration of \appr ($\delta=0.5$)}
\label{fig:divisive_partition}
\vspace{-3mm}
\end{figure}

\stitle{\appr Illustration.}
We describe a version of \appr that
accepts as input a parameter $\delta$,
and then recursively applies partitioning until the overall
$\cc_{avg}<\frac{1}{\delta}\cdot\frac{|E|}{|V|}$;
we will adapt this to Problem~\ref{prob:min_rec} later.
The pseudocode is provided in \papertext{the technical report~\cite{orpheustr}}\techreport{Algorithm~\ref{alg:divisive_partition}},
and we illustrate its execution on an example
in Figure~\ref{fig:divisive_partition}.

As before, we are given a version tree $\tbb=(\vbb,\ebb)$.
We start with all of the versions in one partition.
We first check whether $|R||V| < \frac{|E|}{\delta}$\techreport{~(line 1)}.
If yes, then we terminate; otherwise, we pick one edge $e^*$ with
weight $e^*.w \leq \delta|R|$\techreport{~(lines 5--6)} to cut in order to split the partition into two.
According to Lemma \ref{le:checkout_cost},
if $|R||V| \geq \frac{|E|}{\delta}$,
there must exist some edge whose weight
is no larger than $\delta|R|$.
The algorithm does not prescribe a method for picking
this edge if there are multiple;
the guarantees hold independent of this method.
For instance, we can pick the edge with the smallest weight;
or the one such that after splitting,
the difference in the number of versions in the two partitions is minimized.
In our experiments, we use the latter\techreport{,
and break a tie by selecting the edge that balances the records
between two partitions in addition to the number of versions}.
In our example in Figure~\ref{fig:divisive_partition}(a),
\techreport{we first find that having the entire version tree as a single partition violates the property,
and }we pick the red edge to split the version tree $\tbb$
into two partitions---as shown in Figure~\ref{fig:divisive_partition}(b),
we get one partition $\pp_1$ with the blue nodes (versions) and another $\pp_2$ with the red nodes (versions).

After each edge split, we update the number of records,
versions and bipartite edges\techreport{~(lines 8--9)},
and then we recursively call the algorithm on each partition\techreport{~(lines 10--11)}.
In the example, we terminate for $\pp_2$ but we split the edge $(v_2, v_4)$
for $\pp_1$, and then terminate with three partitions---Figure~\ref{fig:divisive_partition}(c).
We define $\ell$ be the recursion level number.
In Figure~\ref{fig:divisive_partition} (a) (b) and (c), $\ell=0$, $\ell=1$ and $\ell=2$ respectively.
We will use this notation in the performance analysis next.

Now that we have an algorithm for the $\delta$ case,
we can simply apply binary search on $\delta$
and obtain the best $\delta$ for Problem~\ref{prob:min_rec}.
\techreport{We refer readers to Appendix~\ref{sec:deltas} for a detailed analysis of $\delta$.}
\papertext{We can show that for two $\delta$
such that one is smaller than the other,
the edges cut in the former is a superset of the latter.}

\stitle{Performance Analysis.}
Overall, the lowest storage cost is $|R|$ and the lowest checkout cost is $\frac{|E|}{|V|}$ respectively\techreport{~(as formalized in Observation~\ref{obs:min_checkout} and \ref{obs:min_storage})}.
We now analyze the performance  in terms of these quantities:
an algorithm has an approximation ratio of $(X, Y)$ if its storage cost $\calS$
is no larger than $X \cdot R$ while its checkout cost $\cavg$ is no larger than $Y \cdot \frac{|E|}{|V|}$.
We first study the impact of a single split edge.
\vspace{-3pt}
\begin{lemma}\label{le:storage}
Given a bipartite graph $G=(V,R,E)$,
a version tree $\tbb=(\vbb,\ebb)$
and a parameter $\delta$,
let $e^*\in \mathbb{E}$ be the edge that is split
in \appr,
then after splitting the storage cost $\mathcal{S}$
must be within $(1+\delta)|R|$.
\end{lemma}
\vspace{-3pt}
\vspace{-3pt}
\techreport{
\begin{proof}
First according to Lemma~\ref{le:checkout_cost},
if $|R||V|\geq \frac{|E|}{\delta}$,
there must exist some edge $e^*=(v_i,v_j)$
whose weight is less than $\delta|R|$, i.e., $e^*.w \leq \delta|R|$.
Then, we remove one such $e^*$ and split $\gbb$ into two parts $\{\gbb_1, \gbb_2\}$ as
depicted in line 7-9 in Algorithm~\ref{alg:divisive_partition}.
The current storage cost $\mathcal{S}=|R_1|+|R_2|$.
The common records between $\gbb_1$ and $\gbb_2$ is exactly
the common records shared by version $v_i$ and $v_j$, i.e., $e^*.w$. Thus, we have:
\begin{equation*} \small
\begin{split}
|R| & = |R_1\cup R_2| = |R_1| + |R_2| - e^*.w  \geq |R_1| + |R_2| - \delta|R| \\
\implies \calS & =|R_1|+|R_2| \leq (1+\delta)|R|
\end{split}
\end{equation*}
Hence proved.
\end{proof}
}
Now, overall, we have:
\vspace{-3pt}
\begin{theorem}\label{thm:storage_checkout}
Given a parameter $\delta$, \appr
results in a $((1+\delta)^{\ell},\frac{1}{\delta})$-approximation for partitioning.
\end{theorem}
\vspace{-3pt}
\techreport{
\begin{proof}
Let us consider all partitions when Algorithm~\ref{alg:divisive_partition}
terminates at level $\ell$. Each partition (e.g., Figure~\ref{fig:divisive_partition}(c))
corresponds to a subgraph of the version tree (e.g., Figure~\ref{fig:divisive_partition}(a)).
According to Lemma~\ref{le:checkout_cost}, the total checkout cost $\mathcal{C}_k$
in each partition $\mathcal{P}_k=(\mathcal{V}_k,\mathcal{R}_k,\mathcal{E}_k)$
must be smaller than $\frac{|\mathcal{E}_k|}{\delta}$, where $|\mathcal{E}_k|$
is the number of bipartite edges in partition $\pp_k$.
Since $\sum_{k=1}^K |\mathcal{E}_k| = |E|$,
we prove that the overall average
checkout cost $\cavg$ is $\frac{\sum \mathcal{C}_k }{|V|}<\frac{1}{\delta}\cdot\frac{|E|}{|V|}$.
\newline
Next, we consider the storage cost.
The analysis is similar to the complexity analysis for quick sort.
Our proof uses a reduction on the recursive level number $\ell$.
First, when $\ell=0$, all versions are stored in a single partition (e.g. Figure~\ref{fig:divisive_partition}(a)).
Thus, the storage cost is $|R|$.
Next, as the recursive algorithm proceeds,
there can be multiple partitions at each recursive level $\ell$.
For instance, there are two partitions at level $\ell=1$ and three partitions at level $\ell=2$ as shown in Figure~\ref{fig:divisive_partition}(b) and (c).
Assume that there are $\tau$ partitions $\{\pp_1,\pp_2,\cdots,\pp_{\tau}\}$
at level $\ell=\alpha$, and the storage cost for these partitions is
no bigger than $(1+\delta)^{\alpha}\cdot |R|$.
Then according to Lemma~\ref{le:storage},
for each partition $\pp_k$ at level $\ell=\alpha$,
after splitting the storage cost at level $(\alpha+1)$ will be no bigger than $(1+\delta)$ times that at level $\alpha$. Thus, we have the total storage cost at level $(\alpha+1)$ must be no bigger than $(1+\delta)^{\alpha+1}\cdot |R|$.
\end{proof}
}
\vspace{-3pt}

\stitle{Complexity.}
\techreport{At each recursive level of Algorithm~\ref{alg:divisive_partition},
it takes $O(n)$ time for checking the weight of each edge in the version tree (line 5).
The update in line 8--9 can also be done in $O(n)$
using one pass of tree traversal for each partition.}
The total time complexity is $O(n\ell)$, where $\ell$
is the recursion level number when the algorithm terminates.

\stitle{Generalizations.}
We can naturally extend our algorithms for the case where
the version graph is a DAG: in short,
we first construct a version tree $\hat{\mathbb{T}}$
based on the original version graph $\gbb$, then apply \appr
on the constructed version tree $\hat{\mathbb{T}}$.
We describe the details for this algorithm
\techreport{in Appendix~\ref{ssec:general_case}.}
\papertext{in our technical report \cite{orpheustr}.}


\input{incremental.tex}

%% file: incremental.tex
\subsection{Incremental Partitioning}\label{ssec:inc}
\vspace{-3pt}
\appr can be explicitly invoked by users or by \orpheus when there is a need
to improve performance or a lull in activity. We now describe
how the partitioning  identified by \appr is incrementally
maintained during the course of normal operation,
and how we reduce the migration time when \appr identifies
a new partitioning.
\papertext{We only describe the high-level ideas here;
details and guarantees
can be found in the technical report~\cite{orpheustr}.}

\stitle{Online Maintenance.}
When a new version $v_i$ is committed, \orpheus applies the same intuition
as \appr to determine whether
to add $v_i$ to an existing partition, or to create a new partition\papertext{:
if $v_i$ shares a large number of records in common
with one of its parent versions $v_j$, then
$v_i$ is added to the partition $\pp_k$ that parent is in, or else a new partition is created.
Specifically, if $w(v_i,v_j) \leq \delta^* |R|$ and $\calS < \gamma$,
where $\delta^*$ was the splitting parameter used during the last invocation of \appr,
then we create a new version.
This way, the added storage cost is minimized, and the added checkout cost is guaranteed
to be small, as in Lemma~\ref{le:checkout_cost}}.
\techreport{This is again a trade-off between the storage cost and the checkout cost.
Compared to creating a new table, adding $v_i$ to an existing partition has smaller
storage cost but larger checkout cost.
Sharing the same intuition with \appr: if $v_i$ has a large number of common records with one of its parent version $v_j$, we opt to add $v_i$ into the partition $\pp_k$ where $v_j$ is in.
This is because the added storage cost is minimized and the added checkout cost is guaranteed to be small as stated in Lemma~\ref{le:checkout_cost}.
Essentially, the online maintenance is performing
incremental partitioning in the version graph as new versions are coming in.
Specifically, if $w(v_i,v_j) \leq \delta^* |R|$ and $\calS < \gamma$,
where $\delta^*$ was the splitting parameter used during the last invocation of \appr,
then we create a new version; otherwise, $v_i$ is added to partition $\pp_k$. Recall that $\gamma$ is the storage threshold  and $|R|$ is the number of records currently.
}

\smallskip
\noindent
Even with the proposed online maintenance scheme,
the checkout cost tends to diverge from the best checkout cost that \appr can identify
under the current constraints.
This is because \appr performs global partitioning using the full
version graph as input, while online maintenance
makes small changes to the existing partitioning.
To maintain the checkout performance,
\orpheus allows for a tolerance factor
$\mu$ on the current checkout cost (users can also set $\mu$ explicitly).
We let $\cc_{avg}$ and $\cc^*_{avg}$ be the current checkout cost
and the best checkout cost identified by \appr respectively.
If $\cc_{avg} > \mu \cc^*_{avg}$, the {\em migration engine} is triggered,
and we reorganize the partitions by migrating data from the old partitions
to the new ones; until then, we perform online maintenance.
In general, when $\mu$ is small, the migration engine is invoked more frequently.
Next, we discuss how migration is performed.

\stitle{Migration Approach.}
Given the existing partitioning $P=\{\pp_1,\pp_2,$ $\ldots,\pp_{\alpha}\}$
and the new partitioning $P'=\{\pp'_1,\pp'_2,...,\pp'_{\beta}\}$ identified by \appr,
we need an algorithm to efficiently migrate the data from $P$ to $P'$ without
dropping all existing tables and recreating the partitions from scratch, which could be very costly.
\techreport{The question asked here is whether we can make use of the existing tables and only perform some small modifications accordingly.}
To do so, \orpheus needs to identify, for every $\pp'_i\in P'$, the closest partition $\pp_j\in P$,
in terms of modification cost, defined as  $|\rr'_i \setminus \rr_j|+|\rr_j \setminus \rr'_i|$, where
$\rr'_i \setminus \rr_j$ and $\rr_j \setminus \rr'_i$ are the records needed to be inserted and deleted respectively
to transform $\pp_j$ to $\pp'_i$.
\papertext{Since this is expensive to calculate, \orpheus instead approximates this quantity by using the number
of versions in common between $\pp_i$ and $\pp'_i$, along with operating on the version graph to identify
common records, and then greedily identifying the ``closest'' partitions.}
\techreport{This task consists of two main steps: 1) calculate the number of modifications needed for each partition pair $(\pp'_i, \pp_j)$; 2) find the closest partition $\pp_j$ for each $\pp'_i\in P'$.
For step one, if we calculate the modification cost directly based on $\rr'_i$ and $\rr_j$,  it may be very expensive especially when the number of records is large. Instead, we first find the common versions in $\pp'_i$ and $\pp_j$, and then calculate the number of common records based on the version graph $\gbb$ without probing into $\rr'_i$ or $\rr_j$. Next, for step two, we greedily pick the partition pair $(\pp'_i, \pp_j)$ with the smallest modification cost and assign $\pp_j$ to $\pp'_i$. Finally, we perform insertions and deletions on $\pp_j$ accordingly to obtain $\pp'_i$. Note that if the modification cost is larger than $|\rr'_i|$, we would prefer to build partition $\pp'_i$ from scratch rather than modifying the existing partition $\pp_k$. 
}

%% file: evaluation.tex
\section{Partitioning Evaluation}\label{sec:evaluation}


While Section~\ref{ssec:datamodelExp} explores the performance of data models, this section evaluates
the impact of partitioning. 
In Section~\ref{exp:alg_comp} we evaluate if \appr can be more efficient than
existing partitioning techniques; in Section~\ref{exp:partition}, 
we ask whether versioned databases strongly benefit
from partitioning;
and lastly, in Section~\ref{exp:inc} we evaluate how \appr performs for online scenarios.

\subsection{Experimental Setup} \label{exp:setup}
\vspace{-5pt}
\stitle{Datasets.} We evaluated the performance
of \appr using the versioning benchmark
datasets from Maddox et al.~\cite{maddox2016decibel}\papertext{;
see details in~\cite{orpheustr}}.
The versioning model used in the benchmark is similar
to {\tt git},
where a branch is a working copy of a dataset.
\techreport{For simplicity, we can think of branches
as different users' working copies.}
We selected the Science (SCI) and Curation (CUR)
workloads since they are most representative of
real-world use cases.
The SCI workload simulates the working patterns
of data scientists, who often take copies
of an evolving dataset for isolated data analysis.
\techreport{The version graph here can be visualized
as a mainline (i.e., a single linear version chain)
with various branches at different points---both
from different points on the mainline as well as from other already existing branches.}
Thus, the version graph is analogous to a tree with branches.
The CUR workload simulates the evolution of a canonical dataset
that many individuals are contributing to---these individuals
not just branch from the canonical dataset but also periodically
merge their changes back in, resulting in a DAG of versions.
\techreport{Branches can be created from existing branches, and then merged
back into the parent branch.}
We varied the following parameters when we generated
the benchmark datasets: the number of branches $\mathbb{B}$,
the total number of records $|R|$, as well as
the number of inserts (or updates) from parent version(s) $\mathbb{I}$.
We list our configurations
in Table~\ref{table:dataset}.
For instance, dataset \code{SCI\_1M} represents
a SCI workload dataset where the input parameter
corresponding to $|R|$ in the dataset generator is set to $1M$ records.
\techreport{Note that due to the inherent randomness in the dataset generator,
the actual number of records generated does not perfectly
match the value of $|R|$ we input to the generator.
Furthermore, since the version graphs for all \code{CUR\_*} datasets are DAGs (i.e., have multiple merges between versions),
we also list their $|\hat{R}|$, the number of duplicated records
described in Appendix~\ref{ssec:general_case}.
Compared with $|R|$,  $|\hat{R}|$ is about 7 to 10 percent of $|R|$.}
In all of our datasets,
each record contains 100 attributes,
each of which is a 4-byte integer.

\techreport{
\begin{table}[h!]
\centering
\scriptsize
\begin{center}
\vspace{-2mm}
    \begin{tabular}{| c | c | c | c | c|c |c|}
    \hline
    Dataset & $|V|$ & $|R|$ & $|E|$ & $|\mathbb{B}|$ & $|\mathbb{I}|$ & $|\hat{R}|$ \\ \hline \hline
    \code{SCI\_1M} & 1K & 944K & 11M &  100 & 1000 &-\\ \hline
    \code{SCI\_2M} & 1K & 1.9M & 23M &  100 & 2000 &-\\ \hline
    \code{SCI\_5M} & 1K & 4.7M & 57M  & 100 & 5000 &-\\ \hline
    \code{SCI\_8M} & 1K & 7.6M & 91M &  100 & 8000 &-\\ \hline
    \code{SCI\_10M} & 10K & 9.8M & 556M & 1000 & 1000 &-\\ \hline
    \code{CUR\_1M} & 1.1K & 966K & 31M  & 100 & 1000 & 90K\\ \hline
    \code{CUR\_5M} & 1.1K & 4.8M & 157M  & 100 & 5000 & 0.35M\\ \hline
    \code{CUR\_10M} & 11K & 9.7M & 2.34G & 1000 & 1000 & 0.9M \\ \hline
    \end{tabular}
\end{center}
   \vspace{-4mm}
    \caption{Dataset Description}
    \label{table:dataset}
    \vspace{-5mm}
\end{table}}

\papertext{
    \begin{table}[h!]
\centering
\scriptsize
\begin{center}
\vspace{-1mm}
    \begin{tabular}{| c | c | c | c | c|c |}
    \hline
    Dataset & $|V|$ & $|R|$ & $|E|$ & $|\mathbb{B}|$ & $|\mathbb{I}|$  \\ \hline \hline
    \code{SCI\_1M} & 1K & 944K & 11M &  100 & 1000 \\ \hline
    \code{SCI\_2M} & 1K & 1.9M & 23M &  100 & 2000 \\ \hline
    \code{SCI\_5M} & 1K & 4.7M & 57M  & 100 & 5000 \\ \hline
    \code{SCI\_8M} & 1K & 7.6M & 91M &  100 & 8000 \\ \hline
    \code{SCI\_10M} & 10K & 9.8M & 556M & 1000 & 1000\\ \hline
    \code{CUR\_1M} & 1.1K & 966K & 31M  & 100 & 1000 \\ \hline
    \code{CUR\_5M} & 1.1K & 4.8M & 157M  & 100 & 5000 \\ \hline
    \code{CUR\_10M} & 11K & 9.7M & 2.34G & 1000 & 1000 \\ \hline
    \end{tabular}
\end{center}
   \vspace{-5mm}
    \caption{Dataset Description}
    \label{table:dataset}
    \vspace{-4mm}
\end{table}
}

\stitle{Setup.} We conducted our evaluation on a
HP-Z230-SFF workstation with an Intel Xeon E3-1240 CPU
and 16 GB memory running Linux OS (LinuxMint).
We built \orpheus as a wrapper written in C++
over \postgres 9.5\techreport{\footnote{\scriptsize \postgres's version 9.5
added the feature of dynamically
adjusting the number of buckets for hash-join.}},
where we set the memory for sorting and hash operations as $1GB$\techreport{~(i.e., \texttt{work\_mem=1GB})
to reduce external memory sorts and joins}.
In addition, we set the buffer cache size to be minimal\techreport{~(i.e., \texttt{shared\_buffers} \texttt{=128KB}) }
to eliminate the effects of caching on performance.
In our evaluation, for each dataset,
we randomly sampled 100 versions and used them to get an estimate of the checkout time.
Each experiment was repeated 5 times, with the OS page cache being cleaned before each run.
Due to experimental variance,
we discarded the largest and smallest number among the five trials,
and then took the average of the remaining three trials.

\stitle{Algorithms.} We compare \appr against
two partitioning algorithms in the
NScale graph partitioning project~\cite{quamar2014nscale}:
the Agglomerative Clustering-based one (Algorithm 4 in~\cite{quamar2014nscale})
and the K\-Means Clustering-based one (Algorithm 5 in~\cite{quamar2014nscale}),
denoted as \aggl and \kmeans respectively: \kmeans
had the best performance, while \aggl is an intuitive method
for clustering versions.
After mapping their setting into ours,
like \appr, NScale~\cite{quamar2014nscale}'s algorithms group versions
into different partitions while allowing the duplication of records.
However, the NScale algorithms
are tailored for arbitrary graphs,
not for bipartite graphs (as in our case).

We implement \aggl and \kmeans as described.
\aggl starts with each version as
one partition and then sorts these partitions
based on a shingle-based\techreport{\footnote{\scriptsize Shingles are calculated
as signatures of each partition based on a min-hashing based technique.}} ordering.
Then, in each iteration, each partition is merged with a candidate partition
that it shares the largest number of common shingles with.
The candidate partitions have to satisfy two conditions
(1) the number of the common shingles is larger than a threshold $\tau$,
which is set via a uniform sampling-based method, and
(2) the number of records in the new partition after merging is smaller than a constraint $BC$\techreport{, a pre-defined maximum number of records per partition}.
\techreport{Furthermore, based on the shingle ordering, NScale proposes that each partition only
considers its following $l$ partitions as its merging candidates and $l$
is adjusted dynamically. In our experiments, initially $l$ is set to 100.}
To address Problem~\ref{prob:min_rec} with storage
threshold $\gamma$, we conduct a binary search on $BC$
and find the best partitioning scheme under the storage constraint.

For \kmeans, there are two input parameters:
partition capacity $BC$ as in \aggl,
and the number of partitions $K$.
Initially, $K$ random versions are assigned to partitions\techreport{,
the centroid of which is initialized
as the set of records in each partition}.
Next, we assign the remaining versions
to their nearest centroid based on the number of common records,
after which each centroid is updated to
the union of all records in the partition.
In subsequent iterations, each version is moved to a partition,
such that after the movement, the total number of records
across partitions is minimized, while respecting the constraint
that the number of records in each partition is no larger than $BC$.
The number of \kmeans iterations is set to 10.
In our experiment, we vary $K$ and set $BC$ to be infinity.
We tried other values for $BC$;
the results are similar to that when $BC$ is infinity.
Overall, with the increase of $K$,
the total storage cost increases and the checkout cost decreases.
Again, we use binary search to find the best $K$ for \kmeans
and minimize the checkout cost under the storage
constraint $\gamma$ for Problem~\ref{prob:min_rec}.

\subsection{Comparison of Partitioning Algorithms} \label{exp:alg_comp}
In these experiments,
we consider both datasets where the version graph is a tree,
i.e., there are no merges (\papertext{\code{SCI\_5M} and \code{SCI\_10M}}\techreport{\code{SCI\_1M, SCI\_5M} and \code{SCI\_10M}}),
 and datasets where the version graph is a DAG (\papertext{\code{CUR\_5M} and \code{CUR\_10M}}\techreport{\code{CUR\_1M, CUR\_5M} and \code{CUR\_10M}}).
 \papertext{Experiments on additional datasets can be found in~\cite{orpheustr}.}
 We first compare the effectiveness of different
 partitioning algorithms: \appr, \aggl and \kmeans,
 in balancing the storage size and the checkout time.
Then, we compare the efficiency of these algorithms
 by measuring their running time.
\techreport{\begin{figure*}[t]
\centering
\vspace{-4mm}
\includegraphics[width=0.8\linewidth]{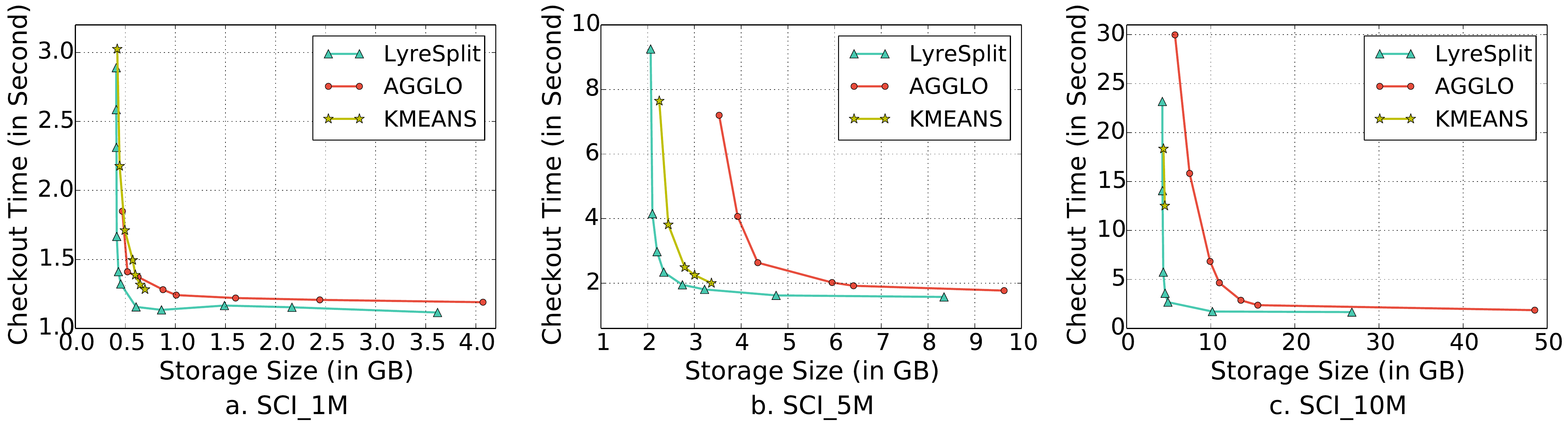}
\vspace{-4mm}
\end{figure*}
\begin{figure*}[t]
\centering
\includegraphics[width=0.8\linewidth]{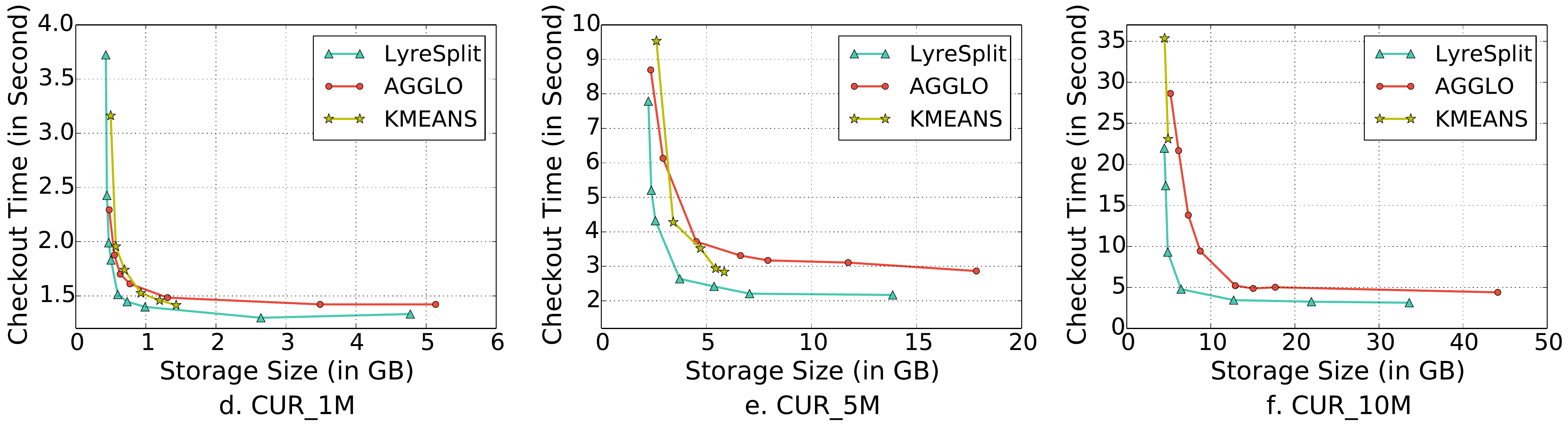}
\vspace{-3mm}
\caption{Storage Size vs. Checkout Time}
\label{fig:real_cur}
\vspace{-5mm}
\end{figure*}
}

\papertext{
\begin{figure}[h]
\centering
\vspace{-3mm}
\includegraphics[width=\linewidth]{fig/real_SCI5M10M.pdf}
\vspace{-11mm}
\end{figure}
\begin{figure}[h]
\centering
\includegraphics[width=\linewidth]{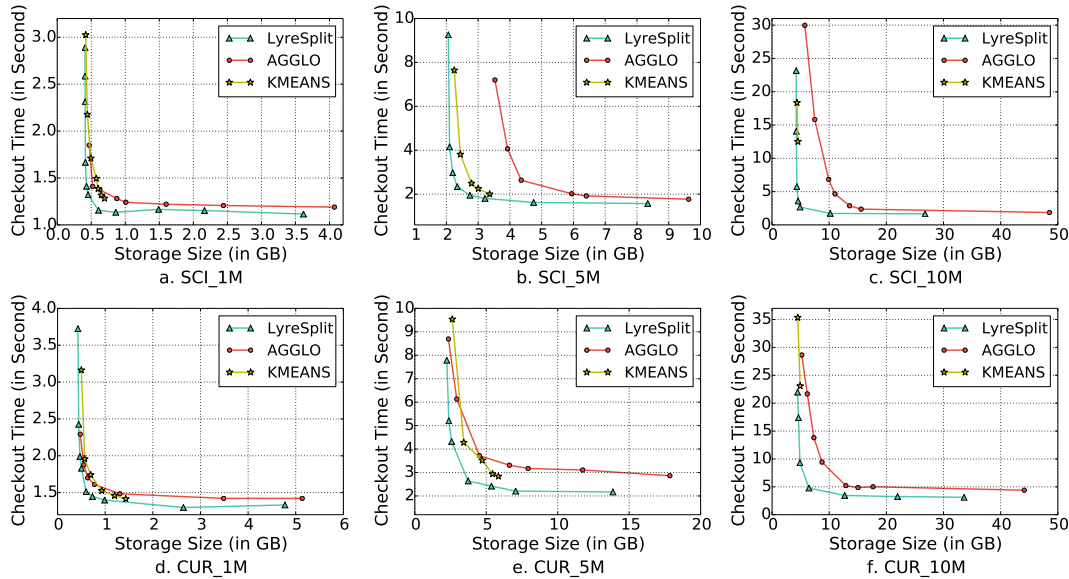}
\vspace{-6mm}
\caption{Storage Size vs. Checkout Time}
\label{fig:real_cur}
\vspace{-5mm}
\end{figure}
}

\stitle{Effectiveness Comparison.}

\vspace{1mm}
\noindent
\fbox{\begin{minipage}{26em} \small
    \textit{Summary of Trade-off between Storage Size and Checkout Time.}   \appr dominates \aggl and \kmeans with respect to the storage size and checkout time after partitioning, i.e., with the same storage size, \appr's partitioning scheme provides a smaller checkout time.
\end{minipage}}
\vspace{0.3mm}

In order to trade-off between $\mathcal{S}$ and $\cc_{avg}$,
we vary $\delta$ for \appr, $BC$ for \aggl and $K$ for \kmeans
to obtain the overall trend between the storage size and the checkout time.
The results are shown in Figure~\ref{fig:real_cur},
where the x-axis depicts the total storage size
for the data table in gigabytes (GB) and the
y-axis depicts the average checkout time in seconds for
the 100 randomly selected versions.
\techreport{Recall that for a \cvd,
its versioning table is of
constant storage size
for different partitioning schemes,
so we do not include this in the storage size computation.}
Each point in Figure~\ref{fig:real_cur} represents a
partitioning scheme obtained by one algorithm with a specific input parameter value.
We terminated the execution of \kmeans when its running time
exceeded 10 hours for each $K$,
which is why there are only two points with star markers
in \techreport{Figure~\ref{fig:real_cur}(c) and \ref{fig:real_cur}(f)}\papertext{Figure~\ref{fig:real_cur}(b) and \ref{fig:real_cur}(d)} respectively.
The overall trend for
\aggl, \kmeans, and \appr
is that with the increase in storage size,
the average checkout time first decreases
and then tends to a constant value---the average checkout time
when each version is stored as a separate table\techreport{,
which in fact corresponds to the smallest possible checkout time}.
\techreport{For instance, in Figure~\ref{fig:real_cur}(f) with \appr,
the checkout time decreases from 22s to 4.8s
as the storage size increases from 4.5GB to 6.5GB,
and then converges at around 2.9s.}

Furthermore, \appr has better performance
than the other two algorithms in both the SCI and CUR datasets
in terms of the storage size and the checkout time,
as shown in Figure~\ref{fig:real_cur}.
For instance, in \techreport{Figure~\ref{fig:real_cur}(b)}\papertext{Figure~\ref{fig:real_cur}(a)},
with 2.3GB storage budget, \appr can provide a
partitioning scheme taking 2.9s for checkout on average,
while both \kmeans and \aggl
give schemes taking more than 7s for checkout.
Thus, with equal or lesser storage size, the partitioning scheme
selected by \appr achieves
much less checkout time than the
ones proposed by \aggl and \kmeans, especially when the storage budget is small.
The reason for this is that
\appr takes a ``global'' perspective to partitioning,
while \aggl and \kmeans take a ``local'' perspective.
Specifically, each split in \appr is decided based
on the derivation structure and similarity between various versions,
as opposed to greedily merging partitions 
with partitions
in \aggl, and moving versions between partitions in \kmeans.


\begin{figure}[h!]
\centering
    \papertext{\vspace{-3mm}}
    \centering
    \includegraphics[width=\linewidth]{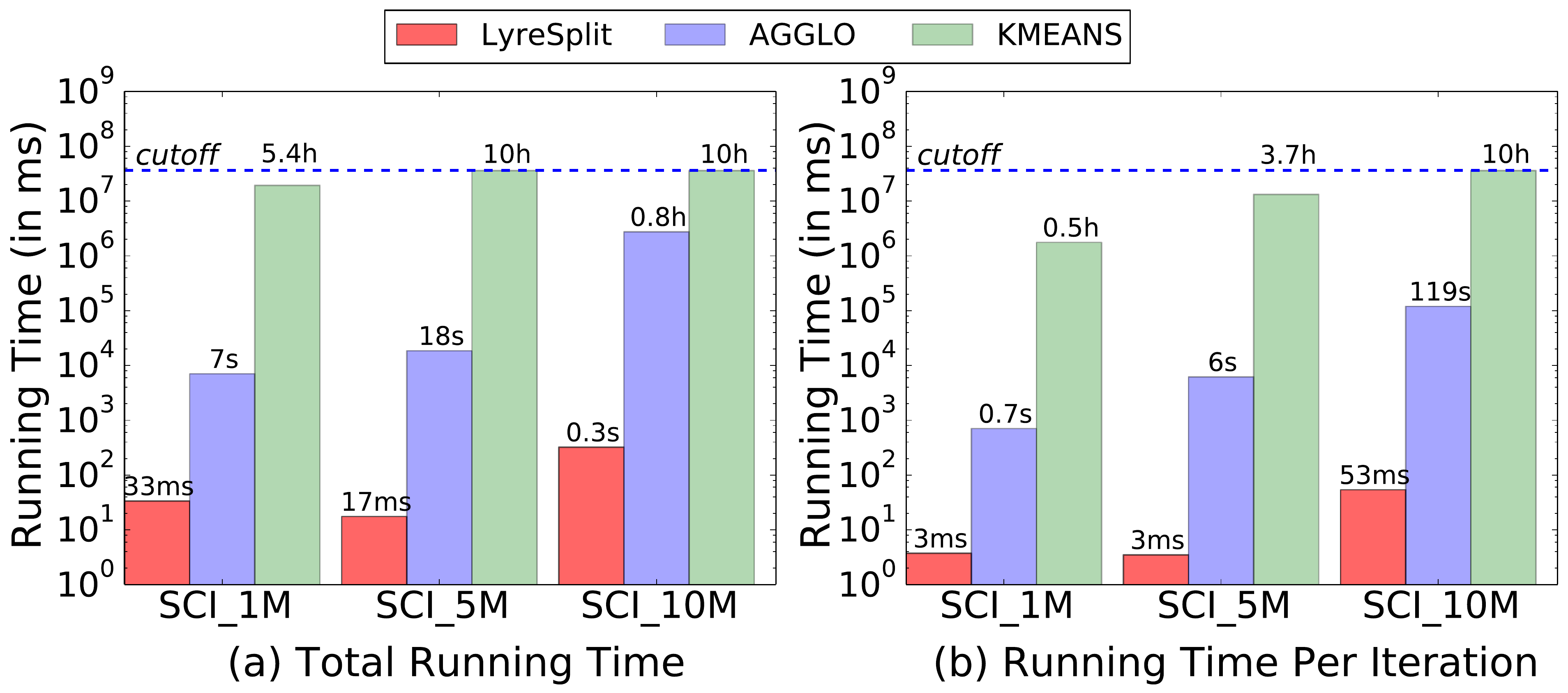}
    \vspace{-6mm}
    \caption{Algorithms' Running Time Comparison (\code{SCI\_*})}
    \label{fig:algT_SCI_all}
    \vspace{-10pt}
\end{figure}

\techreport{
\begin{figure}[h!]
    \centering
    \vspace{-2pt}
    \includegraphics[width=\linewidth]{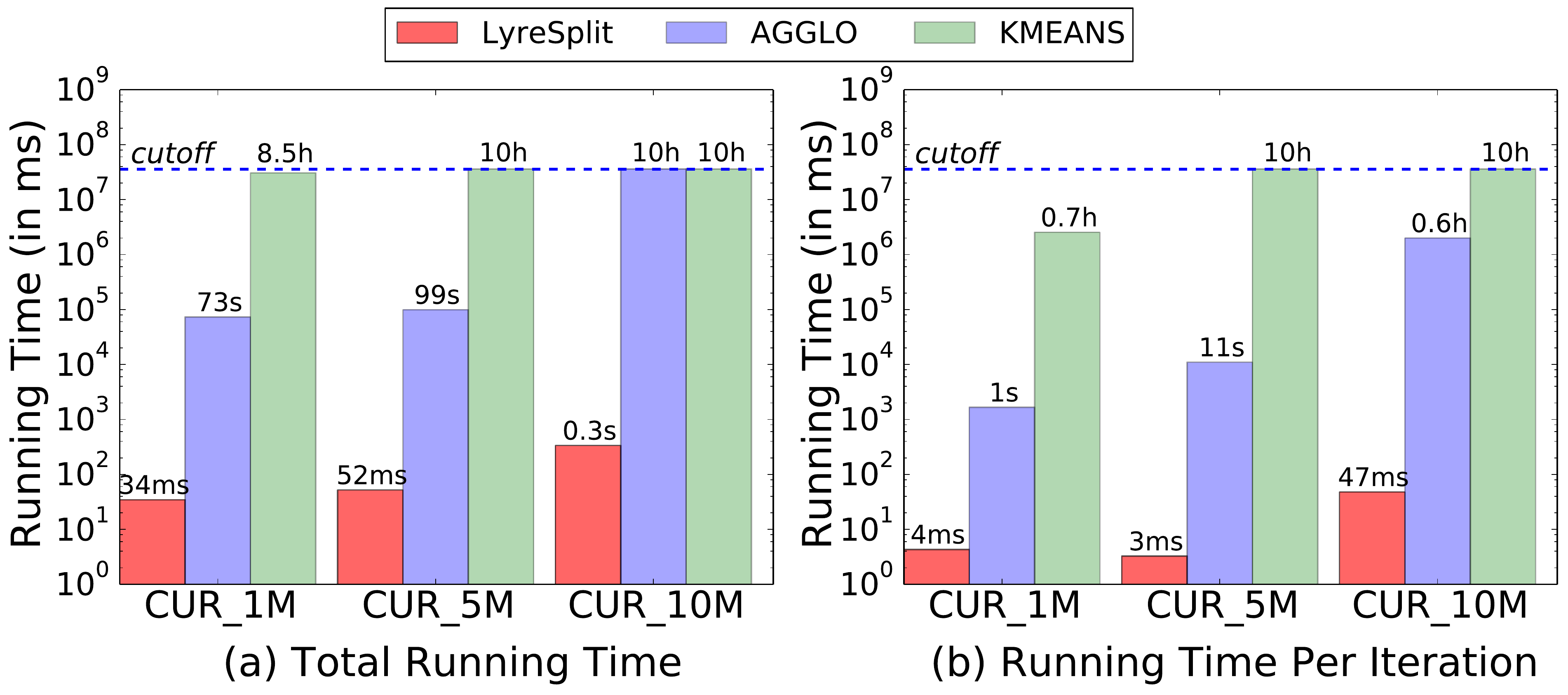}
    \vspace{-6mm}
    \caption{Algorithms' Running Time Comparison (\code{CUR\_*})}
    \label{fig:algT_CUR_all}
    \vspace{-10pt}
\end{figure}
}

\stitle{Efficiency Comparison.}

\vspace{1mm}
\noindent
\fbox{\begin{minipage}{26em}\small

    \textit{Summary of Comparison of Running Time of Partitioning Algorithms.}
    When minimizing the checkout time
    under a storage constraint (Problem~\ref{prob:min_rec}),
    \appr is on average $10^3 \times$ faster than \aggl,
    and more than $10^5 \times$ faster than \kmeans for all \code{SCI\_*} and \code{CUR\_*}  datasets.

\end{minipage}}
\vspace{0.3mm}

\techreport{As discussed, given a storage constraint
in Problem~\ref{prob:min_rec},
we use binary search to find the best $\delta$, $BC$, and $K$ for
\appr, \aggl and \kmeans respectively.}
In this experiment, we set the storage threshold as $\gamma = 2|R|$,
and terminate the binary search process when the
resulting storage cost $\mathcal{S}$ meets the constraint:
$0.99\gamma \leq \mathcal{S}\leq \gamma$.
\papertext{We discuss the results for the \code{SCI} datasets:
the \code{CUR} dataset performance is similar~\cite{orpheustr}.}
Figure~\ref{fig:algT_SCI_all}a\techreport{~and \ref{fig:algT_CUR_all}a}
shows the total running time during the end-to-end binary search process,
while Figure~\ref{fig:algT_SCI_all}b\techreport{~and \ref{fig:algT_CUR_all}b}
shows the running time per binary search iteration.
Again, we terminate \kmeans and \aggl when the running time exceeds 10 hours\techreport{,
thus we cap the running time in Figure~\ref{fig:algT_SCI_all}\techreport{~and \ref{fig:algT_CUR_all}}
at 10 hours.
We can see that \appr takes much less time than \aggl and \kmeans}.
Consider the largest dataset \code{SCI\_10M} in Figure~\ref{fig:algT_SCI_all}
as an example: with \appr the entire binary search procedure
and each binary search iteration took 0.3s and 53ms respectively;
\aggl takes 50 minutes in total;
while \kmeans does not even finish a single iteration
in 10 hours.

Overall, \appr is $10^2 \times$
faster than \aggl for \code{SCI\_1M}, $10^3 \times$ faster for \code{SCI\_5M}, and $10^4 \times$ faster for \code{SCI\_10M} respectively\techreport{~(and is $10^3 \times$
faster than \aggl for \code{CUR\_1M} and \code{CUR\_5M} and
$10^5 \times$ faster for \code{CUR\_10M} respectively)},
and more than $10^5 \times$ faster than \kmeans for all datasets.
This is mainly because \appr only needs to operate
on the version graph while \aggl and \kmeans
operate on the version-record bipartite graph,
 which is much larger than the version graph.
 Furthermore, \kmeans can only finish the binary search process
 within 10 hours for \code{SCI\_1M}\techreport{~and \code{CUR\_1M}}.
 \techreport{This algorithm is extremely slow due to the pairwise comparison
 between each version with each centroid in each iteration,
 especially when the number of centroids $K$ is large.
 Referring back to Figure~\ref{fig:real_cur}(f),
 the running times for the left-most point on the \kmeans line
 takes 3.6h with $K=5$, while the right-most point takes 8.8h with $K=10$.}
 Thus our proposed \appr is much more scalable
 than \aggl and \kmeans.
 Even though \kmeans is closer to \appr in performance (as seen in the previous
 experiments),
 it is impossible to use in practice.



\subsection{Benefits of Partitioning} \label{exp:partition}
\vspace{1mm}
\noindent
\fbox{\begin{minipage}{26em}\small
    \textit{Summary of Checkout Time Comparison with and without Partitioning:} With only
    a $2\times$ increase on the storage,
    we can achieve a substantial $3\times$, $10\times$ and $21\times$
    reduction on checkout time for \code{SCI\_1M}, \code{SCI\_5M}, and \code{SCI\_10M}\techreport{, and $3\times$, $7\times$ and $9\times$ reduction for \code{CUR\_1M}, \code{CUR\_5M}, and \code{CUR\_10M} }respectively.
\end{minipage}}
\vspace{0.5mm}

\begin{figure}[h!]
\centering
    \centering
     \vspace{-6pt}
    \includegraphics[trim=0cm 0cm 0cm 0cm, width=\linewidth]{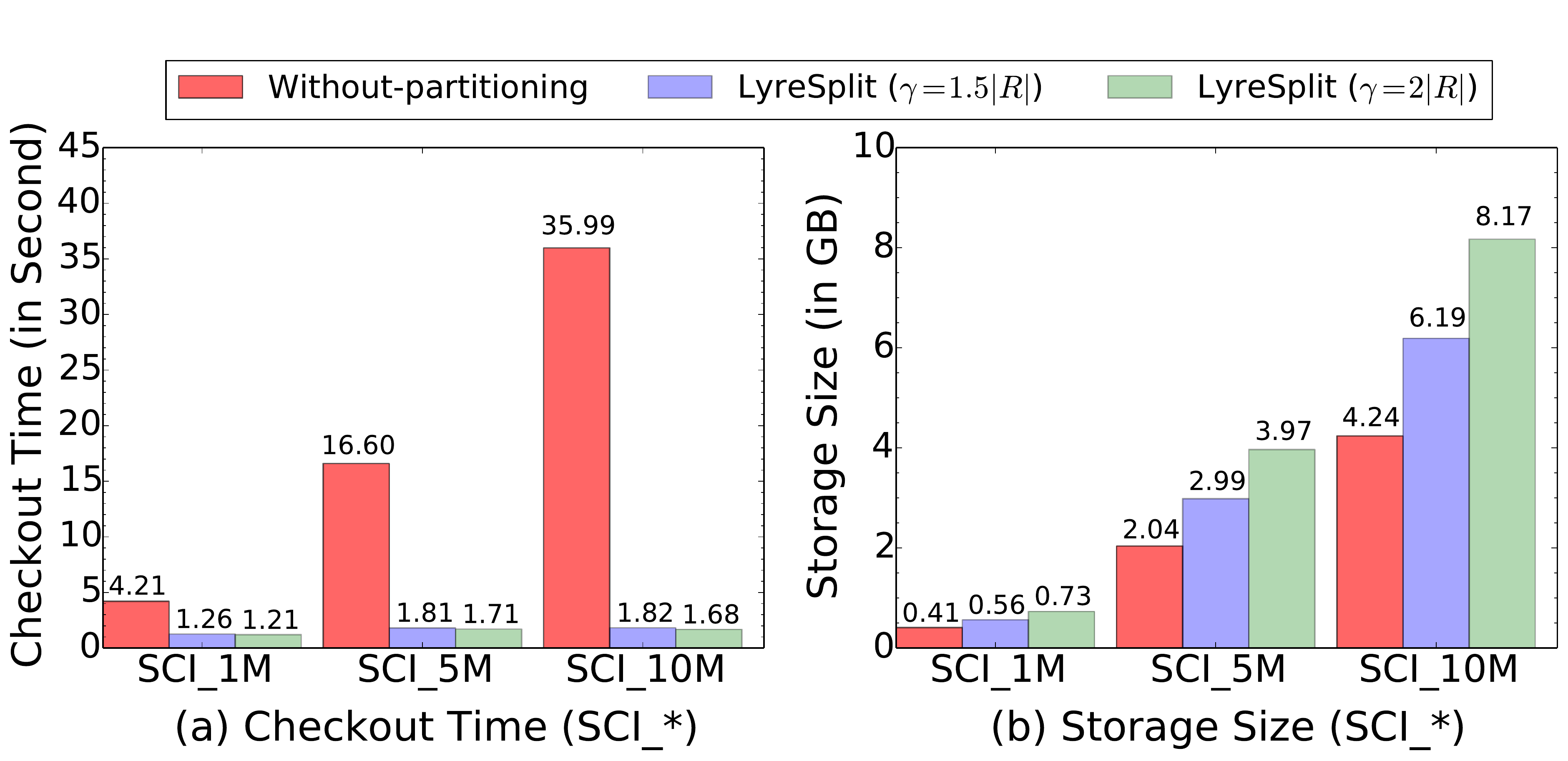}
    \vspace{-2mm}
    \caption{Comparison With and Without Partitioning}
    \label{fig:before_after_partitioning_sci}
    \vspace{-15pt}
\end{figure}

We now study the impact of partitioning and demonstrate
that with a relatively small increase in storage,
the checkout time can be substantially reduced.
We conduct two sets of experiments
with the storage threshold as $\gamma=1.5\times |R|$
and $\gamma=2\times |R|$ respectively,
and compare the average checkout time with and without partitioning.
\papertext{Figure~\ref{fig:before_after_partitioning_sci}(a) illustrates
the comparison on the checkout time for
different datasets,
and Figure~\ref{fig:before_after_partitioning_sci}(b) displays the
corresponding storage size comparison.}\techreport{Figure~\ref{fig:before_after_partitioning_sci} and \ref{fig:before_after_partitioning_cur} illustrate
the comparison on checkout time and storage size for \code{SCI\_*} and \code{CUR\_*} respectively.}
Each collection of bars in {Figure~\ref{fig:before_after_partitioning_sci}}\techreport{~and Figure~\ref{fig:before_after_partitioning_cur}} corresponds to one dataset.
Consider \code{SCI\_5M} in Figure~\ref{fig:before_after_partitioning_sci} as an example: the checkout time without partitioning is 16.6s
while the storage size is 2.04GB;
when the storage threshold is set to be $\gamma=2\times |R|$,
the checkout time after partitioning is 1.71s and the storage size is 3.97GB.
As illustrated in Figure~\ref{fig:before_after_partitioning_sci},
with only $2\times$ increase in the storage size,
we can achieve $3\times$ reduction on \code{SCI\_1M},
$10\times$ reduction on \code{SCI\_5M}, and
$21\times$ reduction on \code{SCI\_10M} for the
average checkout time compared to that without partitioning.
Thus, with partitioning,
we can eliminate the time for accessing irrelevant records.
Consequently, the checkout time remains small even for large datasets.
\techreport{
The results shown in Figure~\ref{fig:before_after_partitioning_cur} are similar to those in
Figure~\ref{fig:before_after_partitioning_sci}:
with $2\times$ increase on the storage size,
we can achieve $3\times$ reduction on \code{CUR\_1M},
$7\times$ reduction on \code{CUR\_5M},
and $9\times$ reduction on \code{CUR\_10M} for average checkout
time compared to that without partitioning.
However, the reduction in Figure~\ref{fig:before_after_partitioning_cur}a
is smaller than that in Figure~\ref{fig:before_after_partitioning_sci}a.
The reason is the following.
We can see that the checkout time without partitioning
is similar for \code{SCI} and \code{CUR} datasets,
but the checkout time after partitioning
for \code{CUR} dataset is greater than the corresponding \code{SCI} dataset.
This is because the average number of records in each version,
i.e., $\frac{|E|}{|V|}$, in \code{CUR}
is around 3 to 4 times greater than that in the corresponding \code{SCI},
as depicted in Table~\ref{table:dataset}.
Recall that $\frac{|E|}{|V|}$ is the minimal checkout
cost $\cavg$ after partitioning as stated in Observation~\ref{obs:min_checkout}.
Thus, the smallest possible {\em checkout} time for \code{CUR},
which is where the blue lines with triangle markers (corresponding to \appr)
in Figure~\ref{fig:real_cur}(d)(e)(f) converges to,
is typically larger than that
for the corresponding \code{SCI} in Figure~\ref{fig:real_cur}(a)(b)(c).
Overall, as demonstrated
in Figure~\ref{fig:before_after_partitioning_sci}
and \ref{fig:before_after_partitioning_cur}, with a small increase in
the storage size, we can reduce the average {\em checkout} time
to within a few seconds even when the number of records
in a \cvd increases dramatically. Referring back to our
motivating experiment in Figure~\ref{fig:model_exp}(c),
we claim that with partitioning the {checkout}
time using split-by-rlist is comparable to that by a-table-per-version.
}


\techreport{
\begin{figure}[h!]
\centering
    \vspace{7pt}
    \centering
    \includegraphics[trim=0cm 0cm 0cm 1cm, width=\linewidth]{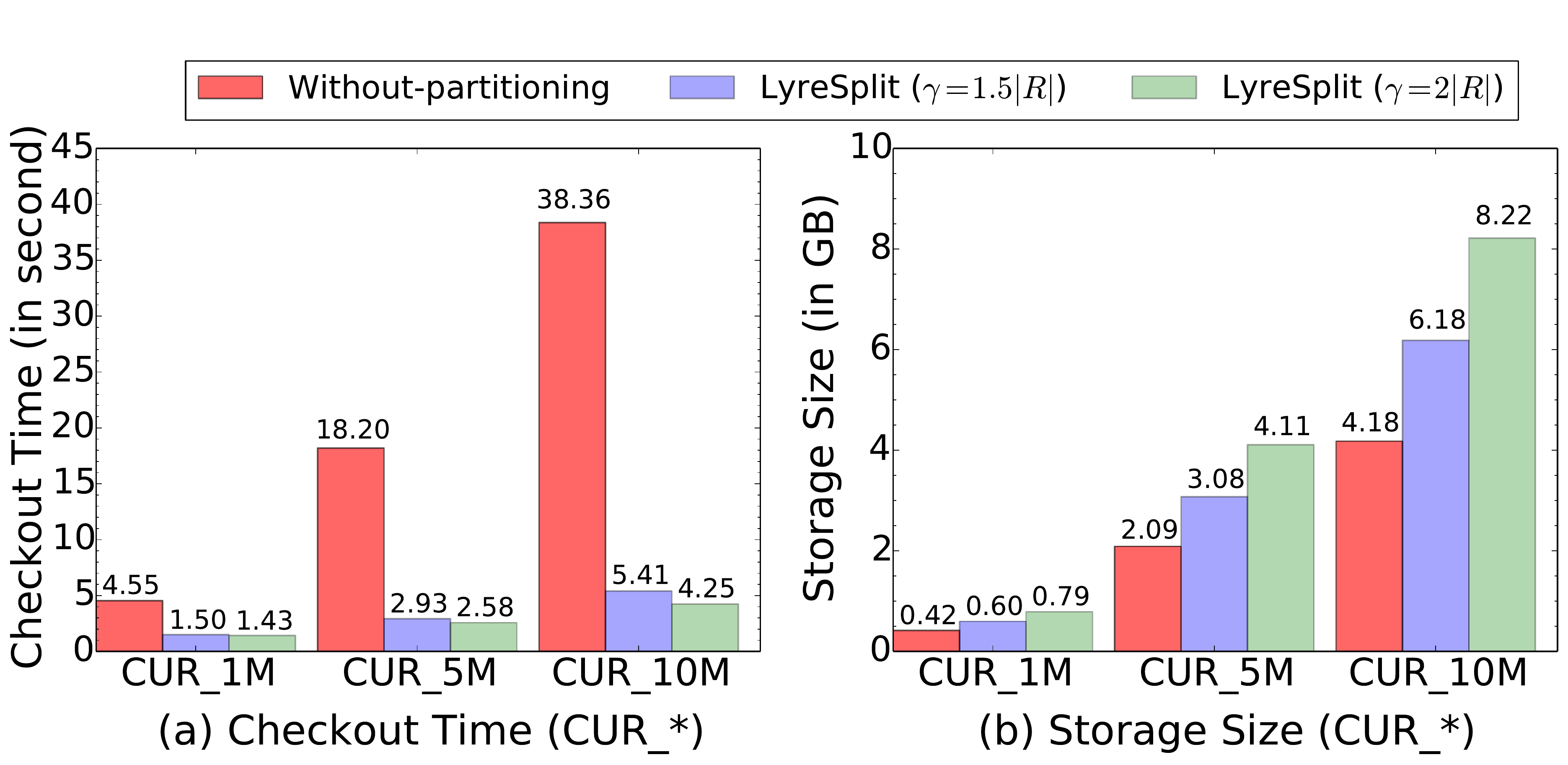}
    \vspace{-2mm}
    \caption{Comparison With and Without Partitioning}
    \label{fig:before_after_partitioning_cur}
    \vspace{-6mm}
\end{figure}
}

\input{exp_incremental.tex}

%% file: exp_incremental.tex
\subsection{Maintenance and Migration}\label{exp:inc}
We now evaluate the performance of \orpheus's partitioning optimizer
over the course of an extended period with many versions being committed to the system.
We employ our \code{SCI\_10M} dataset, which contains the largest number of versions (i.e. $10k$).
Here, the versions are streaming in continuously; as each version commits, 
we perform online maintenance based on the
mechanism described in Section~\ref{ssec:inc}. 
When $\frac{\cc_{avg}}{\cc^*_{avg}}$ reaches the tolerance factor $\mu$, 
the migration engine is automatically invoked, 
and starts to perform the migration of data from the old partitions 
to the new ones identified by \appr. 
We first examine how our online maintenance performs, 
and how frequently migration is invoked. 
Next, we test the latency of our proposed migration approach. 
\papertext{The storage threshold is set to be $\gamma = 1.5|R|$. 
Similar results on $\gamma = 2|R|$ can be 
found in our technical report~\cite{orpheustr}.}\techreport{The storage threshold is set to be $\gamma = 1.5|R|$ and $\gamma = 2|R|$ respectively.}


\stitle{Online Maintenance.}

\vspace{1mm}
\noindent
\fbox{\begin{minipage}{26em}\small

    \textit{Summary of Online Maintenance Compared to \appr.}
    With our proposed online maintenance mechanism, the checkout cost $\cc_{avg}$ diverges slowly from the best checkout cost $\cc^*_{avg}$ identified by \appr. When $\mu=1.5$, our migration engine is triggered only 7\techreport{~and 4} times across a total of 10,000 committed versions \techreport{when $\gamma = 1.5|R|$ and $\gamma = 2|R|$ respectively}. 
\end{minipage}}
\vspace{0.3mm}

\begin{figure}[h!]
\centering
	\vspace{-2mm}
	\begin{subfigure}{0.24\textwidth}
		\centering
		\includegraphics[width=\linewidth]{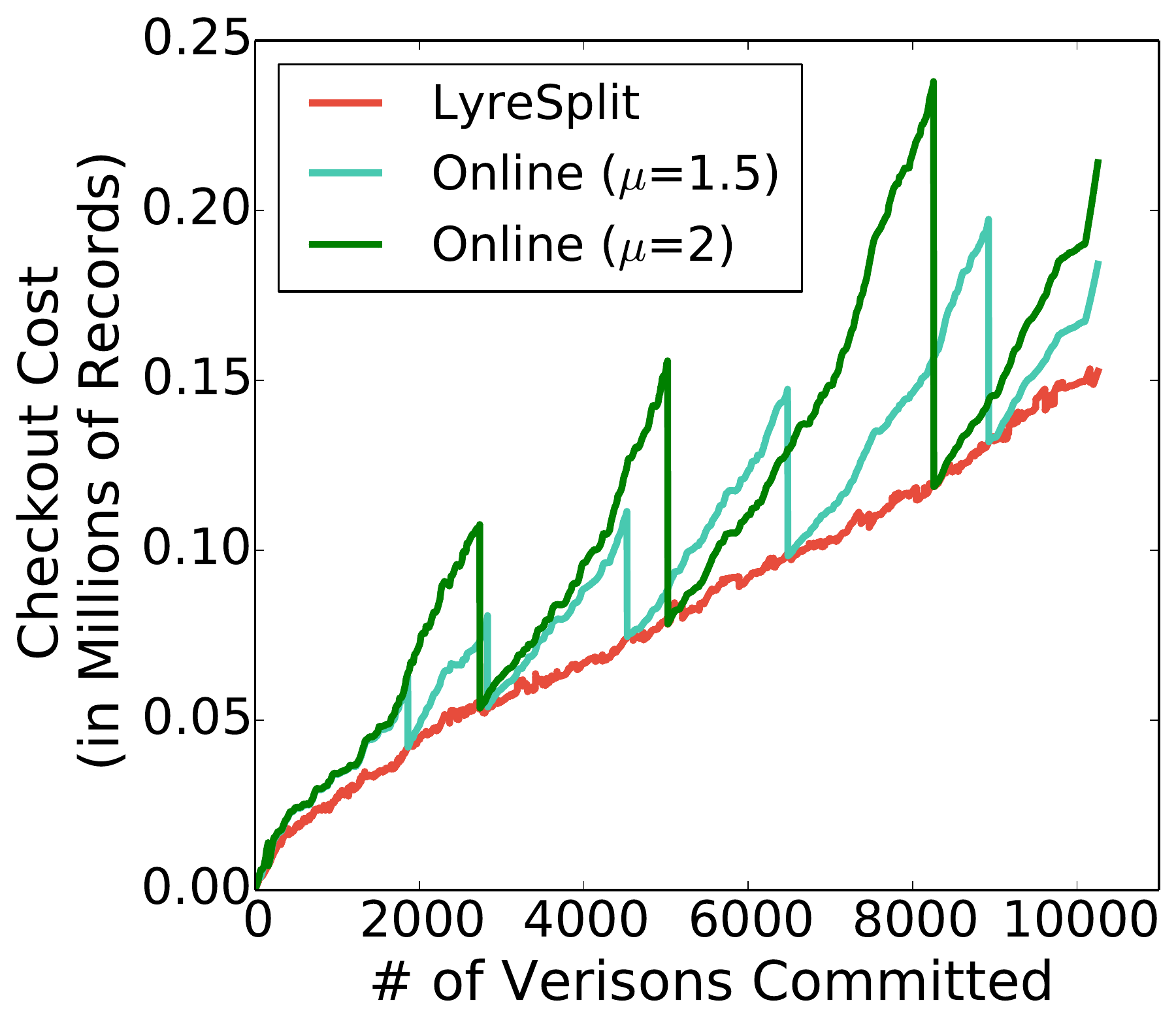}
		\vspace{-5mm}
		\caption{Online Maintenance}
		\label{fig:online}
	\end{subfigure}
	\begin{subfigure}{.225\textwidth}
		\centering
		\includegraphics[width=\linewidth]{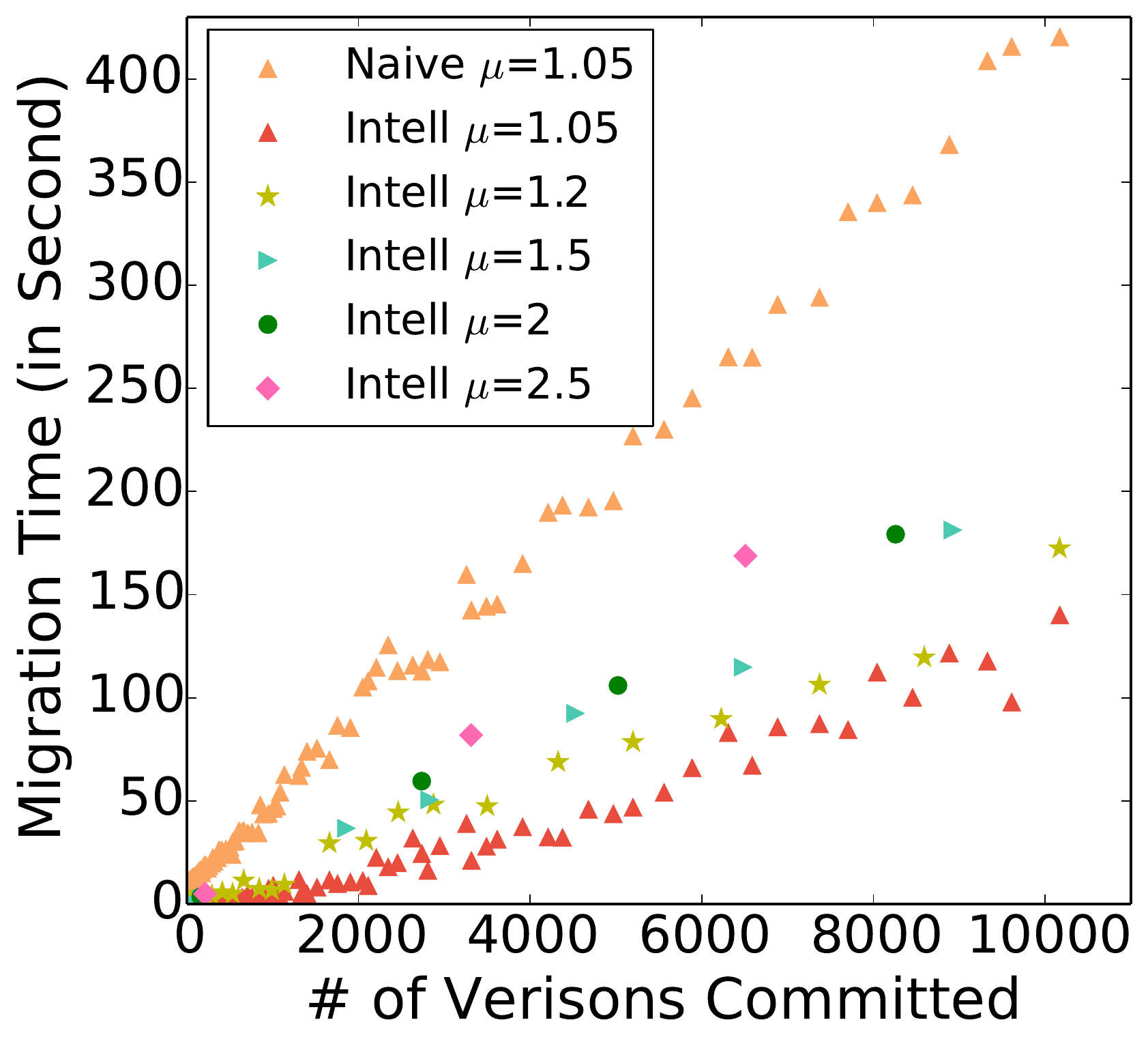}
		\vspace{-5mm}
		\caption{Migration Time}
		\label{fig:migration}
	\end{subfigure}
	\vspace{-2mm}
	\caption{Online Partitioning and Migration \papertext{(\code{SCI\_10M})}\techreport{($\gamma = 1.5|R|$)}}
	\label{fig:mig}
	\vspace{-5mm}
\end{figure}

\noindent As shown in Figure~\ref{fig:online}\techreport{~and \ref{fig:online_2}}, 
the red line depicts the best checkout cost $\cc^*_{avg}$ identified by \appr 
(note that \appr is lightweight and can be run very quickly after every commit),
while the blue and green lines illustrate the current checkout cost 
$\cc_{avg}$ with tolerance factor $\mu=1.5$ and $\mu=2$, respectively.
We can see that with online maintenance, the checkout cost $\cc_{avg}$ 
(blue and green lines) starts to diverge from $\cc^*_{avg}$ (red line).
 When $\frac{\cc_{avg}}{\cc^*_{avg}}$ exceeds the tolerance factor $\mu$, 
 the migration engine is invoked, and the blue and green lines 
 jump back to the red line once migration is complete.
\techreport{With the increase of $\mu$, the frequency of triggering migration
decreases.} 
As depicted in Figure~\ref{fig:online}, when $\mu=1.5$, migration is triggered 7 times,
while it is only triggered 3 times when $\mu=2$, across a total of 10000 versions committed. 
Thus, our proposed online maintenance performs well, 
diverging slowly from \appr. 
\techreport{This can be explained by the same intuition shared by the online maintenance scheme and \appr.}


\techreport{
\begin{figure}[h!]
\centering
	\begin{subfigure}{0.24\textwidth}
		\centering
		\includegraphics[width=\linewidth]{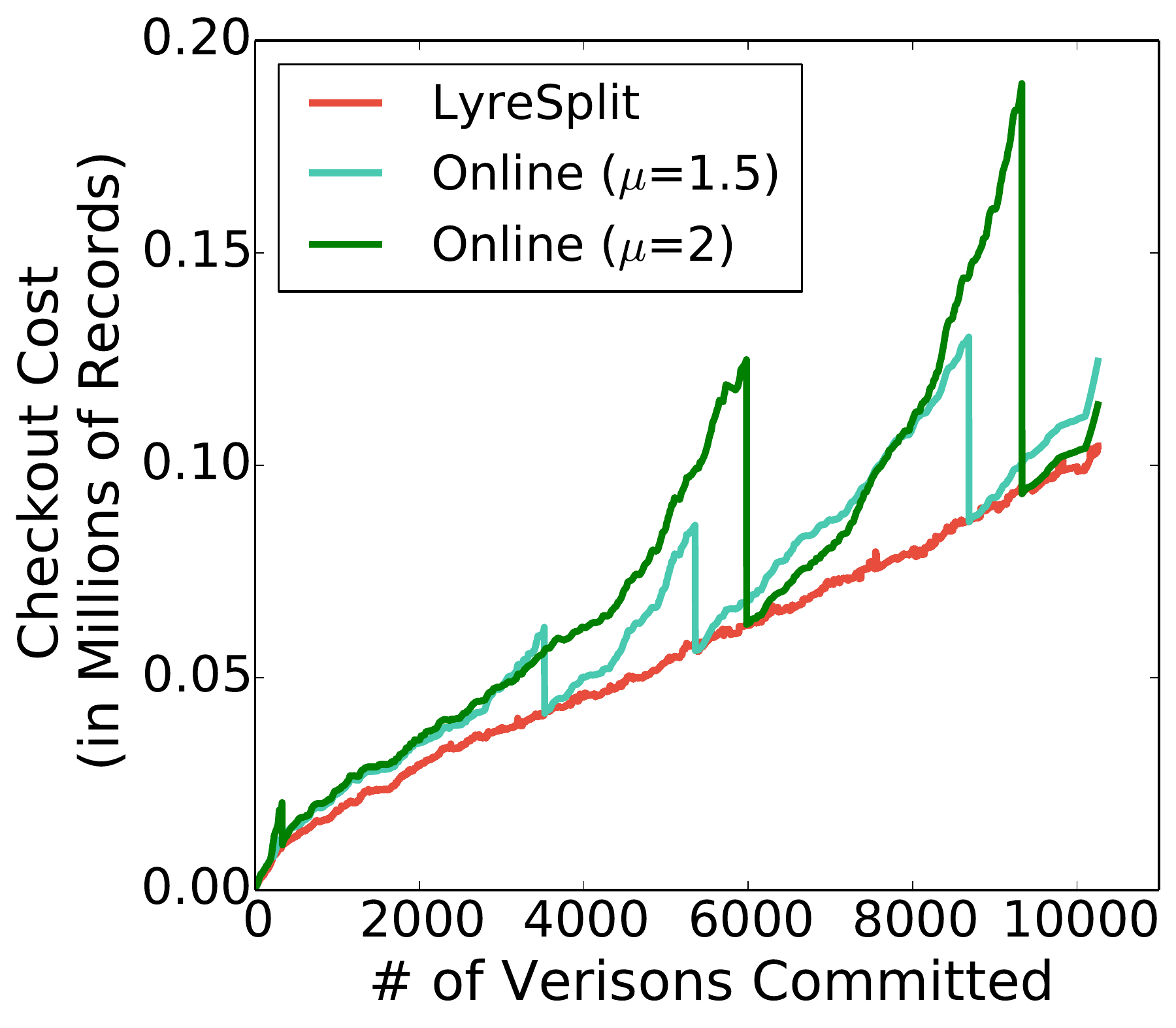}
		\vspace{-5mm}
		\caption{Online Maintenance}
		\label{fig:online_2}
	\end{subfigure}
	\begin{subfigure}{.225\textwidth}
		\centering
		\includegraphics[width=\linewidth]{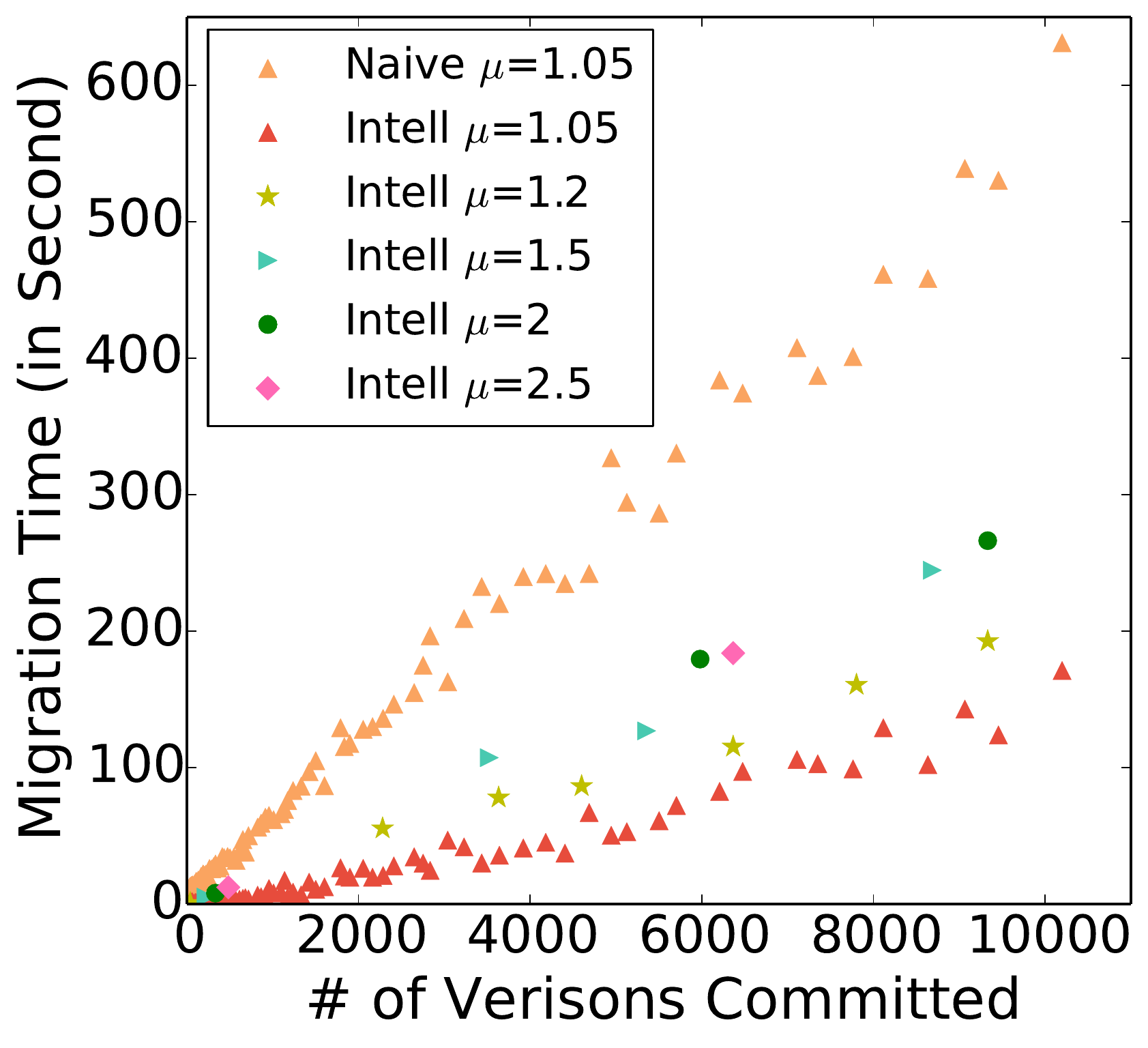}
		\vspace{-5mm}
		\caption{Migration Time}
		\label{fig:migration_2}
	\end{subfigure}
	\vspace{-2mm}
	\caption{Online Partitioning and Migration ($\gamma = 2|R|$)}
	\label{fig:mig_2}
	\vspace{-5mm}
\end{figure}
}

\stitle{Migration Time.}

\vspace{1mm}
\noindent
\fbox{\begin{minipage}{26em}\small
    \textit{Summary of Comparison of Running Time of Migration.}
    When $\mu=1.05$, the migration time with our proposed method is on 
    average $\frac{1}{10}$ of that with naive approach
	of rebuilding the partitions from scratch \techreport{when $\gamma = 1.5|R|$ and $\gamma = 2|R|$}.
    As $\mu$ decreases, the migration time with our proposed method decreases.
\end{minipage}}
\vspace{0.3mm}

Figure~\ref{fig:migration}\techreport{~and \ref{fig:migration_2}} \papertext{depicts}\techreport{depict} the migration time when 
the {\em migration engine} is invoked. 
Figure~\ref{fig:migration} is in correspondence with Figure~\ref{fig:online} 
sharing the same x-axis.
For instance, with $\mu=2$, when the $5024^{th}$ version commits, 
the {\em migration engine} is invoked as shown by the green 
line in Figure~\ref{fig:online}. 
Correspondingly, the migration takes place, 
and we record the migration time with the green circle ($\mu=2$) 
in Figure~\ref{fig:migration}. 
Hence, there are three green circles in Figure~\ref{fig:migration}, 
corresponding to the three migrations in Figure~\ref{fig:online}.
\techreport{Same are Figure~\ref{fig:online_2} and Figure~\ref{fig:migration_2}.}

We now compare our intelligent migration approach from Section~\ref{ssec:inc}, 
denoted {\em intell},
with the naive approach of rebuilding 
partitions from scratch, denoted {\em naive}.
The points with upward triangles 
in Figure~\ref{fig:migration} all have $\mu=1.05$,
with the red points representing {\em intell}, 
and the brown representing {\em naive}:
we see that {\em intell} takes at most $\frac{1}{3}$, 
and on average {\bf $\frac{1}{10}$} of the time of {\em naive}.
For the sake of clarity, we omit the 
migration times for different $\mu$ using {\em naive}, since
they roughly fall on the same line as that of $\mu=1.05$. 
Next, consider the migration time with different $\mu$ using {\em intell}. 
Overall, as $\mu$ decreases, the migration time decreases. 
To see this, one can connect the points corresponding to each $\mu$ 
(denoted using different markers) to form lines in Figure~\ref{fig:migration}. 
When $\mu$ is smaller, migration takes place more frequently, 
due to which the new partitioning scheme identified by \appr 
is more similar to the current one, 
and hence fewer modifications need to be performed. 
Essentially, we are amortizing the migration cost across multiple migrations.
\techreport{Similar results can be found in Figure~\ref{fig:mig_2} when $\gamma=2|R|$.}

%% file: related.tex
\section{Related Work}\label{sec:related}

We now survey work from multiple areas
related to \orpheus.

\stitle{Dataset Version Control.}
A recent vision paper on Datahub~\cite{bhardwaj2014datahub}
acknowledges the need for database
systems to support collaborative
data analytics---we execute on that vision 
by supporting collaborative  analytics
using a traditional relational database\techreport{,
thereby seamlessly leveraging the sophisticated
analysis capabilities}.
Decibel~\cite{maddox2016decibel} describes
a new version-oriented storage engine designed
``from the ground up'' to support versioning.
Unfortunately, the architecture
involves several choices that make it
impossible to support within a traditional relational
database without substantial cha\-nges at all layers of the stack.
For example, the eventual solution
requires the system
to log and query tuple membership on compressed bitmaps,
reason about and operate on ``delta files'',
and execute new and fairly complex
algorithms for even
simple operations such as branch (in our case checkout)
or merge (in our case commit).
It remains to be seen how this storage
engine can be made to interact with other
components, such as the parser,
the transaction manager, and the query optimizer\techreport{, and all
the other benefits that come ``for free'' with a relational database}.
We are approaching the problem from a
different angle---the angle of {\em reuse}: how
do we leverage relational databases
to support versioning without any substantial
changes to existing databases, which have
massive adoption and open-source development
that we can tap into.
Recent work on the principles of dataset
versioning is also relevant~\cite{bhattacherjee2015principles}
in that it shares the concerns of minimizing
storage and recreation cost; however,
the paper considered the unstructured setting from
an algorithmic viewpoint,
and did not aim to build a full-fledged
dataset versioning system.
Lastly, Chavan et al.~\cite{chavan2015towards}
describe a query language for versioning and provenance,
but do not develop a system that can support such a language---our
system can support an important subset of this language
already.

The problem of incremental view maintenance, e.g.,~\cite{ahmad2012dbtoaster},
is also related since it implicitly considers the question of
storage versus query efficiency, which is one of the primary concerns
in data versioning. However,
the considerations and challenges are very different, making the solutions
not applicable to data versioning.
Finally, Buneman et al.~\cite{buneman2004archiving} introduce a range
encoding approach to track the versioning of hierarchical data in scientific
data\-bases, but their method focuses on XML data
and is not applicable to the relational datasets.



\stitle{Temporal Databases.}
There is a rich body of work on time travel (or temporal)
databases, e.g., \techreport{\cite{ahn1986performance,snodgrass1985taxonomy,snodgrass1994tsql2,jensen1999temporal,ozsoyoglu1995temporal,tansel1993temporal}}\papertext{\cite{jensen1999temporal,ozsoyoglu1995temporal,tansel1993temporal}},
focusing on data management
when the state of the data at a specific time is important.
Temporal databases support a linear clock, or
a linear chain of versions, whereas our work focuses on enabling
non-linear histories.
There has been some work on developing temporal databases
by ``bolting-on'' capabilities to a traditional database~\cite{torp1998stratum},
with DB2 \techreport{\cite{chen2003design,saracco2010matter}}\papertext{\cite{saracco2010matter}}
and Teradata~\cite{al2013temporal}
supporting time-travel in this way.
Other systems
adopt an ``in-database'' approach~\cite{kaufmann2013timeline}.
\techreport{For example, the SAP HANA database \cite{farber2012sap} 
maintains a {\em Timeline Index}~\cite{kaufmann2013timeline}
to efficiently support temporal join, 
aggregation, and time travel.
}
Kaufmann et al.~\cite{kaufmann2014benchmarking} provide a good summary
of the temporal features in databases,
while Kulkarni et al.~\cite{kulkarni2012temporal} describe the temporal
features in SQL2011.


The canonical approach to recording time in temporal databases is
via attributes indicating the start and end time, which differs a bit
depending on whether the time is the ``transaction time'' or the ``valid time''.
In either case, if one extends temporal databases to support
arrays capturing versions instead of the start and end time, we will
end up as a solution like the one in Figure~\ref{fig:datamodels}b,
which as shown severely limits performance.
Thus, the techniques we describe in the paper on evaluating efficient data models
and partitioning are still relevant and complement this prior work.

Most work in this area focuses on supporting
constructs that
do not directly apply to \orpheus, \techreport{due to the lack of time-oriented notions}
such as:
(a) queries that probe interval related-properties, such as
which tuples were valid in a specific time interval, via
range indexes~\cite{salzberg1999comparison}, or queries that roll back to specific points~\cite{lee2007flashback};
(b)
temporal aggregation~\cite{kaufmann2013timeline} to aggregate some attributes
for every time interval granularity, and temporal join~\cite{gao2005join} to join tuples if they overlap in time;
(c) queries that involve time-related constructs such as {\sf AS OF, OVERLAPS, PRECEDES}.



There has been limited work on branched temporal
data\-bases~\cite{landau1995historical,salzberg1995branched},
with multiple chains of linear evolution as opposed to
arbitrary branching and merging.
While there has been some work on developing indexing~\cite{lanka1991fully,jiang2000bt}
techniques in that context, these techniques are specifically
tailored for queries that select a specific branch, and a time-window
within that branch, 
which therefore have no correspondences in our context.
Moreover, these techniques require substantial
modifications to the underlying database.

\stitle{Restricted Dataset Versioning.} 
There have been some open-source
projects
on versioning topics related to \orpheus.
For example, LiquiBase~\cite{voxland2014liquibase} 
tracks schema evolution as the only applicable modifications
giving rise to new versions: in our case, we focus
primarily on the data-level modifications, but also support 
schema modifications\papertext{~as described in~\cite{orpheustr}}.
On the other hand, DBV~\cite{dbv} is focused
on recording SQL operations that give rise
to new versions such that these operations can be ``replayed''
on new datasets---thus the emphasis is on
reuse of workflows rather than on efficient versioning.
As other recent projects, Dat~\cite{dat} can be used to
share and sync local copies of dataset across machines,
while Mode~\cite{mode} integrates various analytics tools into a
collaborative data analysis platform.
However, neither of the tools are focused on
providing advanced querying
and versioning capabilities.
In addition, {\tt git} and {\tt svn} can be made to support dataset
versioning, however, recent work shows these
techniques are not efficient~\cite{maddox2016decibel},
and do not support sophisticated querying.

\stitle{Graph Partitioning.}
There has been a lot of work on graph partitioning~\cite{karypis1998fast,liu1996partitioning, feige2001dense, karypis2000multilevel}, with
applications ranging from distributed systems and parallel computing,
to search engine indexing.
The state-of-the-art in this space
is NScale~\cite{quamar2014nscale}, which proposes
algorithms to pack subgraphs into the minimum number of partitions
while keeping the computation load balanced across partitions.
In our setting, the versions are related
to each other in very specific ways;
and by exploiting these properties, our algorithms
are able to beat the NScale ones
in terms of performance, while also
providing a $10^3$$\times$ speedup.
Kumar et al.~\cite{kumar2014sword}
study workload-aware graph partitioning
by performing balanced k-way cuts on the tuple-query
hypergraph for data placement
and replication on the cloud;
in their context, however,
queries are allowed to touch multiple
partitions.

%% file: conclusion.tex
\section{Conclusions}
We presented \orpheus, a dataset version control
system that is ``bolted on'' a relational database,
thereby seamlessly benefiting from advanced querying
as well as versioning capabilities.
We proposed and evaluated four data models for storing
\cvds in a database.
We further optimized the best data model (split-by-rlist) via the \appr
algorithm that applies intelligent but lightweight
partitioning to reduce the amount of irrelevant data
that is read during checkout. We also adapt \appr to operate in an incremental fashion as new versions are introduced.
Our experimental results demonstrate that \appr
is 10$^3\times$ faster in finding the effective partitioning scheme compared to other algorithms,
can improve the checkout performance up to 20$\times$ relative to schemes without partitioning,
 and is capable of operating efficiently (with relatively few and efficient migrations) in a dynamic setting.



%% file: extension.tex

\begin{table}[t!]
\vspace{-10pt}
\centering
\small
\begin{tabular}{c|c|c|c}
   Symb. & Description & Symb. & Description\\
    \hline
    \hline
    $G$ & bipartite graph & $E$ & bipartite edge set in $G$ \\
    \hline
    $V$ & version set in $G$ & $n$ & total number of versions\\
    \hline
    $R$ & record set in $G$ & $m$ & total number of records\\
    \hline
    $v_i$ & version $i$ in $V$ & $r_j$ & record $j$ in $R$\\
    \hline
    $\pp_k$ & $k^{th}$ partition & $\vv_k$ & version set in $\pp_k$\\
    \hline
    $\rr_k$ & record set in $\pp_k$ & $\ee_k$ & bipartite edge set set in $\pp_k$\\
    \hline
    $\mathcal{S}$ & total storage cost & $\gamma$ & storage threshold\\
    \hline
    $\cc_i$ & checout cost for $v_i$ & $\cc_{avg}$ & average checkout cost\\
    \hline
    $\gbb$ & version graph & $\vbb$ & version set in $\gbb$\\
    \hline
    $\ebb$ & edge set in $\gbb$  & $e$ & $e=(v_i,v_j)$: $v_i$ derives $v_j$ \\
    \hline
    $\tbb$ & version tree &  $e.w$ &\# of common records on $e$\\
    \hline
    $l(v_i)$ & level \# of $v_i$ in $\gbb$ & $p(v_i)$ & parent version(s) of $v_i$ in $\gbb$ \\
    \hline
    $R(v_i)$ & record set in $v_i$ & $\ell$ & \# of recursive levels in Alg 1\\
\end{tabular}
\vspace{-6pt}
\caption{Notations}
\label{tbl:notation}
\vspace{-18pt}
\end{table}

\section{Proof of Theorem 1} \label{sec:proof}
\begin{proof}
We reduce the well known {\sc NP-hard} {\sc 3-partition} problem to our Problem~\ref{prob:min_rec}. The {\sc 3-partition} problem is defined as follows:
Given an integer set $\mathcal{A}=\{a_1, \cdots , a_n\}$ where $n$ is divisible by 3, 
partition $\mathcal{A}$ into $\frac{n}{3}$ sets \{$A_1, A_2, A_j \cdots A_{\frac{n}{3}}$\} such that for any $A_j$,  $\sum_{a_i \in A_j} a_i = \frac{B}{n/3}$ where $B=\sum_{a_i\in \mathcal{A}} a_i$. 

To reduce {\sc 3-partition} to our Problem~\ref{prob:min_rec}, we first construct a version-record bipartite graph $G=(V,R,E)$ (Figure~\ref{fig:NP_graph}) that consists of $B$ versions and $(B+D)$ records, where $D$ is the number of dummy records  and can be any positive integer. Specifically:


%
%

\begin{itemize}
\item For each integer $a_i \in \mathcal{A}$: 
\begin{itemize}
\item Create $a_i$ versions $\{v_i^1, v_i^2, \cdots, v_i^{a_i}\}$ in $V$;
\item Create $a_i$ records $\{r_i^1, r_i^2, \cdots ,r_i^{a_i}\}$ in $R$;
\item Connect each $v_i^j$ with $r_i^\tau$ in $E$, where $1\leq j \leq a_i$ and $1\leq \tau \leq a_i$. This forms a biclique between $\{v_i^1, \cdots, v_i^{a_i}\}$ and $\{r_i^1, \cdots ,r_i^{a_i}\}$.
\end{itemize}
\item We also create dummy records $R_D$ and edges $E_D$:
\begin{itemize}
\item $R_D$: create $D$ dummy records $R_D=\{r_0^1,r_0^2, \cdots, r_0^D\}$ in $R$, where $D\geq 1$;
\item $E_D$: connect each dummy record with every version $v\in V$.
\end{itemize}
\end{itemize}


 \begin{figure}[H]
 \centering
  \vspace{-5mm}
 \includegraphics[width=0.5\linewidth,,height=45mm]{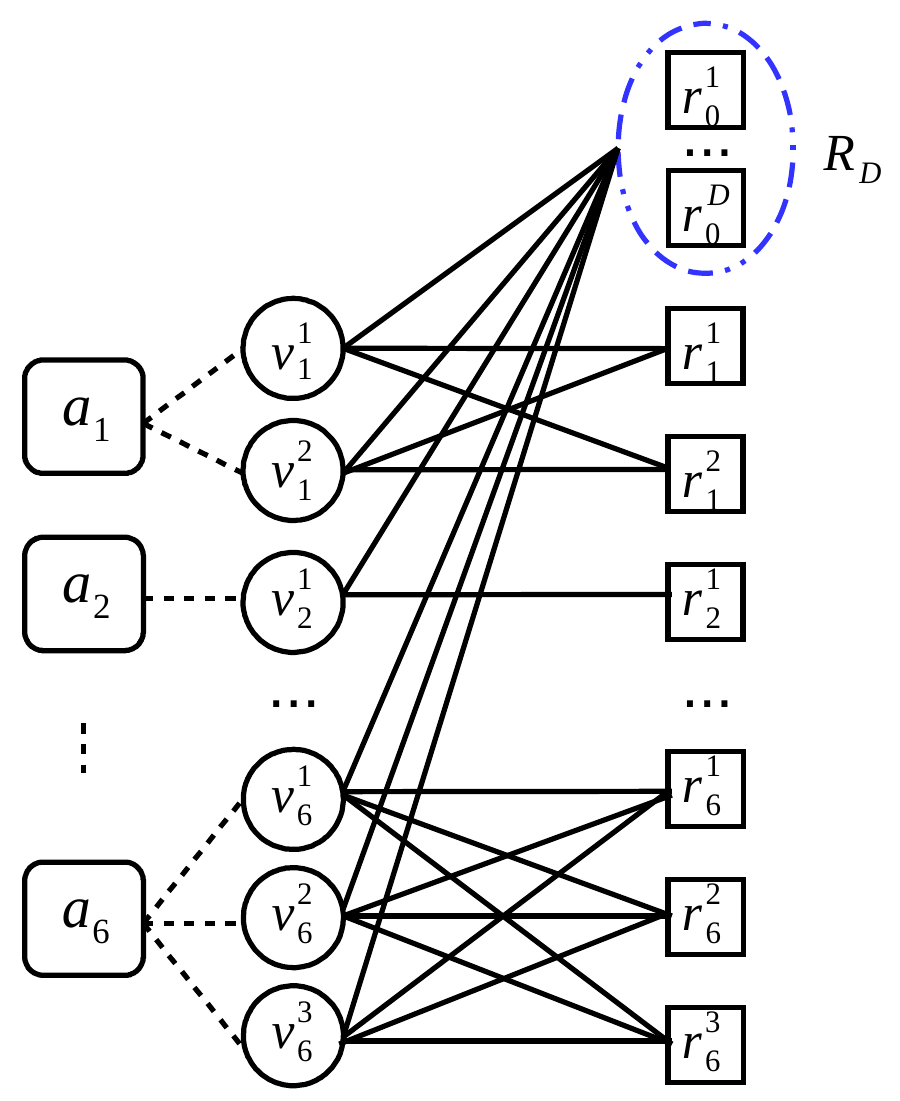}
 \vspace{-2mm}
 \caption{An Example of a Constructed Graph $G$} 
 \label{fig:NP_graph}
 \vspace{-3mm}
 \end{figure}

As inputs to Problem~\ref{prob:min_rec}, we take the constructed graph $G$ and set storage threshold $\gamma = \frac{n}{3}\cdot D+B$. We have the following two claims for the optimal solution to Problem~\ref{prob:min_rec}:

\stitle{Claim 1.} For each $a_i$, its corresponding versions $\{v_i^1, v_i^2, \cdots, v_i^{a_i}\}$ must be in the same partition. 

\stitle{Claim 2.} The optimal solution must have $\frac{n}{3}$ partitions, i.e, $K=\frac{n}{3}$.

\vspace{1mm}
We prove our first claim by contradiction. For a fixed $a_i$, if $\{v_i^1, v_i^2, \cdots, v_i^{a_i}\}$ are in different partitions, denoted as $P'=\{\mathcal{P}_{\tau_1},$
$\mathcal{P}_{\tau_2}, \cdots\}$, we can reduce the average checkout cost while maintaining the same storage cost by moving all these versions into the same partition $\mathcal{P}_{k^*}\in P'$ with the smallest $|\mathcal{R}_{k^*}|$. Furthermore, the only common records between $v_i^x$ and $v_j^y$, where $i\neq j$, are the dummy records in $R_D$, thus only these dummy records will be duplicated across different partitions. Consequently, the total storage cost from records except the dummy record, i.e., $R \setminus R_D$, in all partitions is a constant $B$, regardless of the partitioning scheme. 


Based on the first claim, we have $|\mathcal{R}_k|=|\mathcal{V}_k|+D, \forall k$ and our optimization objective function can be represented as follows: 
\vspace{-5pt}
\begin{equation} \label{eqn:proof} \small
\begin{split}
\mathcal{C}_{avg} &= \frac{1}{B}\sum_{k=1}^K |\mathcal{V}_k|\times(|\mathcal{V}_k|+D) = \frac{1}{B}(\sum_{k=1}^K |\mathcal{V}_k|^2 + B\cdot D)
\end{split}
\vspace{-3pt}
\end{equation}
Next, we prove the correctness of our second claim. First, we show that keeping the total storage cost $\sum_{k=1}^K |\mathcal{R}_k| \leq \frac{n}{3}\times D+B$ is equivalent to keeping the number of partitions $K\leq \frac{n}{3}$. From our first claim, we know that no record in $R \setminus R_D$ will be duplicated and the total number of records that corresponds to $R \setminus R_D$ in all of the partitions is $B$. On the other hand, each partition $\mathcal{P}_k$ must include all dummy records $R_D$, which is of size $D$. Thus, the number of partitions $K$ must be no larger than $\frac{n}{3}$. Furthermore, we claim that the optimal solution must have $\frac{n}{3}$ partitions, i.e., $K=\frac{n}{3}$; otherwise, we can easily reduce the checkout cost by splitting any partition into multiple partitions.

Lastly, we prove that the optimal $\mathcal{C}_{avg}$ equals $B/K+D$ if and only if the decision problem to {\sc 3-partition} is correct. First, since $\sum_{k=1}^K |\mathcal{V}_k| =B$, $\mathcal{C}_{avg}$ in Equation~\ref{eqn:proof} is minimized when all $|\mathcal{V}_k| = B/K, \forall k$. Returning to the {\sc 3-partition} problem, if our decision to {\sc 3-partition} is true, then we can partition the versions in the constructed graph $G$ accordingly and $\mathcal{C}_{avg}=B/K+D$ with each $|\mathcal{V}_k|=\frac{B}{K}=\frac{B}{n/3}$. Second, if the decision problem is false, then $\mathcal{C}_{avg}$ must be larger than $B/K+D$. Otherwise, all $|\mathcal{V}_k|$ must be the same and equal to $B/K$. Subsequently, we can easily partition $\mathcal{A}$ into $\frac{n}{3}$ sets with equal sum for {\sc 3-partition}, which contradicts the assumption that the decision problem is false. 
\end{proof}

\section{Analysis of \large{$\delta$}} \label{sec:deltas}
Given a storage budget $\gamma$ in Problem~\ref{prob:min_rec}, we can simply perform a binary search on $\delta$ and get the best $\delta$ as the input for Algorithm~\ref{alg:divisive_partition}. This claim is evidenced by the fact that the same sequence of edges are snipped for different $\delta$. In general, as $\delta$ increases, there are more partitions, consequently less checkout cost and larger storage cost.

\stitle{Superset property of $\delta$.} Consider two different $\delta$: $\delta_1$ and $\delta_2$, without loss of generality we assume $\delta_1 < \delta_2$, and to simplify the analysis we pick the smallest weight as the splitting edge in each iteration. First we claim that Algorithm~\ref{alg:divisive_partition} takes more iterations when $\delta = \delta_2$ than $\delta = \delta_1$. This is because $\delta_1 < \delta_2$ and the termination constraint is $|R||V| < \frac{|E|}{\delta}$. Next, we assert that the edges cut when $\delta = \delta_1$ is a subset of the same sequence of $\delta = \delta_2$. This is because in each iteration, the edge with the smallest weight is cut for both $\delta_1$ and $\delta_2$, and when $\delta_1$ terminates ($|R||V| < \frac{|E|}{\delta_1}$), $\delta_2$ may still goes on since $|R||V|\geq\frac{|E|}{\delta_2}$. Thus, compared to $\delta_1$, $\delta_2$ has more splits, larger storage cost, and less checkout cost.

\stitle{Binary search on $\delta$.} Initially, the search space for $\delta$ is $[\frac{|E|}{|R||V|},1]$, where each version is stored in a separate partiton(i.e., $\delta=1$) and all versions are in the same partition(i.e., $\delta=\frac{|E|}{|R||V|}$). We first try $\delta = \frac{1}{2}(\frac{|E|}{|R||V|}+1)$ in Algorithm~\ref{alg:divisive_partition} and get the resulting storage cost $\mathcal{S}$ after partitioning. If $\mathcal{S}<\gamma$, then the search space for $\delta$ is reduced to $[\frac{1}{2}(\frac{|E|}{|R||V|}+1), 1]$; otherwise, $[\frac{|E|}{|R||V|}, \frac{1}{2}(\frac{|E|}{|R||V|}+1)]$. Repeat this process until $0.99\gamma\leq \mathcal{S}\leq \gamma$.

%% file: extension_2.tex
\section{Extensions} \label{sec:ext}
\subsection{Version graph is a DAG}\label{ssec:general_case}
When there are merges between versions, 
the version graph $\gbb=(\mathbb{V},\mathbb{E})$ is a DAG. 
We can simply transform the $\gbb$ to a version tree $\hat{\mathbb{T}}$ and 
then apply \appr as before. 
Specifically, for each vertex $v_i \in \mathbb{V}$, 
if there are multiple incoming edges, 
we retain the edge with the highest weight and remove all other incoming edges. 
In other words, for each merge operation in the version graph $\gbb$, 
e.g., where $v_i$ is merged with $v_j$ to obtain $v_k$, 
the corresponding operation in $\hat{\mathbb{T}}$ 
with the removed edge $(v_j,v_k)$ is to inherit records
only from one parent $v_i$ and (conceptually) 
create new records in the \cvd 
for all other records in $v_k$ even though some records 
have exactly the same value as that in $v_j$.

\begin{figure}[ht!]
\centering
\vspace{-3mm}
\includegraphics[width=0.8\linewidth]{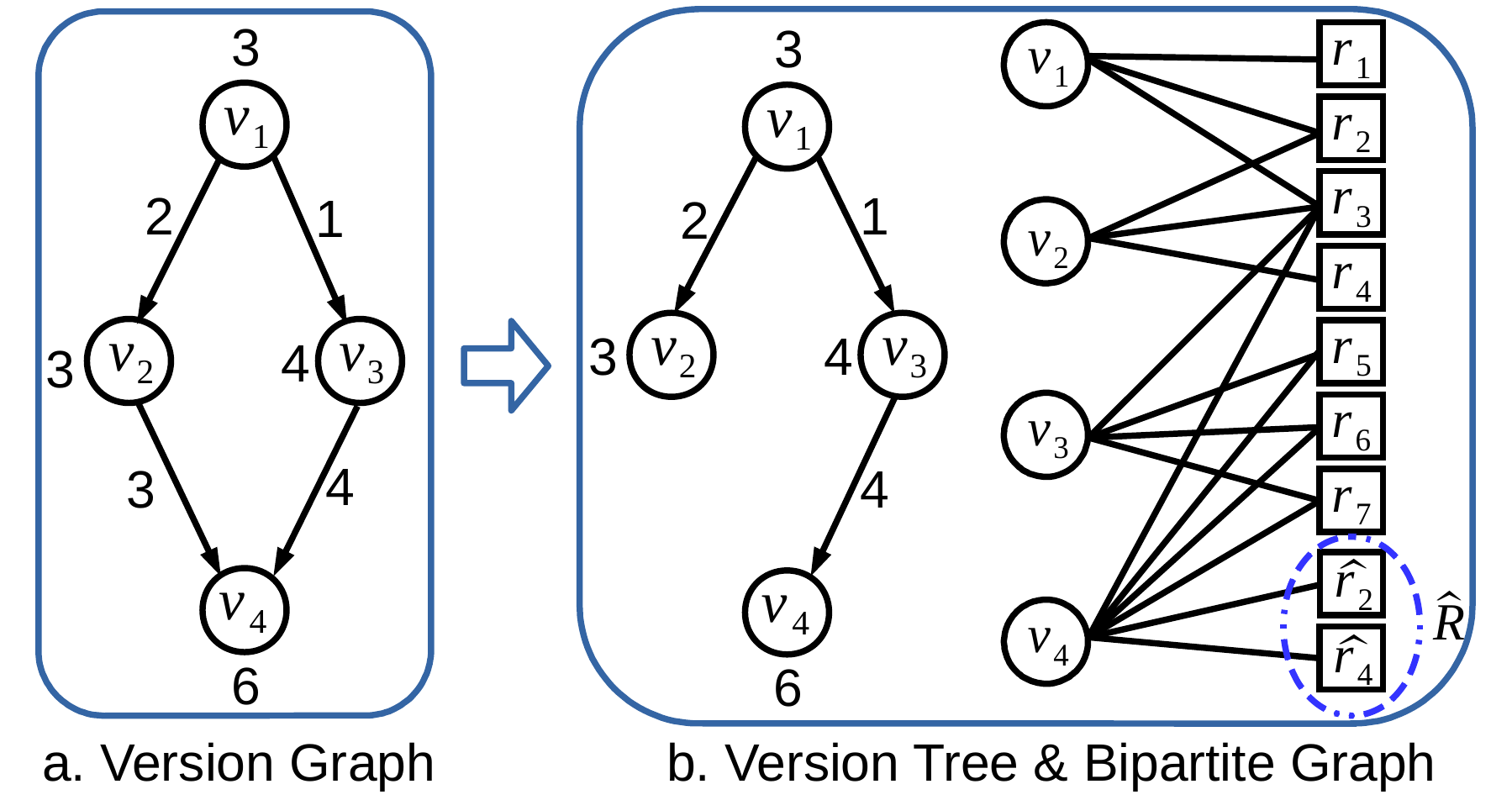}
\vspace{-3mm}
\caption{$\hat{\mathbb{T}}$ and $\hat{G}$ for $\gbb$ in Figure~\ref{fig:metaTable_versionG}}
\label{fig:version_graph}
\vspace{-3mm}
\end{figure}

For example, for the version graph $\gbb$ shown in Figure~\ref{fig:version_graph}(a), its version $v_4$ has two parent versions $v_2$ and $v_3$. Since $3=w(v_2,v_4)<w(v_3,v_4)=4$, we remove edge $(v_2,v_4)$ from $\gbb$ and obtain the version tree $\hat{\tbb}$ in Figure~\ref{fig:version_graph}(b). Moreover, conceptually, we can draw a bipartite graph $\hat{G}$ corresponding to $\hat{\mathbb{T}}$ as shown in Figure~\ref{fig:version_graph}(b) with two duplicated records, i.e., $\{\hat{r_2},\hat{r_4}\}$. That is, $v_4$ in $\hat{\tbb}$ inherits 4 records from $v_3$ and creates two new records $\hat{R}=\{\hat{r_2},\hat{r_4}\}$ even though $\hat{r_2}$ ($\hat{r_4}$) is exactly the same as $r_2$ ($r_4$). Thus, we have 9 records with $|\hat{R}|=2$ and 16 bipartite edges in Figure~\ref{fig:version_graph}(b).

\stitle{Performance analysis.} The number of bipartite edges in the bipartite graph $\hat{G}$ (corresponding to $\hat{\tbb}$) is the same as that in $G$ (corresponding to $\gbb$), i.e., $|E|$. However, compared to $G$, the number of records in $\hat{G}$ is larger, i.e., $|R|+|\hat{R}|$, where $R$ is the set of records in the original version-record bipartite graph $G$ and $\hat{R}$ is the set of duplicated records. According to Theorem \ref{thm:storage_checkout}, given $\delta$, \appr provides a partitioning scheme with the checkout cost within $\frac{1}{\delta}\cdot\frac{|E|}{|V|}$ and the storage cost within $(1+\delta)^{\ell}(|R|+|\hat{R}|)$. We formally state the performance guarantee in Theorem~\ref{thm:storage_checkout_g}. Moreover, this analysis is obtained by treating $\hat{R}$ as different from $R$ when calculating the storage cost and checkout cost. In post-processing, we can combine $\hat{R}$ with $R$ when calculating the real storage cost and checkout cost, making the real $\mathcal{S}$ and $\cc_{avg}$ even smaller.

\begin{theorem}\label{thm:storage_checkout_g}
Given a version graph $\gbb$ with merges and a parameter $\delta$, \appr
results in a $(\frac{|R|+|\hat{R}|}{|R|}(1+\delta)^{\ell},\frac{1}{\delta})$-approximation for partitioning.
\end{theorem}

\subsection{Weighted Checkout Cost} \label{ssec:weighted}
In this section, we focus on the weighted checkout cost case, 
where versions are checked out with different frequencies. 

 \stitle{Problem formulation.} Let $\cc_{w}$ denote the weighted checkout cost;
 say version $v_i$ is checked out with probability or frequency $f_i$ 
Then the weighted checkout cost $\cc_{w}$ 
can be represented as $\cc_{w}=\frac{\sum_{i=1}^n (f_i\times \cc_i)}{\sum_{i=1}^n f_i}$.
With this weighted checkout cost, we can modify the problem formulation for
Problem~\ref{prob:min_rec} by 
simply replacing $\cavg$ with $\cc_w$.

\stitle{Proposed Algorithm.} Without the loss of generality, we assume that $f_i$ for any version $v_i$ is an integer. Given a version tree\footnote{\scriptsize if the version graph is a DAG instead, we first transform it into a version tree as discussed in Appendix~\ref{ssec:general_case}.} $\tbb=(\vbb, \ebb)$ and the frequency $f_i$ for each version $v_i$, we construct a version tree $\tbb'=(\vbb', \ebb')$ in the following way:

\begin{itemize}
\item For each version $v_i \in \vbb$: 
\begin{itemize}
\item $\vbb'$: Create $f_i$ versions $\{v_i^1, v_i^2, \cdots, v_i^{f_i}\}$ in $\vbb'$;
\item $\ebb'$: Connect $v_i^j$ with $v_i^{j+1}$ to form a chain in $\ebb'$, where $1\leq j<f_i$
\end{itemize}
\item For each edge $(v_i,v_j) \in \ebb$: 
\begin{itemize}
\item $\ebb'$: Connect $v_i^{f_i}$ with $v_j^1$ in $\ebb'$
\end{itemize}
\end{itemize}

The basic idea of constructing $\tbb'$ is to duplicate each version $v_i\in \vbb$ $f_i$ times. Afterwards, we apply \appr directly on $\tbb'$ to obtain the partitioning scheme. However, after partitioning, $v_i^j\in \vbb'$ with the same $i$ may be assigned to different partitions, denoted as $P'$. Thus, as a post process, we move all $v_i^j$ ($1\leq j\leq f_i$) into the same partition $\pp \in P'$ that has the smallest number of records. Correspondingly, we get a partitioning scheme for $\vbb$, i.e., for each $v_i\in \vbb$, assign it to the partition where $v_i^j\in \vbb'$ ($1\leq j\leq f_i$) is in.  

\stitle{Performance analysis.} At one extreme, when each version is stored in a separate table, the checkout cost $\cc_w$ for $\tbb$ is the lowest with each $\cc_i=|R(v_i)|$, the number of records in version $v_i$; thus, $\cc_w=\frac{\sum_{i=1}^n (f_i\times |R(v_i)|)}{\sum_{i=1}^n f_i}$, denoted as $\zeta$. At the other extreme, when all versions are stored in a single partition, the total storage cost is the smallest, i.e., $|R|$. In the following, we study the performance of the extended algorithm in the weighted case, and compare the storage cost and weighted checkout cost with $|R|$ and $\zeta$ respectively.


First, consider the bipartite graph $G'=(V',R',E')$ corresponding to the constructed version tree $\tbb'$. The number of versions $|V'|$ equals $\sum_{i=1}^n f_i$, since there are $f_i$ replications for each version $v_i$; the number of records $|R'|$ is the same as $|R|$, since there are no new records added; the number of bipartite edges $|E'|$ is $\sum_{i=1}^n (\sum_{j=1}^{f_i} |R(v_i^j)|) = \sum_{i=1}^n (f_i \times |R(v_i)|)$, since the number of records in each version $v_i^j$ with the same $i$ is in fact $|R(v_i)|$. Next, based on Theorem~\ref{thm:storage_checkout}, the average checkout cost after appyling Algorithm~\ref{alg:divisive_partition} is within $\frac{1}{\delta}\cdot\frac{|E'|}{|V'|} = \frac{1}{\delta}\cdot \frac{\sum_{i=1}^n (f_i\times |R(v_i)|)}{\sum_{i=1}^n f_i} = \frac{1}{\delta}\cdot \zeta$, while the storage cost is within $(1+\delta)^{\ell}\cdot |R'| = (1+\delta)^{\ell}\cdot |R|$, where $\ell$ is the termination level in Algorithm~\ref{alg:divisive_partition}. After post-processing, the total storage cost as well the average checkout cost decreases since we pick the partition with the smallest number of records for all $v_i^j$ with a fixed $i$. At last, note that after mapping the partitioning scheme from $\tbb'$ to $\tbb$, the total storage cost and the average (unweighted) checkout cost for $\tbb'$ are in fact the total storage cost and the weighted checkout cost for $\tbb$ respectively. Thus, with the extended algorithm, we achieve the same approximation bound as in Theorem~\ref{thm:storage_checkout} with respect to the lowest storage cost and weighted checkout cost, i.e., in the weighted checkout case, our algorithm also results in $((1+\delta)^{\ell}, \frac{1}{\delta})$-approximation for partitioning.

\begin{figure}[ht!]
\centering
\vspace{-3mm}
\includegraphics[width=0.7\linewidth]{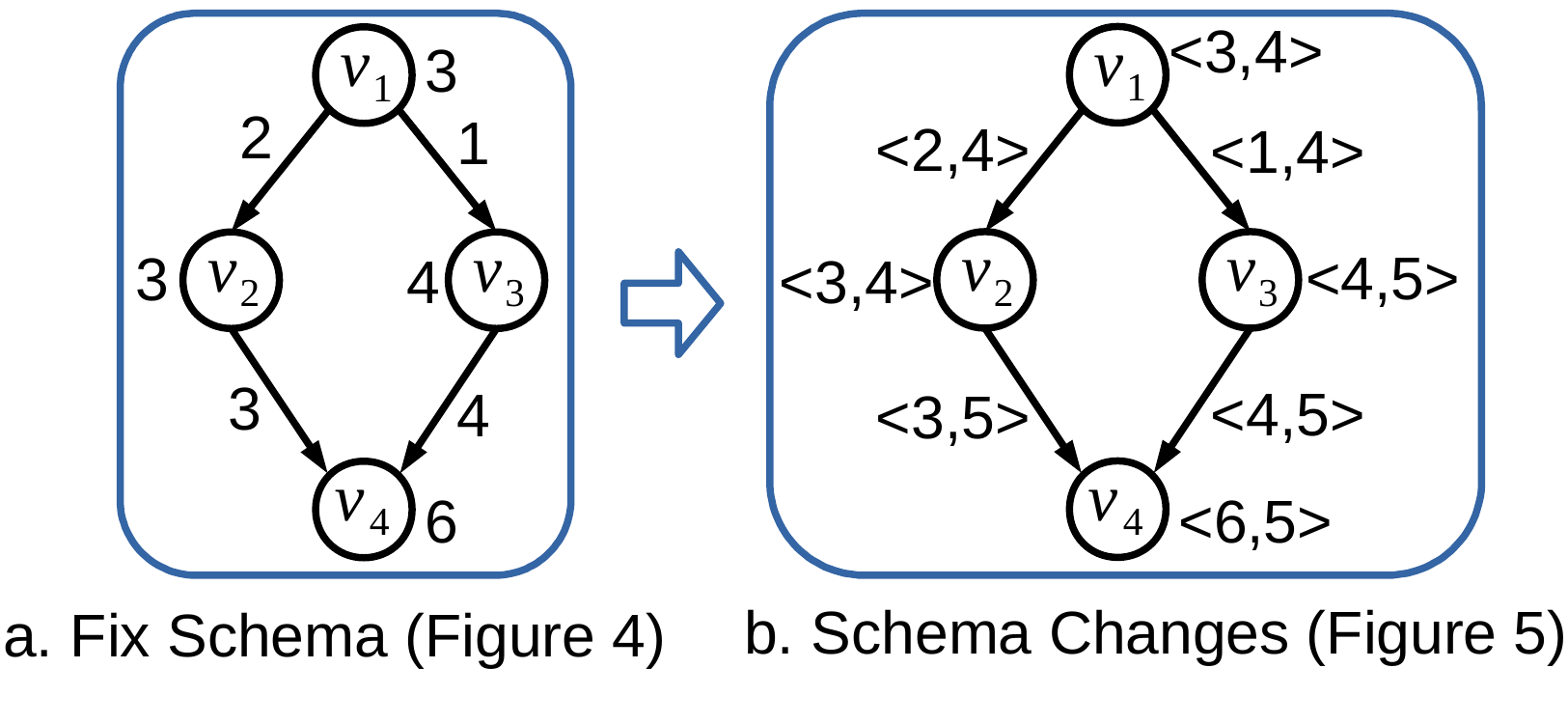}
\vspace{-3mm}
\caption{Version Graph $\gbb$ with/without Schema Changes}
\label{fig:version_graph_with_attribute}
\vspace{-3mm}
\end{figure}

\begin{figure*}[t!]
\centering
\vspace{-10mm}
\includegraphics[width=0.8\linewidth]{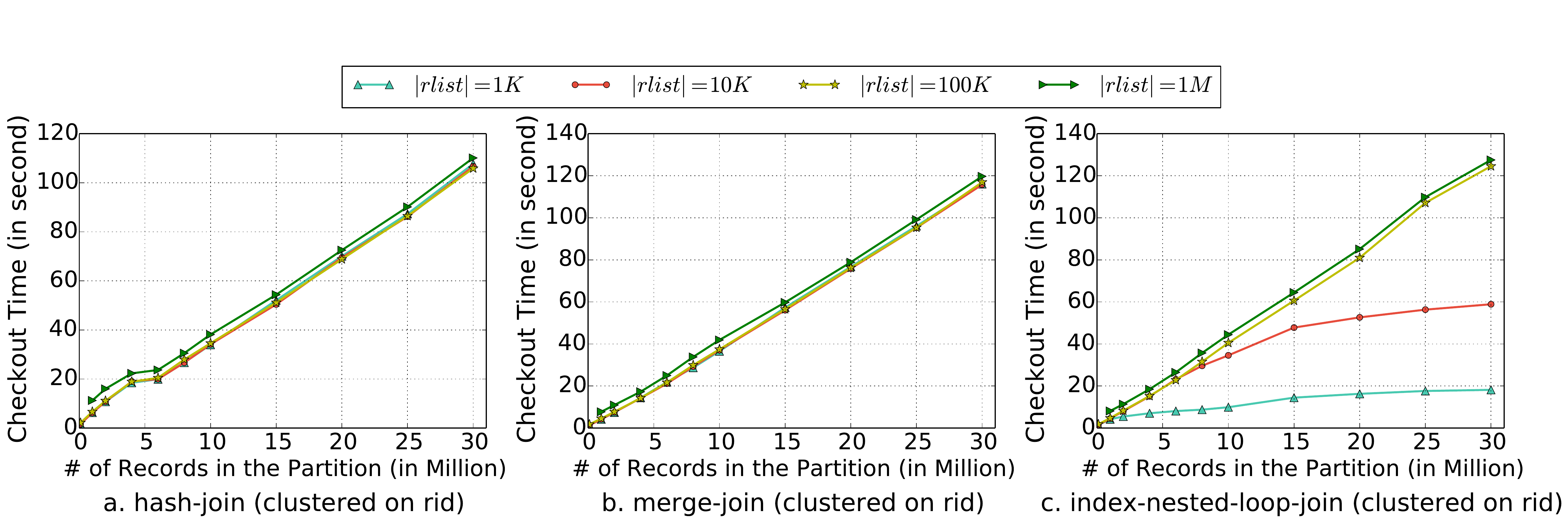}
\vspace{-3mm}
\end{figure*}

\begin{figure*}[t!]
\centering
\includegraphics[width=0.8\linewidth]{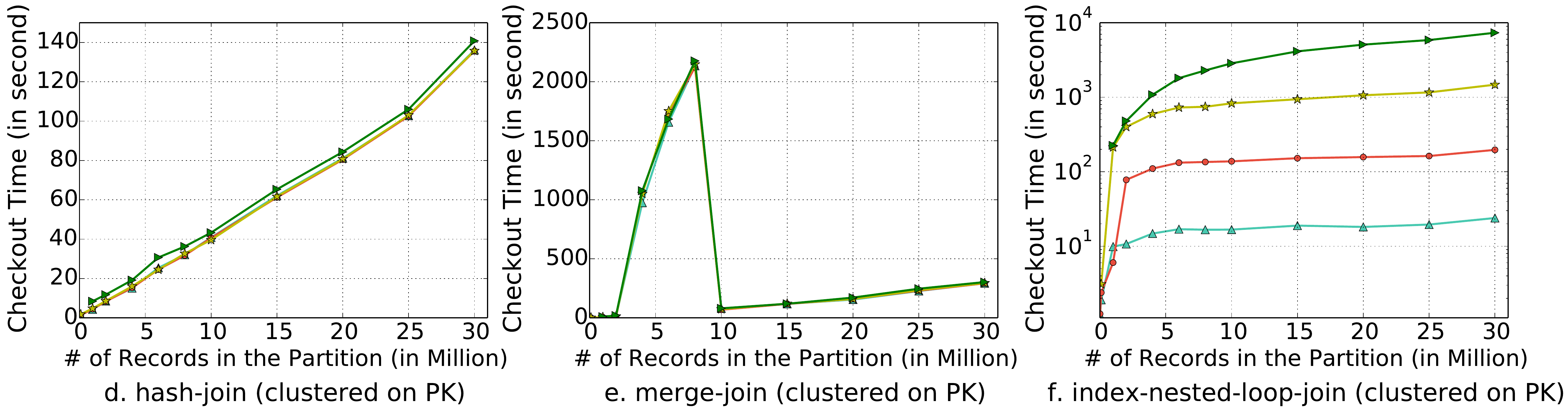}
\vspace{-3mm}
\caption{Checkout Cost Model Validation}
\label{fig:cost_model}
\vspace{-3mm}
\end{figure*}
\subsection{Schema Changes}\label{ssec:schema_change}
Our algorithm can be adapted to the single-pool setting described in Section~\ref{ssec:version_graph} with schema changes.
Recall the examples in Figure~\ref{fig:metaTable_versionG} and \ref{fig:attributeTable}, corresponding to the fixed and dynamic schema settings.
Figure~\ref{fig:version_graph_with_attribute}a only maintains the number of records in each node (version), and the number of common records between two versions for each edge. In addition, Figure~\ref{fig:version_graph_with_attribute}b also records the number of attributes and common attributes for each node and edge respectively. For instance, $v_3$ has five attributes and shares four common attributes with $v_1$.

Given a version graph, let $A$ be the total number of attributes in all versions. For instance, Figure~\ref{fig:version_graph_with_attribute}b, corresponding to Figure~\ref{fig:attributeTable}, has five attributes in total. Without partitioning, the storage cost and the checkout cost can be represented as $\calS=|A||R|$ and $\cavg=|A||R|$ respectively, where $|R|$ is the number of records. Next, let $a(v_i)$ and $a(v_i,v_j)$ denote the number of attributes in version $v_i$ and the number of common attributes between version $v_i$ and $v_j$, respectively. Recall that $w(v_i,v_j)$ denotes the number of common records between version $v_i$ and $v_j$, disregarding the schema. For instance, if version $v_j$ is obtained by deleting an attribute from $v_i$, then $a(v_i,v_j)=a(v_i)-1$ and $w(v_i,v_j)=|R(v_i)|$. 

The high-level idea is similar to \appr: split an edge if its "weight" is smaller than some threshold. However, the weight here not only depends on the number of common records $w(v_i,v_j)$, but also the number of common atrributes $a(v_i,v_j)$. Specifically, if $a(v_i,v_j)\times w(v_i,v_j) \leq \delta\times|A||R|$, edge $(v_i,v_j)$ is considered as a candidate splitting edge\footnote{\scriptsize If each attribute is of different size, we can simply replace "the number of attributes" with "the number of bytes" in the whole algorithm.}. Note that when there is no schema change, $a(v_i,v_j)=|A|$, and the constraint is reduced to $w(v_i,v_j) \leq \delta|R|$ (line 5 in Algorithm~\ref{alg:divisive_partition}). The remaining algorithm is the same as Algorithm~\ref{alg:divisive_partition}.

%% file: evaluation_cost_model.tex
\subsection{Verification of Checkout Cost Model}\label{ssec:exp_cost_model}
In this section, we both analyze and experimentally evaluate the checkout cost model proposed in Section~\ref{ssec:prob}. We demonstrate that the checkout cost $\cc_i$ of a version $v_i$ grows linearly with the number of records in the partition $\pp_{k}$ that contains $v_i$, i.e., $\cc_i \propto |\rr_{k}|$.

As depicted in the \sql query in Table~\ref{table:SQL}, the checkout cost is impacted by the cost of two operations: (a) obtaining the list of records \rlist associated with $v_i$; (b) joining data table with \rlist to get all valid records. The cost from part (a) is a constant regardless of the partitioning scheme we use, and it is small since \rlist can be obtained efficiently using a physical primary key index on \vid. Thus, we focus our analysis on the cost from part (b). 

We focus on three important types of join operations: hash-join, merge-join and nested-loop-join. In the following, we evaluate the checkout cost model for all these join algorithms and provide a detailed analysis.  
We vary the number of records in the checkout version ($|rlist|$) and the number of records in its corresponding partition ($|\rr_{k}|$) in our experiments. The parameter $|\rr_{k}|$ is varied 
from 1K to 30M and $|rlist|$ is varied from 1K to 1M, 
where \rlist is a sorted list of randomly sampled \rids from $\rr_{k}$. 
In addition, we have two different physical layouts for the data table, 
one clustered on \rid and another clustered on its original relation 
primary key (PK)--- <protein1, protein2> in Figure~\ref{fig:datamodels}. 
For each of the three join types, 
we compare the checkout time (in seconds) vs.~the estimated checkout cost (in millions of records).
Note that we build an {\em index} on \rid in the data table, 
otherwise, the nested-loop-join would be very time-consuming 
since each outer loop requires a full scan on the inner table. 
The results are presented in Figure~\ref{fig:cost_model}, 
where each line is plotted with a fixed $|rlist|$ (1K, 10K, 100K, and 1M respectively) 
and varying $|\rr_{k}|$. We now describe the performance of the individual join algorithms below.


 \stitle{Hash-join.} No matter which physical layout is used, the query plan for a hash-join based approach is to first build a hash table for \rlist and then sequentially scan the data table with each record probing the hash table. 
 By benefiting from the optimized implementation of the hash-join in \postgres, 
 the cost of probing each \rid in the hash table is almost a constant. 
 With fixed $|rlist|$, the building phase in hash-join is the same, 
 while the running time in the probing phase is proportional to $|\rr_{k}|$. 
 Hence,  as depicted in Figure~\ref{fig:cost_model}(a) and (d), with a fixed $|rlist|$, 
 the running time increases linearly with the growth of $|\rr_{k}|$. 


 \stitle{Merge-join.} When the data table is clustered on \rid, 
 the query plan for a merge-join based approach 
 is to first sort \rlist obtained from the versioning table, 
 then conduct an index scan using \rid index on the data table 
 and merge with the \rlist from the versioning table. 
 First, since \rlist from the versioning table has already been sorted, 
 quicksort can immediately terminate after the first iteration. Second, since the data table is physically clustered on \rid, an index scan on \rid is equivalent to a sequential scan in the data table. Thus, with fixed $|rlist|$, the running time grows linearly with the increase of $|\rr_{k}|$, which is experimentally verified in Figure~\ref{fig:cost_model}(b). 

On the other hand, when the data table is clustered on the relation primary key, \postgres gives different query plans for different $|\rr_{k}|$. When $|\rr_{k}|$ is equal to 4M, 6M and 8M, 
the query plan is the same as the above---sort \rlist,
 conduct an index scan on \rid and merge with \rlist. 
 However, since the physical layout is no longer clustered on \rid, 
 having an index scan on \rid is equivalent to performing random access $|\rr_{k}|$ times  
 into the data table, which is very time-consuming as illustrated in Figure~\ref{fig:cost_model}(e). 
 For other $|\rr_{k}|$ except 4M, 6M and 8M, 
 the query plan is to first sort \rlist from the versioning table, 
 conduct a sequantial scan on the data table, sort the \rids, 
 and then finally merge \rids with \rlist. 
 Thus, with fixed $|rlist|$, the running time is proportional to $|\rr_{k}|$, 
 but greater than the hash-join based approach due to the overhead of sorting, 
 as shown by the last five points in Figure~\ref{fig:cost_model}(e). 



 \stitle{Index-nested-loop-join.} No matter which physical layout is used, the query plan for an index-nested-loop-join based approach is to perform a random I/O in the data table for each \rid in \rlist from the versioning table. Consider the scenario where $|rlist|$ is fixed and the data table is clustered on \rid. When $|rlist|$ is much smaller than $|\rr_{k}|$, the running time is almost the same since each random I/O is a constant and $|rlist|$ is fixed. This is also verified by the right portion of the blue line ($|rlist|$=1K) and red line ($|rlist|$=10K) in Figure~\ref{fig:cost_model}(c). However, when $|rlist|$ is comparable to $|\rr_{k}|$, the running time is proportional to $|\rr_{k}|$ as illustrated in the green ($|rlist|$=1M) and yellow ($|rlist|$=100K) line in Figure~\ref{fig:cost_model}(c). This is because hundreds of thousands of random I/Os are eventually reduced to a full sequential scan on the data table when $\rr_{k}$ is clustered on \rid. Returning to the checkout cost model, since partitioning algorithms tend to group similar versions together, after partitioning, $|rlist|$ is very likely to be comparable to $|\rr_{k}|$ and thus the checkout time can be quantified by $|\rr_{k}|$. Furthermore, the yellow line ($|rlist|$=100K) in Figure~\ref{fig:cost_model}(c) indicates that even when $\frac{|rlist|}{|\rr_{k}|}=\frac{1}{300}$, random I/Os will still be reduced to a sequential scan, consequently the running time grows linearly with $|\rr_{k}|$. 

However, note that when the data table is not clustered on $rid$, each random I/O takes almost constant time as shown in Figure~\ref{fig:cost_model}(f). Since random I/O is more time-consuming than sequential I/O, the index-nested-loop-join performs much worse than hash-join as shown in Figure~\ref{fig:cost_model}(d) and (f). 


\stitle{Overall Takeaways.} When the data table is clustered on \rid, the checkout cost can be quantified by $|\rr_{k}|$ for hash-join and merge-join based approaches; while for index-nested-loop-join, the checkout cost can also be quantified by $|\rr_{k}|$ when $\frac{|rlist|}{|\rr_{k}|}\geq\frac{1}{300}$, which is typically the case in the partitions after partitioning especially for latest versions. On the other hand, when the data table is not clustered on \rid, the checkout cost for the hash-join based approach can still be quantified by $|\rr_{k}|$, while the merge-join and the index-nested-loop-join based approaches perform worse than that of hash-join for most cases. Overall, a hash-join based approach has the following advantages:\begin{inparaenum}[\itshape (a)\upshape]
\item the checkout time using hash-join does not rely on any index on \rid;
\item hash-join based approach has good and stable performance regardless of the physical layout;
\item the checkout cost using hash-join is easy to model, laying foundation for further optimization on {\em checkout} time.
\end{inparaenum}
Thus, throughout our paper we focus on hash-join for the 
{checkout} command and model the checkout cost $\cc_i$ 
as linear in the number of records $|\rr_{k}|$ in the partition that contains $v_i$. 

%% file: estimated-vs-actual.tex
\begin{figure*}[t]
\centering
\includegraphics[width=0.8\linewidth]{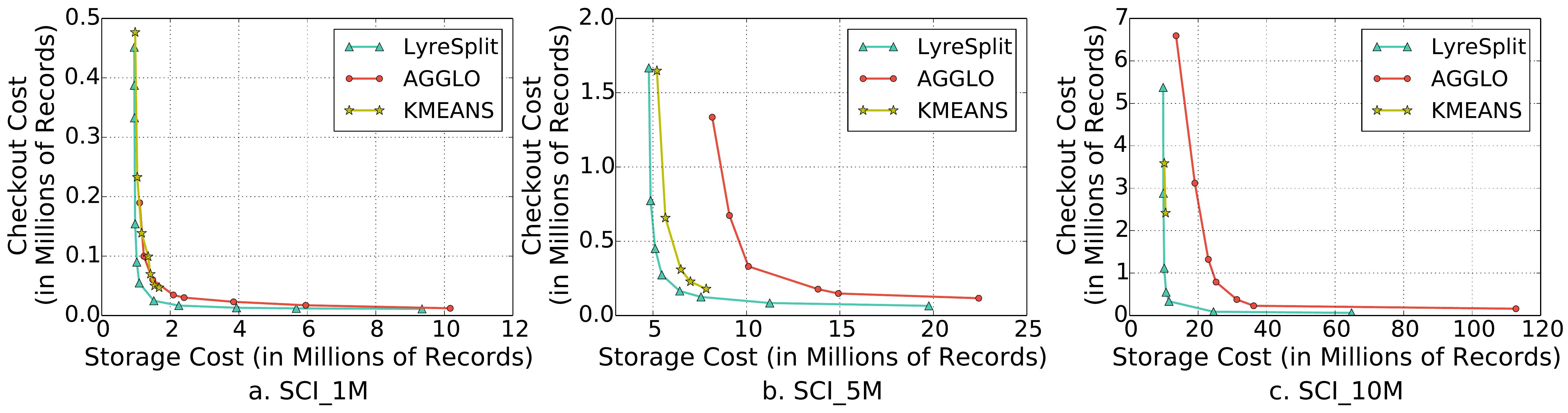}
\vspace{-3mm}
\caption{Estimated total Storage Cost vs. Estimated Checkout Cost (\code{SCI\_*})}
\vspace{-5mm}
\label{fig:est_d3_d8_d6}
\end{figure*}

\begin{figure*}[t]
\centering
\includegraphics[width=0.8\linewidth]{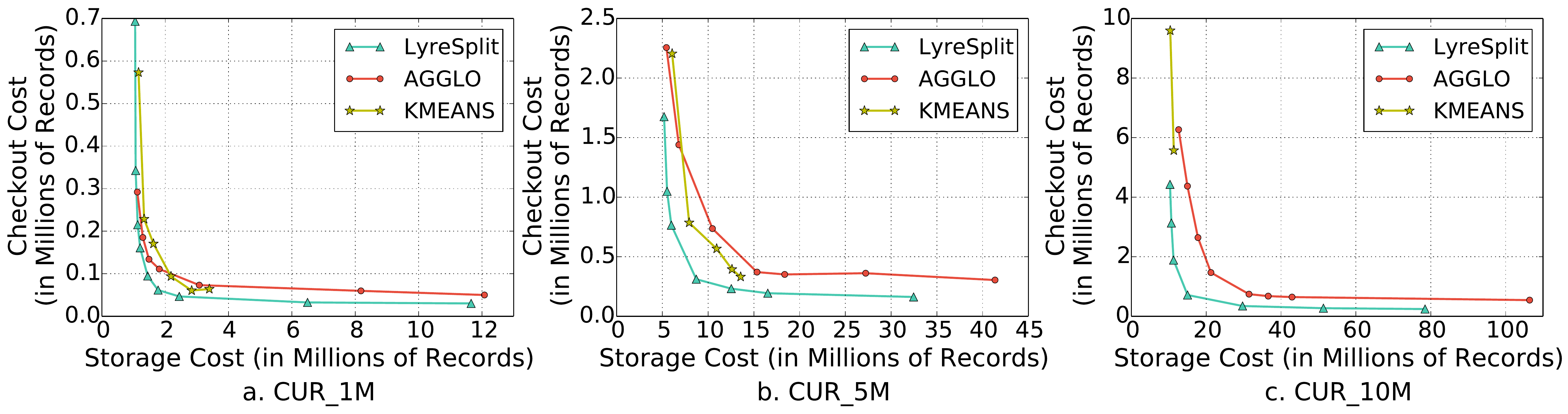}
\vspace{-3.2mm}
\caption{Estimated Storage Cost vs. Estimated Checkout Cost (\code{CUR\_*})}
\label{fig:est}
\end{figure*}

\begin{figure*}[t]
\centering
\vspace{-3mm}
\includegraphics[width=0.8\linewidth]{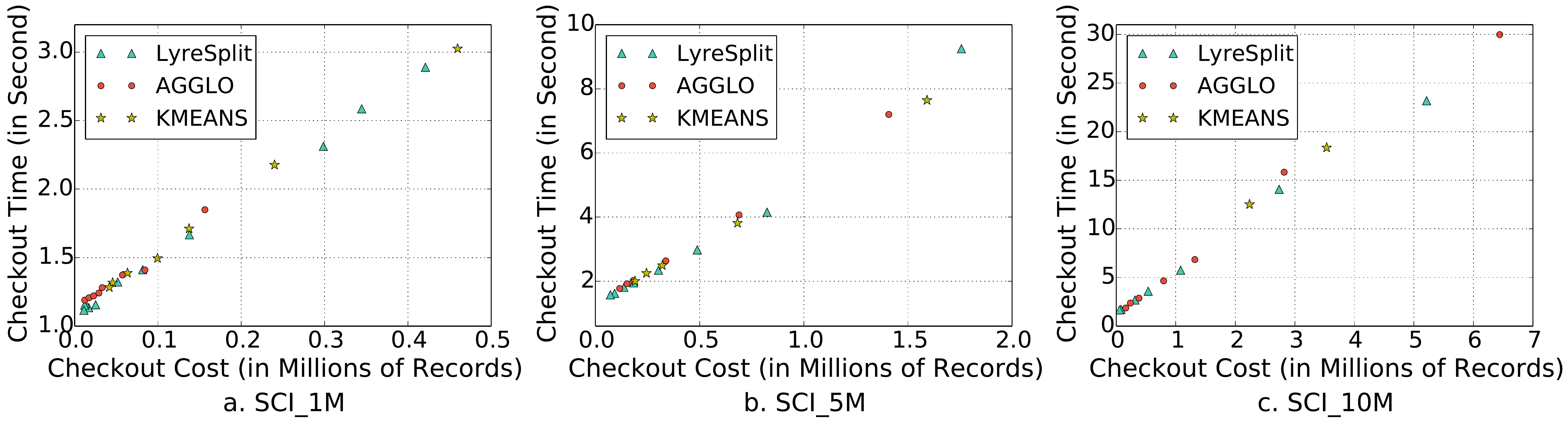}
\caption{Estimated Checkout Cost vs. Real Checkout Time (\code{SCI\_*})}
\label{fig:real_est_sci}
\vspace{-3mm}
\end{figure*}

\begin{figure*}[t]
\centering
\includegraphics[width=0.8\linewidth]{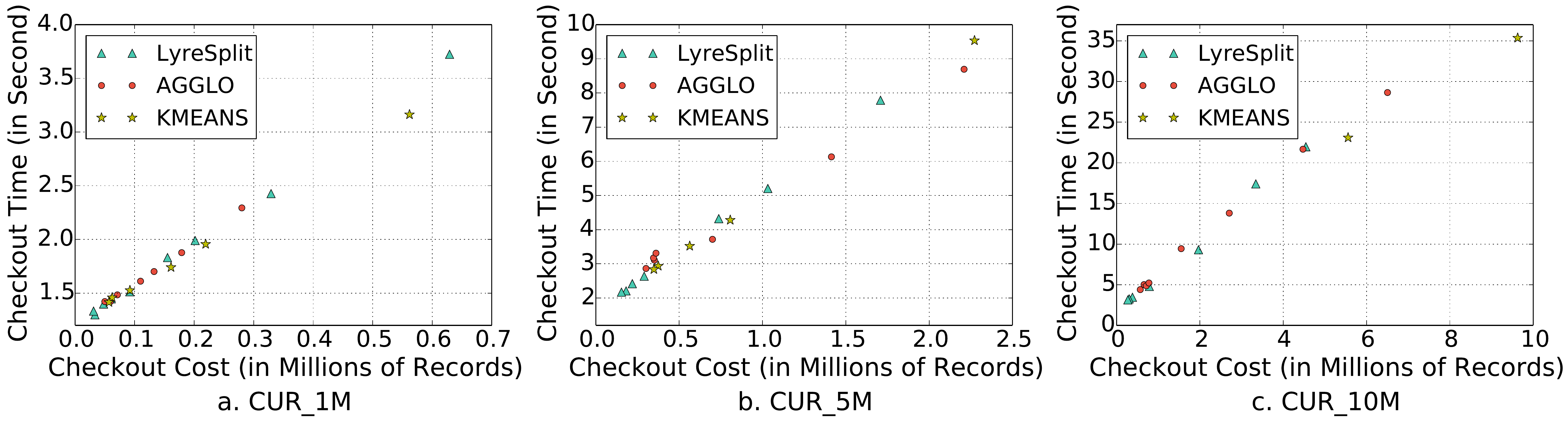}
\caption{Estimated Checkout Cost vs. Real Checkout Time (\code{CUR\_*})}
\label{fig:real_est_cur}
\end{figure*}

\subsection{Estimated Storage Cost and Checkout Cost}

In the main body of the paper, we performed
an experimental evaluation of 
the trade-off between the actual checkout time
and the storage size in Figure~\ref{fig:real_cur}.
One question we may have is whether our cost model
has a similar trade-off.
In Figure~\ref{fig:est}, we report
the estimated checkout cost versus the estimated
storage cost according to our model.
 We can see that the trend in Figure~\ref{fig:est} 
 is very similar to that in Figure~\ref{fig:real_cur}. 
 However, the absolute reduction on checkout cost 
 in Figure~\ref{fig:est} is typically greater 
 than that in Figure~\ref{fig:real_cur}. 
 This is because we do not count the constant overhead 
 in the estimated checkout cost. 
 We again check the correctness of our checkout 
 cost model by comparing the estimated checkout 
 cost with the actual checkout time.
 We present the result in Figure~\ref{fig:real_est_sci} and \ref{fig:real_est_cur}.
 We can see that the points in Figure~\ref{fig:real_est_sci} and \ref{fig:real_est_cur}(a)(b)(c) 
 roughly form a straight line, 
 validating that the checkout time is linearly correlated to our cost model. 
 Hence, we conclude that our cost modeling and problem formulation are accurate.